\documentclass[pdflatex,sn-chicago]{sn-jnl}

\usepackage{graphicx}%
\usepackage{multirow}%
\usepackage{amsmath,amssymb,amsfonts}%
\usepackage{amsthm}%
\usepackage{mathrsfs}%
\usepackage[title]{appendix}%
\usepackage{xcolor}%
\usepackage{textcomp}%
\usepackage{manyfoot}%
\usepackage{booktabs}%
\usepackage{algorithm}%
\usepackage{algorithmicx}%
\usepackage{algpseudocode}%
\usepackage{listings}%
\usepackage{caption}
\usepackage{subcaption}
\usepackage{bbm}

\newtheorem{theorem}{Theorem}
\newtheorem{proposition}[theorem]{Proposition}%
\newtheorem{remark}{Remark}%

\DeclareMathOperator*{\argmin}{arg\,min} 
\DeclareMathOperator*{\argmax}{arg\,max}
\DeclareMathOperator{\tr}{tr}

\definecolor{DarkRed}{rgb}{0.5,0,0}
\definecolor{DarkBlue}{rgb}{0,0,.545}
\hypersetup{
    colorlinks=true,
    linkcolor=DarkBlue,
    filecolor=DarkBlue,      
    urlcolor=DarkRed,
    citecolor=DarkBlue,
}

\usepackage{multibib}
\newcites{Main}{References}
\newcites{Supp}{References}

\begin{document}

\title[Article Title]{BOB: Bayesian Optimized Bootstrap for Uncertainty Quantification in Gaussian Mixture Models}

\author[1]{\fnm{Santiago} \sur{Marin}}

\author[1]{\fnm{Bronwyn} \sur{Loong}}

\author[1,2]{\fnm{Anton H.} \sur{Westveld}}

\affil[1]{\orgdiv{{Research School of Finance,
Actuarial Studies and Statistics}}, \orgname{{The Australian National University}}, \orgaddress{\city{{Canberra}}, \state{{ACT}}, \country{{Australia}}}}

\affil[2]{\orgdiv{Department of Statistical Sciences and Operations Research}, \orgname{Virginia Commonwealth University}, \orgaddress{\city{Richmond}, \state{VA}, \country{USA}}}

\abstract{A natural way to quantify uncertainties in Gaussian mixture models (GMMs) is through Bayesian methods. That said, sampling from the joint posterior distribution of GMMs via standard Markov chain Monte Carlo (MCMC) imposes several computational challenges, which have prevented a broader full Bayesian implementation of these models. A growing body of literature has introduced the Weighted Likelihood Bootstrap and the Weighted Bayesian Bootstrap as alternatives to MCMC sampling. The core idea of these methods is to repeatedly compute maximum \textit{a posteriori} (MAP) estimates on many randomly weighted posterior densities. These MAP estimates then can be treated as approximate posterior draws. Nonetheless, a central question remains unanswered: How to select the random weights under arbitrary sample sizes. We, therefore, introduce the Bayesian Optimized Bootstrap (BOB), a computational method to automatically select these random weights by minimizing, through Bayesian Optimization, a black-box and noisy version of the reverse Kullback–Leibler (KL) divergence between the Bayesian posterior and an approximate posterior obtained via random weighting. Our proposed method outperforms competing approaches in recovering the Bayesian posterior, it provides a better uncertainty quantification, and it retains key asymptotic properties from existing methods. BOB’s performance is demonstrated through extensive simulations, along with real-world data analyses.}


\maketitle

\section{Introduction}\label{sec:intro}

Gaussian Mixture Models (GMMs) are powerful and flexible tools, which turn up naturally when the population of sampling units consists of homogeneous clusters or subgroups \citep{van2015overfitting}, and can be applied in a wide range of scientific problems including model-based clustering \citep{reftery_model_clust}, density estimation \citep{izenman1988}, or as flexible semi-parametric model-based approaches for analyzing complex data \citep{OMORI2007425}. It is no surprise, then, that uncertainty quantification to infer properties or make predictions about a population of interest based on mixture models is now a crucial and important area of statistical research \citep{s_wade_cluster_inference}, and a natural way to quantify uncertainty in mixture models is through Bayesian methods by specifying a sampling model and a prior distribution \citep{ni2020scalable}.
\par
\textbf{Motivation}. Despite the flexibility and wide applicability of GMMs, the large computational costs of Bayesian analysis have prevented a broader full Bayesian implementation of these models. For instance, one could sample from the posterior distribution of GMMs via standard Markov chain Monte Carlo (MCMC). However, since the log-likelihood function is non-concave, the resulting posterior might be multimodal. The multimodality of the posterior, combined with the serial nature of MCMC algorithms, results in a sampler that might get trapped in local regions of high posterior density, failing to explore the entire parameter space and leading to mixing limitations between posterior modes \citep{celeux2000computational, fong2019scalable}. Moreover, in the context of mixture models, MCMC algorithms scale poorly to large datasets \citep{ni2020scalable} and their serial nature has prevented the adoption of recent developments in parallel computing \citep{newtonBoots, pompe2021introducing}, exacerbating even further the computational limitations of these methods. Consequently, new and refined posterior samplers are still required.
\par
\textbf{Related Work}. This research joins a growing body of literature, dating back to \citeyear{raftery_boostrap_likelihood} when \citeauthor{raftery_boostrap_likelihood} proposed the Weighted Likelihood Bootstrap (WLB) as an alternative to MCMC. The main idea behind WLB is to generate approximate posterior draws by independently maximizing a series of randomly weighted log-likelihood functions. WLB, however, does not naturally incorporate any prior information in the sampling procedure. With this in mind, researchers have been recently extending the WLB framework to accommodate prior information by weighting not only the likelihood but also the prior. A few examples include the Weighted Bayesian Bootstrap, or WBB \citep{newtonBoots, ng2022random} and the Bayesian Bootstrap spike-and-slab Lasso \citep{rockova2022}. In the context of generalized Bayesian analysis and model misspecification, we can find the loss-likelihood Bootstrap \citep{lyddon2019general}, the Posterior Bootstrap \citep{fong2019scalable, pompe2021introducing} and the Deep Bayesian Bootstrap \citep{deep_boots}.
\par
\textbf{Our Contribution}. In this paper, we first illustrate how to carry out a Bayesian implementation of GMMs within a WBB framework by developing an algorithm to optimize a randomly weighted and non-concave posterior distribution associated with a GMM. Secondly, we address the problem of selecting adequate random weights in order to obtain better posterior approximations. Up until now, one key question has not been addressed: \textit{how to select the random weights under arbitrary---especially under fixed---sample sizes}. Few analyses have been proposed from the lenses of asymptotic theory but, to the best of our knowledge, the core of this question remains unanswered. In fact, as discussed by \cite{raftery_boostrap_likelihood} and \cite{deep_boots}, ``no general recipe (to select the random weights) yet exists." Thus, in this work we develop the Bayesian Optimized Bootstrap (BOB), a computational approach to automatically select the random weights by minimizing a black-box and noisy version of the reverse Kullback–Leibler (KL) divergence between the Bayesian posterior and an approximate posterior obtained via random weighting. Since this divergence is expensive to evaluate and we do not have access to a closed-form expression for the divergence or its derivatives, the minimization is carried out via Bayesian Optimization, or BO (and thus the name BOB), as BO is one of the most efficient methods for optimizing a black-box objective function, requiring only a few evaluations of the objective itself \citep{jones1998efficient, jones2001taxonomy}. We show that BOB leads to a better approximation of the Bayesian posterior, it allows for uncertainty quantification, and, unlike standard MCMC, it is able to adopt recent developments in parallel computing.
\par
\textbf{Article Outline.} The rest of this article is organized as follows. Section \ref{sec:RWBB_EM} revisits WBB and describes how a Bayesian implementation of GMMs can be carried out within a WBB framework. In Section \ref{sec:BOB} we formally introduce BOB and draw connections with existing Bootstrap-based posterior samplers. Simulation exercises are carried out in Section \ref{sec:sims}. In section \ref{sec:data_analysis}, we demonstrate the applicability of BOB on real-world data. We conclude with a discussion in Section \ref{sec:conc}.

\section{Revisiting Weighted Bayesian Bootstrap}
\label{sec:RWBB_EM}

\subsection{Problem Setup}
\label{subsec:Model_setup}

For our sampling model, let $\left\{\mathbf{y}_{1},\dots,\mathbf{y}_{n}\right\}$ be a random sample from a Gaussian Mixture with $K$ components. More precisely, let the density of $\mathbf{y}_{i}\in\mathbb{R}^{d}$, for $i\in\{1\,\dots,n\}$, be
\begin{equation}
    \label{model}
    p\bigl(\mathbf{y}_{i}\bigl|\left\{\pi_{k}, \boldsymbol{\mu}_{k}, \boldsymbol{\Sigma}_{k}\right\}_{k=1}^{K}\bigr) = \sum_{k=1}^{K}\pi_{k}\varphi\left(\mathbf{y}_{i};\, \boldsymbol{\mu}_{k}, \boldsymbol{\Sigma}_{k}\right),
\end{equation}
where $\pi_{k}\in[0,1]$ is the weight of component $k$ so that $\sum_{k=1}^{K}\pi_{k}=1$, $K\geq2$ is a known integer, $\varphi(\cdot\,;\,\boldsymbol{\mu}_{k},\boldsymbol{\Sigma}_{k})$ denotes the Gaussian density with mean vector $\boldsymbol{\mu}_{k}\in\mathbb{R}^{d}$ and covariance matrix $\boldsymbol{\Sigma}_{k}\in\mathbb{M}_{+}^{d}$, and $\mathbb{M}_{+}^{d}$ denotes the set of  symmetric positive semidefinite matrices of size $d\times d$. Additionally, let us define the latent indicator variables $z_{ik} = \mathbbm{1}\{i\text{-th obs. is drawn from the }k\text{-th component}\}$ so that $\mathbf{z}_{i} = (z_{i1},\dots,z_{iK})'\sim\mathcal{M}(1; \left\{\pi_{k}\right\}_{k=1}^{K})$, where $\mathcal{M}$ denotes the multinomial distribution. Then, the joint distribution of $\mathbf{Y}=(\mathbf{y}_{1}',\dots,\mathbf{y}_{n}')'\in\mathbb{R}^{n\times d}$ and $\mathbf{Z}=(\mathbf{z}_{1}',\dots,\mathbf{z}_{n}')'\in\{0,1\}^{n\times K}$ would be given by
\begin{equation}
    p\bigl(\mathbf{Y},\mathbf{Z}\bigl|\left\{\pi_{k}, \boldsymbol{\mu}_{k}, \boldsymbol{\Sigma}_{k}\right\}_{k=1}^{K}\bigr)=\prod_{i=1}^{n}\prod_{k=1}^{K}\left(\pi_{k}\varphi\left(\mathbf{y}_{i};\,\boldsymbol{\mu}_{k},\boldsymbol{\Sigma}_{k}\right)\right)^{z_{ik}}.
\end{equation}
\par
For our prior specification, let $\boldsymbol{\mu}_{k}|\boldsymbol{\Sigma}_{k}\sim\mathcal{N}(\boldsymbol{\beta}_{k},\,\boldsymbol{\Sigma}_{k}/\lambda_{k})$, $\boldsymbol{\Sigma}_{k}\sim\mathcal{IW}(\nu_{k},\,\boldsymbol{\Psi}_{k}^{-1})$, and $\boldsymbol{\pi}\sim\mathcal{D}(a_{1},\dots,a_{K})$,
which corresponds to a normal-inverse-Wishart prior for $\boldsymbol{\mu}_{k}$ and $\boldsymbol{\Sigma}_{k}$ with scale matrix $\boldsymbol{\Psi}_{k}$, and a Dirichlet prior for the mixture proportions $\boldsymbol{\pi}=(\pi_{1},\dots,\pi_{K})'$. Additionally, let $\boldsymbol{\theta}=\left\{\pi_{k}, \boldsymbol{\mu}_{k}, \boldsymbol{\Sigma}_{k}\right\}_{k=1}^{K}\in\boldsymbol{\Theta}$ be the collection of all model parameters so that, under the sampling distribution and priors introduced above, the joint Bayesian posterior would be, up to a proportionality constant, given by
\begin{equation}
    \label{eq:posterior}
    p(\boldsymbol{\theta}, \mathbf{Z}|\mathbf{Y}) \propto p(\boldsymbol{\pi})\prod_{k=1}^{K}\left\{p(\boldsymbol{\Sigma}_{k})p(\boldsymbol{\mu}_{k}|\boldsymbol{\Sigma}_{k})\prod_{i=1}^{n}p(\mathbf{y}_{i}|\boldsymbol{\mu}_{k},\boldsymbol{\Sigma}_{k},z_{ik})p(z_{ik}|\pi_{k})\right\}.
\end{equation}

\subsection{Posterior Sampling via Random Weighting}
\label{subsec:random_weighting_method}

We now illustrate how one can approximately sample from the joint posterior in \eqref{eq:posterior} via WBB. Following \cite{newtonBoots}, WBB can be summarized in three simple steps: 
\par
\textbf{Step 1:} Start by sampling non-negative random weights $\mathbf{u}=(u_{1},\dots,u_{n})'$ from some weight distribution $F_{u}$ and construct the re-weighted log-density of $\mathbf{Y}$ and $\mathbf{Z}$ as
\begin{equation}
    \label{step1_wbb}
    \log p_{u}\left(\mathbf{Y},\mathbf{Z}|\boldsymbol{\theta}\right) = \sum_{i=1}^{n}u_{i}\left\{\sum_{k=1}^{K}z_{ik}\left[\log\pi_{k} + \log\varphi(\mathbf{y}_{i};\,\boldsymbol{\mu}_{k},\boldsymbol{\Sigma}_{k})\right]\right\}.
\end{equation}
\par
\textbf{Step 2:} Sample $\Tilde{\mathbf{u}}=(\Tilde{u}_{\pi},\Tilde{u}_{\mu_{1}},\dots,\Tilde{u}_{\mu_{K}}, \Tilde{u}_{\Sigma_{1}},\dots,\Tilde{u}_{\Sigma_{K}})'$ from some weight distribution $F_{\Tilde{u}}$ and construct the re-weighted log-prior density as
\begin{equation}
    \begin{gathered}
    \log p_{\Tilde{u}}(\boldsymbol{\theta}) \propto 
    \sum_{k=1}^{K}\Biggl\{\Tilde{u}_{\pi}\left(a_{k}-1\right)\log\pi_{k} - \Tilde{u}_{\Sigma_{k}}\left[\left(\frac{\nu_{k} +d}{2} + 1\right)\log|\boldsymbol{\Sigma}_{k}| + \frac{\tr\left(\boldsymbol{\Psi}_{k}\boldsymbol{\Sigma}_{k}^{-1}\right)}{2}\right] \\
     - \Tilde{u}_{\mu_{k}}\left[\frac{\lambda_{k}}{2}(\boldsymbol{\mu}_{k} - \boldsymbol{\beta}_{k})'\boldsymbol{\Sigma}_{k}^{-1}(\boldsymbol{\mu}_{k} - \boldsymbol{\beta}_{k}) \right]\Biggr\}.
    \end{gathered}
    \label{step2_wbb}
\end{equation}
\par
\textbf{Step 3:} By marginalizing $\mathbf{Z}$ out of equation (\ref{step1_wbb}), one would have that 
$\log p_{u}\left(\mathbf{Y}\left|\boldsymbol{\theta}\right.\right) = 
\log\left(\sum_{Z}p_{u}\left(\mathbf{Y},\mathbf{Z}\left|\boldsymbol{\theta}\right.\right)\right)$
and the re-weighted log-posterior of $\boldsymbol{\theta}$ would be, up to an additive constant, given by $\log p_{u}(\boldsymbol{\theta}|\mathbf{Y})\propto \log p_{u}\left(\mathbf{Y}\left|\boldsymbol{\theta}\right.\right)  +  \log p_{\Tilde{u}}(\boldsymbol{\theta})$. WBB proceeds to maximize this posterior density seeking
\begin{equation}
    \label{max_step_wbb}
    \boldsymbol{\theta}^{*} = \argmax_{\boldsymbol{\theta}\in\boldsymbol{\Theta}}\left\{\log p_{u}(\boldsymbol{\theta}|\mathbf{Y})\right\}.
\end{equation}
\par
Under a traditional Bayesian framework, the randomness in $\boldsymbol{\theta}$ arises from treating it as a random variable with a prior distribution. On the other hand, as discussed in \cite{deep_boots}, $\boldsymbol{\theta}^{*}$ can be seen as an \textit{estimator}, where, given the data, the only source of randomness comes from the random weights. Thus, by fixing the data $\mathbf{Y}$ and repeating steps 1 - 3 many times, one would obtain approximate posterior draws $\{\boldsymbol{\theta}^{*}_{(s)}\}_{s\in\mathbb{N}}$. This idea is attractive as optimizing can be easier and less computationally intensive than sampling from an intractable posterior. Note, additionally, that the random weights are independent across iterations. Consequently, $\boldsymbol{\theta}^{*}_{(s)}$ and $\boldsymbol{\theta}^{*}_{(s')}$, for $s\neq s'$, are also independent and could be sampled in parallel. Moreover, Bootstrap-based posterior samplers do not require costly tuning runs, \textit{burn-in} periods, or convergence diagnostics \citep{fong2019scalable, pompe2021introducing}, making random weighting an attractive alternative to MCMC sampling.
\par
Nonetheless, performing the joint optimization in (\ref{max_step_wbb}) is still a challenge on its own, especially in light of the non-concavity of $\log p_{u}(\boldsymbol{\theta}|\mathbf{Y})$. Thus, we treat the latent indicator variables, $\mathbf{Z}$, as missing data and develop a randomly weighted expectation-maximization (EM) algorithm \citep{dempster1977EM} to solve (\ref{max_step_wbb}).

\subsection{Randomly Weighted Expectation-Maximization}
\label{subsec:EM}

\subsubsection{Expectation Step}
\label{subsubsec:E_step}

We start by computing the expected value of $\mathbf{Z}$, conditional on $\mathbf{Y}$, $\mathbf{u}$ and $\boldsymbol{\theta}^{(t)}$, where $\boldsymbol{\theta}^{(t)}$ is the current value of $\boldsymbol{\theta}$ within the EM algorithm. By Bayes' rule, we have that 
\begin{equation}
    \begin{split}
        \label{e_step_raw}
        \mathbb{E}\left[z_{ik}|\boldsymbol{\theta}^{(t)},\mathbf{Y},\mathbf{u}\right] & = 
        \frac{\exp\left\{u_{i}\left[\log\mathbb{P}\left(z_{ik}=1|\pi_{k}^{(t)}\right) + \log\varphi\left(\mathbf{y}_{i};\,\boldsymbol{\mu}_{k}^{(t)},\boldsymbol{\Sigma}_{k}^{{(t)}}\right)\right]\right\}}{\sum_{r=1}^{K}\exp\left\{u_{i}\left[\log\mathbb{P}\left(z_{ir}=1|\pi_{r}^{(t)}\right) + \log\varphi\left(\mathbf{y}_{i};\,\boldsymbol{\mu}_{r}^{(t)},\boldsymbol{\Sigma}_{r}^{{(t)}}\right)\right]\right\}}\\
        & = \frac{\left[\pi_{k}^{(t)}\varphi\left(\mathbf{y}_{i};\,\boldsymbol{\mu}_{k}^{(t)},\boldsymbol{\Sigma}_{k}^{{(t)}}\right)\right]^{u_{i}}}{\sum_{r=1}^{K}\left[\pi_{r}^{(t)}\varphi\left(\mathbf{y}_{i};\,\boldsymbol{\mu}_{r}^{(t)},\boldsymbol{\Sigma}_{r}^{{(t)}}\right)\right]^{u_{i}}} = q_{ik}.
    \end{split}
\end{equation}

\subsubsection{Maximization Step}
\label{subsubsec:M_step}

In the maximization step, we then maximize the following surrogate objective function with respect to $\boldsymbol{\theta}\in\boldsymbol{\Theta}$.
\begin{equation}
    \begin{gathered}
        \mathbb{E}\left[\log p_{u}(\boldsymbol{\theta},\mathbf{Z}|\mathbf{Y})|\boldsymbol{\theta}^{(t)}\right] = -\frac{1}{2}\sum_{i=1}^{n}u_{i}\left\{\sum_{k=1}^{K}q_{ik}\left[\log|\boldsymbol{\Sigma}_{k}| + \left(\mathbf{y}_{i} - \boldsymbol{\mu}_{k}\right)'\boldsymbol{\Sigma}_{k}^{-1}\left(\mathbf{y}_{i} - \boldsymbol{\mu}_{k}\right)\right]\right\}\\
    -\sum_{k=1}^{K}\Biggl\{\Tilde{u}_{\mu_{k}}\left[\frac{\lambda_{k}}{2}(\boldsymbol{\mu}_{k}- \boldsymbol{\beta}_{k})'\boldsymbol{\Sigma}_{k}^{-1}(\boldsymbol{\mu}_{k}- \boldsymbol{\beta}_{k})\right] + \Tilde{u}_{\Sigma_{k}}\left[\left(\frac{\nu_{k}+d}{2}+1\right)\log|\boldsymbol{\Sigma}_{k}| + \frac{\tr\left(\boldsymbol{\Psi}_{k}\boldsymbol{\Sigma}^{-1}_{k} \right)}{2}\right] \\
    - \left[\Tilde{a}_{k} - 1 + \sum_{i=1}^{n}u_{i}q_{ik}\right]\log\pi_{k} \Biggr\},
    \end{gathered}
  \label{surrogate_objective}
\end{equation}
where $\Tilde{a}_{k} = (a_{k} - 1)\Tilde{u}_{\pi} + 1$. As we will illustrate shortly, the objective in \eqref{surrogate_objective} leads to well-known maximization problems, which can be solved relatively cheaply.
\par
\begin{proposition}\label{Prop_joint_mu_Omega}
Let us define $\Tilde{n}_{k} = \sum_{i=1}^{n}u_{i}q_{ik}$, $\Tilde{\lambda}_{k} = \Tilde{u}_{\mu_{k}}\lambda_{k}$, $\Tilde{\boldsymbol{\Psi}}_{k} = \Tilde{u}_{\Sigma_{k}}\boldsymbol{\Psi}_{k}$, $\Tilde{\nu}_{k} = \Tilde{u}_{\Sigma_{k}}(\nu_{k} + d + 2) -2 -d$, $\Tilde{\Bar{\mathbf{y}}}_{k} = (\Tilde{n}_{k})^{-1}\sum_{i=1}^{n}(u_{i}q_{ik})\mathbf{y}_{i}$ and $\Tilde{\mathbf{S}}_{k} = \sum_{i=1}^{n}(u_{i}q_{ik})(\mathbf{y}_{i} - \Tilde{\Bar{\mathbf{y}}}_{k})(\mathbf{y}_{i} - \Tilde{\Bar{\mathbf{y}}}_{k})'$, for all $k\in\{1,\dots,K\}$. Additionally, let $\Bar{\boldsymbol{\beta}}_{k} = \Tilde{\lambda}_{k}/(\Tilde{\lambda}_{k} + \Tilde{n}_{k})\boldsymbol{\beta}_{k} + \Tilde{n}_{k}/(\Tilde{\lambda}_{k} + \Tilde{n}_{k})\Tilde{\Bar{\mathbf{y}}}_{k}$, $\Bar{\boldsymbol{\Psi}}_{k} = (\Tilde{\boldsymbol{\Psi}}_{k} + \Tilde{\mathbf{S}}_{k} + \Tilde{\lambda}_{k}\Tilde{n}_{k}/(\Tilde{\lambda}_{k} + \Tilde{n}_{k})(\Tilde{\Bar{\mathbf{y}}}_{k} - \boldsymbol{\beta}_{k})(\Tilde{\Bar{\mathbf{y}}}_{k}- \boldsymbol{\beta}_{k})')$, $\Bar{\lambda}_{k} = \Tilde{\lambda}_{k} + \Tilde{n}_{k}$, and $\Bar{\nu}_{k} = \Tilde{\nu}_{k} + \Tilde{n}_{k}$. Then, the update of $\boldsymbol{\mu}_{k}$ and $\boldsymbol{\Sigma}_{k}$ would be given by
\begin{gather}
    \label{joint_mu_omega}
    \bigl(\boldsymbol{\mu}_{k}^{(t+1)}, \boldsymbol{\Sigma}_{k}^{(t+1)}\bigr) = \argmax_{\substack{\boldsymbol{\mu}_{k},\, \boldsymbol{\Sigma}_{k}}}\bigl\{h(\boldsymbol{\mu}_{k}, \boldsymbol{\Sigma}_{k}) \bigr\},
\end{gather}
where 
\begin{equation*}
    h(\boldsymbol{\mu}_{k}, \boldsymbol{\Sigma}_{k}) = -\left(\frac{\Bar{\nu}_{k}+d}{2}+1\right)\log|\boldsymbol{\Sigma}_{k}| - \frac{1}{2}\tr\left(\Bar{\boldsymbol{\Psi}}_{k}\boldsymbol{\Sigma}_{k}^{-1}\right) -\frac{\Bar{\lambda}_{k}}{2}(\boldsymbol{\mu}_{k} - \Bar{\boldsymbol{\beta}}_{k})'\boldsymbol{\Sigma}_{k}^{-1}(\boldsymbol{\mu}_{k} - \Bar{\boldsymbol{\beta}}_{k}).
\end{equation*}
\end{proposition}
Details on the derivation of $h:\mathbb{R}^{d}\times\mathbb{M}_{+}^{d}\rightarrow\mathbb{R}$ are presented in the supplementary materials. To optimize $h$, we make use of a two-stage procedure in which we first update $\boldsymbol{\Sigma}_{k}$ and then update $\boldsymbol{\mu}_{k}$.
\par
\textbf{Update $\boldsymbol{\Sigma}_{k}$:} Start by noting that $\boldsymbol{\Sigma}_{k}\mapsto\log\int_{\mathbb{R}^{d}}\exp(h(\boldsymbol{\mu}_{k}, \boldsymbol{\Sigma}_{k}))d\boldsymbol{\mu}_{k}$ is a monotone transformation of $\boldsymbol{\Sigma}_{k}\mapsto h(\boldsymbol{\mu}_{k}, \boldsymbol{\Sigma}_{k})$ \cite[Ch. 9]{schilling2017measures}. Thus, optimizing (\ref{joint_mu_omega}) with respect to $\boldsymbol{\Sigma}_{k}$ yields
\begin{equation}
    \label{precision_update}
    \boldsymbol{\Sigma}_{k}^{(t+1)} = \argmax_{\boldsymbol{\Sigma}_{k}\in\mathbb{M}_{+}^{d}}\left\{
    -\frac{\Bar{\nu}_{k} + d + 1}{2}\log|\boldsymbol{\Sigma}_{k}| - \frac{1}{2}\tr\left(\Bar{\boldsymbol{\Psi}}_{k}\boldsymbol{\Sigma}_{k}^{-1}\right)\right\}.
\end{equation}
One can identify the solution to \eqref{precision_update} as the mode of an $\mathcal{IW}(\Bar{\nu}_{k},\, \Bar{\boldsymbol{\Psi}}_{k} ^ {-1})$ distribution. This is,
\begin{equation}
    \boldsymbol{\Sigma}_{k}^{(t+1)} = \frac{\Bar{\boldsymbol{\Psi}}_{k}}{\Bar{\nu}_{k} + d + 1}.
\end{equation}
\par
\textbf{Update $\boldsymbol{\mu}_{k}$:} Again, note that $\boldsymbol{\mu}_{k}\mapsto\int_{\mathbb{M}_{+}^{d}}\exp(h(\boldsymbol{\mu}_{k},\boldsymbol{\Sigma}_{k}))d\boldsymbol{\Sigma}_{k}$ is a monotone transformation of $\boldsymbol{\mu_{k}}\mapsto h(\boldsymbol{\mu}_{k},\boldsymbol{\Sigma}_{k})$, so optimizing (\ref{joint_mu_omega}) with respect to $\boldsymbol{\mu}_{k}$ reduces to
\begin{equation}
    \label{ridge_update}
    \boldsymbol{\mu}_{k}^{(t+1)} = \argmax_{\boldsymbol{\mu}_{k}\in\mathbb{R}^{d}}\left\{ 
    \left[1 + \Bar{\lambda}_{k}(\boldsymbol{\mu}_{k} - \Bar{\boldsymbol{\beta}}_{k})'\Bar{\boldsymbol{\Psi}}_{k}^{-1}(\boldsymbol{\mu}_{k} - \Bar{\boldsymbol{\beta}}_{k}) \right]^{-(\Bar{\nu}_{k}+1)/2}\right\},
\end{equation}
which corresponds to the mode of a $t_{(\Bar{\nu}_{k}-d+1)}(\Bar{\boldsymbol{\beta}}_{k},\,\Bar{\boldsymbol{\Psi}}_{k}/(\Bar{\lambda}_{k}(\Bar{\nu}_{k}-d+1)))$ distribution, where $t_{\nu}(\mathbf{m}, \boldsymbol{\Lambda})$ denotes the \textit{t}-distribution with $\nu$ degrees of freedom, location vector $\mathbf{m}$ and scale matrix $\boldsymbol{\Lambda}$ \cite[Ch. 2]{bishop2006pattern}. More precisely, the update of $\boldsymbol{\mu}_{k}$ is given by $\boldsymbol{\mu}_{k}^{(t+1)} = \Bar{\boldsymbol{\beta}}_{k}$.

\textbf{Update $\boldsymbol{\pi}$:} Lastly, note from \eqref{surrogate_objective} that the update of $\boldsymbol{\pi}$ corresponds to the mode of a $\mathcal{D}\left(\Bar{a}_{1},\dots,\Bar{a}_{K}\right)$ distribution, where $\Bar{a}_{k} = \Tilde{a}_{k}+\Tilde{n}_{k}$. Namely, for $k\in\{1,\dots,K\}$,
\begin{equation}
    \pi_{k}^{(t+1)} = \frac{\Bar{a}_{k} -1}{\left( \sum_{r=1}^{K}\Bar{a}_{r} \right) - K}.
\end{equation}

Our EM algorithm, then, iterates between the E and the M steps until convergence.

\subsubsection{Dealing with Suboptimal Modes: Tempered EM}
\label{subsubsec:Tempering}

Despite the simplicity of our EM algorithm, a well-known drawback of non-convex optimization methods is that they might converge to a local optima (i.e., a suboptimal mode). Random restarts (RRs) have been widely used to increase the parameter space exploration and escape suboptimal modes. However, this approach is too computationally intensive and yet, does not guarantee that one would reach the global mode.
\par
Tempering and annealing \citep{kirkpatrick1983optimization, Sambridge_2014_Parallel}, on the other hand, are optimization techniques which also increase the parameter space exploration at a lower computational cost.
More precisely, let $\left\{T_{t}\right\}_{t\in\mathbb{N}}$ be a \textit{tempering profile},
i.e., a sequence of positive numbers such that $\lim_{t\rightarrow\infty}T_{t}=1$, where $t$ is the $t$-th iteration of the EM algorithm. Then, a tempered EM algorithm modifies the E step in equation \eqref{e_step_raw} as $\Tilde{q}_{ik,t} = q_{ik}^{1/T_t}/(\sum_{r=1}^{K}q_{ir}^{1/T_t})$,
where larger values of $T_{t}$ allow the algorithm to explore more of the target posterior by flattening the distribution of $\Tilde{q}_{ik,t}$, and as $t\rightarrow\infty$ one would recover the original objective function, progressively attracting the solution towards the global mode \citep{allassonniere2021temp, lartigue2022deterministic}.
Thus, we let
\begin{equation}
    T_{t} = 1 + a^{\tau} + b\frac{\sin{(\tau)}}{\tau},\qquad\text{where}\qquad\tau=\frac{t+cr}{r},\;\;t\in\mathbb{N},
\end{equation}
with $a\in[0,1)$, $b\in\mathbb{R}$, and $c,r > 0$, which corresponds to the oscillatory tempering profile with gradually decreasing amplitudes from \cite{allassonniere2021temp}. To select a suitable combination of $a,b,c$ and $r$, we use a grid-search over a range of possible values, choose the combination that yields the largest objective using an unweighted EM algorithm, and fix such a solution throughout the entire sampler. As discussed in \cite{lartigue2022deterministic}, the tempering hyper-parameters can remain fixed across different optimization problems with similar characteristics, as in the case of randomly weighted EM.

\section{Introducing BOB}
\label{sec:BOB}

So far, we have developed and presented a randomly weighted EM algorithm to approximately sample from the posterior distribution of GMMs within a WBB framework, but we have not yet answered a key and important question: \textit{How to select the random weights}. \cite{raftery_boostrap_likelihood} showed that under low-dimensional settings and assuming the square of Jeffreys prior, uniform Dirichlet weights for the likelihood yield approximate posterior draws that are first order correct, i.e., consistent (they tend to concentrate around a small neighbourhood of the maximum likelihood estimator---MLE) and asymptotically normal. More recently, \cite{ng2022random} established first order correctness and model selection consistency for a wide range of weight distributions in linear models with Lasso priors, and \cite{rockova2022} showed that, for a number of weight distributions, the approximate posterior from Bayesian Bootstrap spike-and-slab Lasso concentrates at the same rate as the actual Bayesian posterior. 
Note, however, that these results are all based on the assumption that $n\rightarrow\infty$. Considering the case of a fixed $n$ is important because, under a small to medium sample size, the prior, $p(\boldsymbol{\theta})$, would have more influence on the posterior, and as the effect of $p(\boldsymbol{\theta})$ gets bigger, the relationship between $p_{u}(\boldsymbol{\theta}|\mathbf{Y})$ and the random weights becomes less clear \citep{deep_boots}, making the choice of adequate weights a much more important and difficult task. Thus, we propose BOB as an alternative methodology for automatically selecting the random weights under arbitrary sample sizes.
\par
Before illustrating our proposed method, though, let us present a brief summary of Variational Bayes (VB) methods, as we draw inspiration from them. VB aims to obtain an approximation, $g_{_{VB}}(\boldsymbol{\theta}|\hat{\boldsymbol{\nu}})$, with variational parameters $\hat{\boldsymbol{\nu}}\in\boldsymbol{\Theta}$ such that $\hat{\boldsymbol{\nu}}= \argmin_{\boldsymbol{\nu}\in\boldsymbol{\Theta}}\mathbb{E}_{g_{_{VB}}(\boldsymbol{\theta}|{\boldsymbol{\nu}})}[\log g_{_{VB}}(\boldsymbol{\theta}|{\boldsymbol{\nu}}) - \log p(\boldsymbol{\theta}|\mathbf{Y})]$, i.e., VB aims to minimize the reverse KL divergence between the Bayesian posterior and a variational approximation \citep{blei2017variational}. 
\par
With this in mind, we propose the following weighting scheme: Draw the likelihood weights as $u_{i} = \frac{w_{i}^{\alpha}}{\sum_{i=1}^{n}w_{i}^{\alpha}}$, with $w_{i}\sim\mathcal{E}(1)$ and $\alpha\sim\delta_{\alpha}(x_{\alpha})$, where $\delta_{\alpha}(x_{\alpha})$ denotes the Dirac measure so that $\mathbb{P}(\alpha = x_{\alpha}) = 1$, $x_{\alpha}\geq1$, and $\mathcal{E}(1)$ denotes the exponential distribution with mean 1. If $x_{\alpha}=1$, the distribution for the likelihood weights would be the uniform Dirichlet, while $x_{\alpha}>1$ would yield an overdispersed distribution \citep{raftery_boostrap_likelihood, gelman1992inference}. For the weights associated with the prior, we set, for $k\in\{1,\dots,K\}$, $\Tilde{u}_{\mu_{k}}\sim\delta_{\Tilde{u}_{\mu_{k}}}(x_{\mu_{k}})$, $\Tilde{u}_{_{\Sigma_{k}}}\sim\delta_{\Tilde{u}_{\Sigma_{k}}}(x_{_{\Sigma_{k}}})$, and $\Tilde{u}_{\pi}\sim\delta_{\Tilde{u}_{\pi}}(x_{\pi})$, where $x_{\mu_{k}}, x_{_{\Sigma_{k}}}, x_{\pi} \geq 0$. Hence, our problem is reduced to finding appropriate values for $\boldsymbol{x}=(x_{\alpha}, x_{\mu_{1}}, \dots, x_{\mu_{K}}, x_{_{\Sigma_{1}}}, \dots, x_{_{\Sigma_{K}}}, x_{\pi})'\in\boldsymbol{\mathcal{X}}\subset\mathbb{R}^{2(K+1)}$, where $\boldsymbol{\mathcal{X}}$ is a compact search space. Inspired by VB, we propose to select the optimal $\boldsymbol{x}$ as
\begin{equation}
    \label{kl_div_optim}
    \boldsymbol{x}^{*} = \argmin_{\boldsymbol{x}\in\boldsymbol{\mathcal{X}}}\left\{\mathcal{L}(\boldsymbol{x})\right\},
\end{equation}
with
\begin{equation*}
    \begin{split}
        \mathcal{L}(\boldsymbol{x}) & = \int_{\boldsymbol{\Theta}}g_{{u}}(\boldsymbol{\theta}|{\boldsymbol{x}},\mathbf{Y})\log\left(\frac{g_{{u}}(\boldsymbol{\theta}|{\boldsymbol{x}},\mathbf{Y})}{p(\boldsymbol{\theta}|\mathbf{Y})}\right) d\boldsymbol{\theta} \\
        & =  \mathbb{E}_{g_{_{u}}(\boldsymbol{\theta}|{\boldsymbol{x}},\mathbf{Y})}\left[\log g_{{u}}(\boldsymbol{\theta}|\boldsymbol{x}, \mathbf{Y}) - \log\left(\frac{p(\boldsymbol{\theta})p(\mathbf{Y}|\boldsymbol{\theta})}{p(\mathbf{Y})}\right) \right] \\
        & \propto \mathbb{E}_{g_{_{u}}(\boldsymbol{\theta}|{\boldsymbol{x}},\mathbf{Y})}\left[\log g_{{u}}(\boldsymbol{\theta}|\boldsymbol{x}, \mathbf{Y}) \right] -  \mathbb{E}_{g_{_{u}}(\boldsymbol{\theta}|{\boldsymbol{x}},\mathbf{Y})}\left[ \log p(\boldsymbol{\theta}) + \log p(\mathbf{Y}|\boldsymbol{\theta})\right],
    \end{split}
\end{equation*}
where the latter line is known as the negative evidence lower bound (ELBO) \citep{blei2017variational}. In other words, we propose to select the optimal $\boldsymbol{x}^{*}$ so that we minimize the reverse KL divergence between the Bayesian posterior (up to a proportionality constant) and the approximate posterior induced by random weighting, denoted as  $g_{{u}}(\boldsymbol{\theta}|{\boldsymbol{x}}, \mathbf{Y})$. As discussed in section \ref{sec:RWBB_EM}, given the data $\mathbf{Y}$, the only source of variation in $\boldsymbol{\theta}^{*}$ comes from the random weights, which, in our weighting scheme, depend on $\boldsymbol{x}$.
\par
Ideally, we would like to compute $\boldsymbol{x}^*$ as in \eqref{kl_div_optim}, but we cannot optimize $\mathcal{L}(\boldsymbol{x})$ directly because: (a) we do not know the form of the joint density $g_{{u}}(\boldsymbol{\theta}|{\boldsymbol{x}},\mathbf{Y})$, and (b) the expectation $\mathbb{E}_{g_{_{u}}(\boldsymbol{\theta}|{\boldsymbol{x}}, \mathbf{Y})}[\cdot]$ is analytically intractable. By Sklar's theorem \citep{sklar1959fonctions}, the approximate joint posterior density has the form 
\begin{equation*}
g_{{u}}(\boldsymbol{\theta}|{\boldsymbol{x}},\mathbf{Y}) = c(G_{u,1}(\theta_{1}|\boldsymbol{x}, \mathbf{Y}),\dots,G_{u,M}(\theta_{M}|\boldsymbol{x}, \mathbf{Y}))\prod_{i=1}^{M} g_{u,j}(\theta_{j}|\boldsymbol{x}, \mathbf{Y}),
\end{equation*}
where $c(G_{u,1}(\theta_{1}|\boldsymbol{x}, \mathbf{Y}),\dots,G_{u,M}(\theta_{M}|\boldsymbol{x}, \mathbf{Y}))$ is the copula density of $\boldsymbol{\theta}$ (i.e., $c$ represents the dependence structure in the approximate posterior), $g_{u,j}(\theta_{j}|\boldsymbol{x}, \mathbf{Y})$ denotes the marginal density of the $j$-th entry of $\boldsymbol{\theta}$ with cumulative distribution function $G_{u,j}(\theta_{j}|\boldsymbol{x}, \mathbf{Y})$, and $M=\text{dim}(\boldsymbol{\theta})$. Given posterior draws $\{\boldsymbol{\theta}^{*}_{(s)}\}_{s\in\mathbb{N}}$, one could think about estimating such a copula density (see e.g. \cite{provost2024nonparametric} for a review on copula density estimation in the bivariate case); however, obtaining accurate copula density estimates in high dimensional settings is remarkably difficult. Note that the number of unique parameters in our GMM is given by $dK + \frac{d(d+1)}{2}K + (K-1) = \mathcal{O}(d^{2})$, which grows quadratically with $d$. For instance, letting $d=15$ and $K=4$ would feature 543 unique parameters, leading to inaccurate and unstable copula density estimates. Thus, let us assume that the approximate posterior parameters are mutually independent. More precisely, let $c(G_{u,1}(\theta_{1}|\boldsymbol{x}, \mathbf{Y}),\dots,G_{u,M}(\theta_{M}|\boldsymbol{x}, \mathbf{Y})) \approx 1$, for all $\boldsymbol{\theta}\in\boldsymbol{\Theta}$, so that
\begin{equation*}
    \log g_{{u}}(\boldsymbol{\theta}|{\boldsymbol{x}},\mathbf{Y}) \approx \sum_{i=1}^{M} \log g_{u,j}(\theta_{j}|\boldsymbol{x}, \mathbf{Y}).
\end{equation*}
\par
That being so, to overcome (a), we propose to obtain a batch of approximate posterior draws, denoted by $\{\boldsymbol{\theta}^{*}_{(s)}\}_{s=1}^{S_b}$, and compute univariate kernel density estimates (KDEs) of each marginal density $g_{u,j}$, denoted by $\hat{g}_{u,j}$, which can be evaluated efficiently as in \cite{FKSUM_2021, FKSUM_2022}. To overcome (b), we propose to use the sample mean as an unbiased estimator of the expected value. 
More formally, for any $\boldsymbol{x}\in\boldsymbol{\mathcal{X}}$, we obtain a batch of approximate posterior draws and estimate $\mathcal{L}(\boldsymbol{x})$ with
\begin{equation}
    \label{approx_loss}
     \hat{\mathcal{L}}(\boldsymbol{x})=\frac{1}{S_b}\sum_{s=1}^{S_b}\left\{\left(\sum_{j=1}^{M}\log\hat{g}_{u,j}(\theta^{*}_{j,(s)}|\boldsymbol{x}, \mathbf{Y})\right) - \log p(\boldsymbol{\theta}^{*}_{(s)}) - \log p(\mathbf{Y}|\boldsymbol{\theta}^{*}_{(s)}) \right\}.
\end{equation}
\par
Hence, we can think about $\hat{\mathcal{L}}(\boldsymbol{x})$ as a noisy and black-box approximation of ${\mathcal{L}}(\boldsymbol{x})$ in the sense that, given an input $\boldsymbol{x}\in\boldsymbol{\mathcal{X}}$, we can evaluate the objective, but we do not have access to a closed-form expression for neither the objective nor its derivatives. One could use grid-search methods to minimize $\hat{\mathcal{L}}(\boldsymbol{x})$; however, evaluating $\hat{\mathcal{L}}(\boldsymbol{x})$ is tremendously expensive because, for each $\boldsymbol{x}$, we need to sample a batch of posterior draws, compute univariate KDEs of the approximate marginals, and compare those to the Bayesian posterior. As a consequence, exhaustive grid-search approaches would be infeasible to implement. To overcome this, we use Bayesian Optimization (BO) to minimize $\hat{\mathcal{L}}(\boldsymbol{x})$, as it is one of the most efficient approaches to optimize a noisy black-box objective with little evaluations \citep{jones1998efficient, jones2001taxonomy}.
\par
\textit{Comments on the independence assumption between approximate posterior parameters:} To construct $\hat{\mathcal{L}}(\boldsymbol{x})$, we assume that the approximate posterior parameters are independent, as this is a widely implemented technique in the Bayesian literature. For instance, mean-field variational Bayes (MFVB) assumes that the variational distribution factorizes as $g_{_{VB}}(\boldsymbol{\theta}|{\boldsymbol{\nu}}) = \prod_{j=1}^{M}g_{_{VB,j}}(\theta_{j}|\nu_{j})$. In such cases, the MFVB objective function is given by
\begin{equation*}
    \mathbb{E}_{g_{_{VB}}(\boldsymbol{\theta}|{\boldsymbol{\nu}})}\left[\left(\sum_{j=1}^{M} \log g_{_{VB,j}}(\theta_{j}|\nu_{j})\right) - \log p(\boldsymbol{\theta}) - \log p(\mathbf{Y}|\boldsymbol{\theta}) \right],
\end{equation*}
which closely reassembles our noisy black-box objective in equation \eqref{approx_loss}. In a GMM context, however, one could argue that the independence assumption between model parameters might be unreasonable, as the latent indicator variables $\mathbf{Z}$ depend on each other (i.e., if an observation $\mathbf{y}_{i}$ belongs to a cluster, it automatically cannot belongs to any other cluster), as well as on the mean vectors and covariance matrices. In such cases, the MFVB distribution, $\prod_{j=1}^{M}g_{_{VB,j}}(\theta_{j}|\nu_{j})$, might not contain the true posterior, as it is not able to capture such a complicated dependence structure, negatively affecting the entire analysis. Fortunately, this is not our case. Unlike MFVB, our goal is not to obtain a variational distribution to approximate the true posterior, but to obtain adequate random weights within a WBB framework. Thus, the negative effects would be less severe if the joint density $\sum_{j=1}^{M}\log \hat{g}_{u,j}(\theta^{*}_{j}|\boldsymbol{x},\mathbf{Y})$ used in equation \eqref{approx_loss} does not contain the true posterior, as we do not use $\hat{\mathcal{L}}(\boldsymbol{x})$ to construct an entire variational distribution to approximate $p(\boldsymbol{\theta}|\mathbf{Y})$. We obtain our approximate posterior draws through a randomly weighted EM algorithm, where the latent variables $\mathbf{Z}$ are integrated out in the E-step, incorporating those dependencies in the posterior draws. Moreover, recall from the M-step that the updates of $\boldsymbol{\beta_{k}}$, $\boldsymbol{\Sigma}_{k}$, and $\boldsymbol{\pi}$ do not depend on each other, giving room to assume that $\log \hat{g}_{{u}}(\boldsymbol{\theta}|{\boldsymbol{x}},\mathbf{Y}) \approx \sum_{i=1}^{M} \log \hat{g}_{u,j}(\theta_{j}|\boldsymbol{x}, \mathbf{Y})$.

\subsection{Minimizing $\hat{\mathcal{L}}$ via Bayesian Optimization}
\label{subsec:BO}

We now illustrate our BO approach to minimize $\hat{\mathcal{L}}$. First up, though, note that   maximizing $\Upsilon(\boldsymbol{x}) = - \hat{\mathcal{L}}(\boldsymbol{x})$ is equivalent to minimizing $\hat{\mathcal{L}}(\boldsymbol{x})$. Thus, we consider the problem
\begin{equation*}
    \boldsymbol{x}^{*} = \argmax_{\boldsymbol{x}\in\boldsymbol{\mathcal{X}}}\left\{\Upsilon(\boldsymbol{x})\right\}.
\end{equation*}
Namely, given the black-box function $\Upsilon:\boldsymbol{\mathcal{X}}\rightarrow\mathbb{R}$, we would like to find its global maximum using repeated evaluations of $\Upsilon$. However, since each evaluation is expensive, our goal is to maximize $\Upsilon$ using as few evaluations as possible. 

By combining the current evidence, i.e., the evaluations $\{(\boldsymbol{x}_{m}, {\Upsilon}(\boldsymbol{x}_{m}))\}_{m=1}^{l}$, with some prior information over ${\Upsilon}$, i.e., $Q({\Upsilon})$, BO efficiently determines future evaluations. More precisely, BO uses the posterior $Q(\Upsilon\,|\,\{(\boldsymbol{x}_{m}, {\Upsilon}(\boldsymbol{x}_{m}))\}_{m=1}^{l})$ to construct an acquisition function $\xi:\boldsymbol{\mathcal{X}}\rightarrow\mathbb{R}$ and choose its maximum, $\boldsymbol{x}_{l+1}=\argmax_{\boldsymbol{x}\in\boldsymbol{\mathcal{X}}}\xi(\boldsymbol{x})$, as our next point for evaluation \citep{dragonfly_2020}.
\par
For our prior on $\Upsilon$, we assume a Gaussian Process ($\mathcal{GP}$). This is,
\begin{equation*}
    \Upsilon\sim \mathcal{GP}(\phi, \kappa),
\end{equation*}
where $\phi:\boldsymbol{\mathcal{X}}\rightarrow\mathbb{R}$ and $\kappa:\boldsymbol{\mathcal{X}}\times\boldsymbol{\mathcal{X}}\rightarrow\mathbb{R}$ are a mean and a kernel (covariance) function, respectively, so that $\Upsilon(\boldsymbol{x})\sim\mathcal{N}(\phi(\boldsymbol{x}), \kappa(\boldsymbol{x}, \boldsymbol{x}))$, for all $\boldsymbol{x}\in\boldsymbol{\mathcal{X}}$. Given the evidence $\{(\boldsymbol{x}_{m}, \hat{y}_{m})\}_{m=1}^{l}$, with $\boldsymbol{x}_{m}\in\boldsymbol{\mathcal{X}}$, $\hat{y}_{m} = \Upsilon(\boldsymbol{x}_{m}) + \epsilon_{m}\in\mathbb{R}$, and $\epsilon_{m}\sim\mathcal{N}(0, \eta^{2})$, the resulting posterior $\Upsilon\,|\,\{(\boldsymbol{x}_{m}, \hat{y}_{m})\}_{m=1}^{l}$ would also be a $\mathcal{GP}$ with mean and covariance given by
\begin{gather}
    \hat{\phi}(\boldsymbol{x}) = \boldsymbol{k}'(\mathbf{K} + \eta^{2}\mathbf{I}_{l})^{-1}\hat{\boldsymbol{y}},\\
    \hat{\kappa}(\boldsymbol{x}, \boldsymbol{\vartheta}) = {\kappa}(\boldsymbol{x}, \boldsymbol{\vartheta}) - \boldsymbol{k}'(\mathbf{K}+ \eta^{2}\mathbf{I}_{l})^{-1}\boldsymbol{v},
\end{gather}
where $\boldsymbol{x}, \boldsymbol{\vartheta} \in \boldsymbol{\mathcal{X}}$, $\hat{\boldsymbol{y}} = (\hat{y}_{1},\dots,\hat{y}_{l})'$, $k_{m} = \kappa(\boldsymbol{x}, \boldsymbol{x}_{m})$, $v_{m} = \kappa(\boldsymbol{\vartheta}, \boldsymbol{x}_{m})$, $\boldsymbol{k}=(k_{1},\dots,k_{l})'$, $\boldsymbol{v}=(v_{1},\dots,v_{l})'$, and $(\mathbf{K})_{m,m'} = \kappa(\boldsymbol{x}_{m}, \boldsymbol{x}_{m'})$ \citep{rasmussen2006gaussian}.
\par
As suggested by \cite{snoek2012practical}, we use a Mat\`ern 2.5 kernel, given by
\begin{gather*}
    \kappa(\boldsymbol{x}, \boldsymbol{\vartheta}) = \zeta_{0}\left(1 + \sqrt{5r^{2}(\boldsymbol{x}, \boldsymbol{\vartheta})} + \frac{5}{3}r^{2}(\boldsymbol{x}, \boldsymbol{\vartheta})\right) \exp\left\{-\sqrt{5r^{2}(\boldsymbol{x}, \boldsymbol{\vartheta})}\right\}, 
\end{gather*}
where $r^{2}(\boldsymbol{x}, \boldsymbol{\vartheta})=\sum_{r=1}^{2(K+1)}\left(x_{r}-\vartheta_{r}\right)^{2}/\zeta_{r}^{2}$, $\zeta_{0}$ is a covariance amplitude, and $\zeta_{r}$, for $r\in\{1,\dots,2(K+1)\}$ are $\mathcal{GP}$ length scales. For our acquisition function, we use the Expected Improvement (EI) over the best current value \citep{jones1998efficient}, as it has been shown to be efficient in the number of evaluations required to find the global maximum of a wide range of black-box functions \citep{bull2011convergence, snoek2012practical}. More precisely, given the best current value---denoted by $\boldsymbol{x}^{+}$, we set
\begin{equation*}
    \xi(\boldsymbol{x})=\mathbb{E}\left[\max\left\{0,\,\Upsilon(\boldsymbol{x})-\Upsilon(\boldsymbol{x}^{+})\right\}\,|\,\{(\boldsymbol{x}_{m}, \hat{y}_{m})\}_{m=1}^{l}\right].
\end{equation*}
\par
That being said, note that the BO procedure is carried out just once. Then, we use the learned random weights throughout the entire sampling process, as described in section \ref{sec:RWBB_EM}. Additionally, obtaining the batch of approximate posterior draws, computing the univariate KDEs, and evaluating the Bayesian posterior, can all be trivially implemented in parallel, reducing the cost of evaluating and minimizing $\hat{\mathcal{L}}$. Thus BOB, as WLB and WBB, also embraces recent developments in parallel computing. BOB, however, let us select the optimal $\boldsymbol{x}^{*}$, and hence, adequate random weights. We summarize our proposed method in algorithm \ref{alg:bob_algorithm}.

\begin{algorithm}
\caption{Bayesian Optimized Bootstrap}\label{alg:bob_algorithm}
\textbf{Input:} \\
\hspace*{\algorithmicindent} Data: $\mathbf{Y}$\\
\hspace*{\algorithmicindent} Total number of posterior draws: $S$\\
\hspace*{\algorithmicindent} Prior hyper-parameters: $\left\{a_{k},\,\lambda_{k},\,\nu_{k},\,\boldsymbol{\beta}_{k},\, \boldsymbol{\Psi}_{k}\right\}_{k=1}^{K}$\\
\hspace*{\algorithmicindent} Batch size: $S_{b}$\\
\hspace*{\algorithmicindent} Compact search space: $\boldsymbol{\mathcal{X}}$\\
\hspace*{\algorithmicindent} Tempering hyper-parameters: $\{a,\,b,\,c,\,r\}$\\
\textbf{Output:} \\
\hspace*{\algorithmicindent} Posterior draws: $\left\{\boldsymbol{\theta}^{*}_{\text{BOB},(s)}\right\}_{s=1}^{S}$
\begin{algorithmic}[1]
\State Compute $\boldsymbol{x}^{*}\leftarrow\argmin_{\boldsymbol{x}\in\boldsymbol{\mathcal{X}}}\hat{\mathcal{L}}(\boldsymbol{x})$ via Bayesian Optimization
\State Set $(x_{\alpha}^{*}, x_{\mu_{1}}^{*}, \dots, x_{\mu_{K}}^{*}, x_{{\Sigma_{1}}}^{*}, \dots, x_{{\Sigma_{K}}}^{*}, x_{\pi}^{*})'\leftarrow\boldsymbol{x}^{*}$
\For{$s\in\{1\,\dots,S\}$} \Comment{in parallel}
    \State Set $\Tilde{u}_{\pi} \leftarrow x_{\pi}^{*}$
    \State Set $\Tilde{u}_{\mu_{k}}\leftarrow x_{\mu_{k}}^{*}$, $\forall k\in[K]$
    \State Set $\Tilde{u}_{\Sigma_{{k}}} \leftarrow x_{\Sigma_{k}}^{*}$, $\forall k\in[K]$
    \State Construct $u_{i} \leftarrow \frac{w_{i}^{x_{\alpha}^{*}}}{\sum_{i=1}^{n}w_{i}^{x_{\alpha}^{*}}}$, where $w_{i}\sim\mathcal{E}(1)$, $\forall i\in[n]$
    \State Compute $\boldsymbol{\theta}^{*}_{\text{BOB},(s)}\leftarrow\argmax_{\boldsymbol{\theta}\in\boldsymbol{\Theta}}\log p_{u}(\boldsymbol{\theta};\mathbf{Z}|\mathbf{Y})$ via Tempered EM
\EndFor
\end{algorithmic}
\end{algorithm}

To construct our Bayesian optimizer, we make use of the well-established ``\texttt{DiceOptim}" \texttt{R} package \citep{DiceOptim2012, DiceOptim2021}. Our implementation of BOB, as well as of WBB, are all available in the ``\texttt{BOBgmms}" \texttt{R} package, which can be found in the supplementary materials or online at \href{https://github.com/marinsantiago/BOBgmms}{\texttt{github.com/marinsantiago/BOBgmms}}. Source code to reproduce the results from this article can also be found in the supplementary materials or online at \href{https://github.com/marinsantiago/BOBgmms-examples}{\texttt{github.com/marinsantiago/BOBgmms-examples}}.

\subsection{Asymptotic Properties}
\label{subsec:theory}

Throughout this subsection, we want to show that BOB retains key asymptotic properties from WLB and WBB. To that end, let $\hat{\boldsymbol{\theta}}_{\text{MLE}}\in\boldsymbol{\Theta}$ be the MLE for $\boldsymbol{\theta}$ and let $\boldsymbol{\theta}^{*}_{\text{BOB}}\in\boldsymbol{\Theta}$ be a draw from BOB. It is assumed that the ordering of $\hat{\boldsymbol{\theta}}_{\text{MLE}}$ and $\boldsymbol{\theta}^{*}_{\text{BOB}}$ is the same. Let also, for $k\in\{1,\dots,K\}$, $a_{k}=a_{k}(n)$, $\lambda_{k}=\lambda_{k}(n)$, $\boldsymbol{\Psi}_{k}=\boldsymbol{\Psi}_{k}(n)$, and $\nu_{k}=\nu_{k}(n)$, such that, as $n\rightarrow\infty$, $a_{k}(n)\rightarrow1$, $\lambda_{k}(n)\rightarrow0$, $\boldsymbol{\Psi}_{k}(n)\rightarrow\mathbf{0}_{d\times d}$, and $\nu_{k}(n)\rightarrow-(d+2)$.
Then, note that BOB can be seen as a generalization of WLB and WBB. In fact, by setting $\boldsymbol{x}^{*}=(1,0,0,\dots,0)'\in\boldsymbol{\mathcal{X}}$ or $\boldsymbol{x}^{*}=(1,1,1,\dots,1)'\in\boldsymbol{\mathcal{X}}$, one would recover WLB or WBB (with fixed prior weights \citep{ng2022random}), respectively. Thus, if $\{\mathbf{y}_{i}\}_{i=1}^{n}$ are random samples from a sufficiently regular model $p(\mathbf{y}|\boldsymbol{\theta}_{0})$, as described in section \ref{subsec:Model_setup}, and if the BOB solution converges to the WLB or WBB solutions, as $n\rightarrow\infty$, then $\boldsymbol{\theta}^{*}_{\text{BOB}}$ would retain key asymptotic properties such as consistency and asymptotic normality. More precisely, following theorems 1 and 2 from \cite{raftery_boostrap_likelihood}, along with the results from \cite{newtonBoots},  we have that if $\boldsymbol{x}^{*}\rightarrow(1,0,0,\dots,0)'\in\boldsymbol{\mathcal{X}}$ or $\boldsymbol{x}^{*}\rightarrow(1,1,1,\dots,1)'\in\boldsymbol{\mathcal{X}}$, as $n\rightarrow\infty$, then:
\begin{enumerate}
    \item For any scalar $\varepsilon>0$, as $n\rightarrow\infty$, 
         \begin{equation*}
            \mathbb{P}_{u}(\lVert \boldsymbol{\theta}^{*}_{\text{BOB}} -  \hat{\boldsymbol{\theta}}_{\text{MLE}} \rVert > \varepsilon\, |\,\{\mathbf{y}_{i}\}_{i=1}^{n})\rightarrow 0,
         \end{equation*}
         for almost every infinite sequence of data $\left\{\mathbf{y}_{i}\right\}_{i\in\mathbb{N}}$. 
         
         \item For all measurable $\mathbf{B}\in\mathcal{B}(\boldsymbol{\Theta})$, as $n\rightarrow\infty$,
         \begin{equation*}
            \mathbb{P}_{u}(\sqrt{n}(\boldsymbol{\theta}^{*}_{\text{BOB}} -  \hat{\boldsymbol{\theta}}_{\text{MLE}})\in\mathbf{B}\,|\,\{\mathbf{y}_{i}\}_{i=1}^{n})\rightarrow\mathbb{P}(\mathbf{T}\in\mathbf{B}),
        \end{equation*}
        for almost every infinite sequence of data $\left\{\mathbf{y}_{i}\right\}_{i\in\mathbb{N}}$. In this case, $\mathbf{T}\sim\mathcal{N}(\mathbf{0}, \boldsymbol{\mathcal{I}}(\boldsymbol{\theta}_{0})^{-1})$, $ \boldsymbol{\mathcal{I}}(\boldsymbol{\theta})$ is the Fisher information, and $\mathcal{B}(\boldsymbol{\Theta})$ denotes the Borel field on $\boldsymbol{\Theta}$.
\end{enumerate}

\begin{remark}
The probabilities, $\mathbb{P}_{u}$, from above, refer to the distribution of $\boldsymbol{\theta}^{*}_{\text{BOB}}$ induced by random weights given the sequence of data $\{\mathbf{y}_{i}\}_{i=1}^{n}$.
\end{remark}

\begin{remark}
The Bernstein–von Mises theorem states that, under regularity conditions, the actual Bayesian posterior converges to a normal distribution. More formally, as $n\rightarrow\infty$,
$\{\boldsymbol{\theta}|\mathbf{Y}\}\rightarrow\mathcal{N}(\boldsymbol{\theta}_0, [n\,\boldsymbol{\mathcal{I}}(\boldsymbol{\theta}_{0})]^{-1})$. Comparing this result with our previous results, one have that it is possible to approximate posterior credible sets with BOB as $\mathbb{P}( \boldsymbol{\theta}^{*}_{\text{BOB}}\in\mathbf{B}{}\,|\,\mathbf{Y})\approx\int_{\mathbf{B}}p(\boldsymbol{\theta}|\mathbf{Y})d\boldsymbol{\theta}$, for all $\mathbf{B}\in\mathcal{B}(\boldsymbol{\Theta})$, and the approximation would get better with a growing $n$.
\end{remark}

On the whole, we have that BOB provides an automatic and much more informed approach to select the random weights, while retaining the asymptotic first order correctness from existing Bootstrap-based posterior samplers.

\section{Simulations}
\label{sec:sims}
We now evaluate the performance of different approximate posterior samplers trough various simulation experiments.
\subsection{Simulations Setup}
\label{subsec:sim_setup}
To generate the simulated data, we start by sampling $\mathbf{z}_{i}$, for $i\in\{1,\dots,n\}$, from $p(\mathbf{z}_{i}|K)\propto\prod_{k=1}^{K}(\frac{1}{K})^{z_{ik}}$, where $K\in\{2,3,4\}$. For the dimension $d$, we are going to consider low $(d=5)$, medium $(d=10)$, and high $(d=15)$ dimensional problems. To generate each $\boldsymbol{\mu}_{k}$, we set, for $k\in\{1,\dots,K\}$,
\begin{equation*}
    \boldsymbol{\mu}_{k} = \left((5k - 4)\mathbf{1}_{\lceil 0.6d \rceil}', \mathbf{0}_{(d-\lceil 0.6d \rceil)}' \right)'\in\mathbb{R}^{d},
\end{equation*}
where only 60\% of the entries of $\boldsymbol{\mu}_{k}$ are important parameters in separating the clusters, while the remaining entries are set to zero. Lastly, we set each $\boldsymbol{\Sigma}_{k}$ to be an identity matrix of size $d\times d$, i.e., $\boldsymbol{\Sigma}_{k} = \mathbf{I}_{d}$.
In total, we consider nine different simulations settings, which are summarized in table \ref{tab:sim_setings}. In all cases, we standardize the data so that each feature has mean 0 and variance 1.

\begin{table}[h]
    \caption{Simulation Settings}\label{tab:sim_setings}
    \begin{tabular*}{\textwidth}{@{\extracolsep\fill}cccccccccc}
    \toprule%
    & \multicolumn{9}{@{}c@{}}{Setting} \\
    \cmidrule{2-10}
    Simulation parameters & 1 & 2 & 3 & 4 & 5 & 6 & 7 & 8 & 9 \\
    \midrule
    $n$ & 50 & 50 & 50 & 100 & 100 & 100 & 150 & 150 & 150\\
    $d$ & 5  & 10 & 15 & 5   & 10  & 15  & 5   & 10  & 15 \\
    $K$ & 2  & 2  & 2  & 3   & 3   & 3   & 4   & 4   & 4  \\
    \botrule
    \end{tabular*}
\end{table}

The prior hyper-parameters are set as $\boldsymbol{\beta}_{k}=\mathbf{0}_{d}$, $\boldsymbol{\Psi}_{k}=\mathbf{I}_{d}$, $a_{k}=1.1$, $\lambda_{k}=\lambda$, and $\nu_{k}=\nu$, for $k\in\{1,\dots,K\}$. To select the regularization parameters, $\lambda$ and $\nu$, we use a likelihood-based cross-validation approach, as in \cite{friedman_glasso}. The idea is to consider a grid of values for $\lambda$ and $\nu$, run an unweighted EM algorithm using each pair of $\lambda$ and $\nu$ on a training subset of the data, and choose the pair of $\lambda$ and $\nu$ that maximizes the log-likelihood over the validation set. We then use the same values of $\lambda$ and $\nu$ across all the posterior samplers. That being said, the goal of these experiments is not to choose the ``best" regularization parameters. Instead, as pointed out by \cite{deep_boots}, the goal is to compare different posterior approximations for given values of $\lambda$ and $\nu$.
\par
For our approximate posterior samplers we consider BOB as well as two versions of WBB, namely WBB1 (with random prior weights) and WBB2 (with fixed prior weights). To implement WBB1 we set $\mathbf{u}\overset{iid}{\sim}\mathcal{E}(1)$ and $\Tilde{\mathbf{u}}\overset{iid}{\sim}\mathcal{E}(1)$, and to implement WBB2 we set $\mathbf{u}\overset{iid}{\sim}\mathcal{E}(1)$ and $\Tilde{\mathbf{u}}\overset{iid}{\sim}\delta_{\Tilde{u}}(1)$. We also consider a mean-field Automatic Differentiation Variational Inference (ADVI) algorithm \citep{kucukelbir2017automatic} and for our MCMC algorithm we use a No-U-Turn sampler, or NUTS \citep{hoffman2014no}. Both NUTS and ADVI are executed in Stan \citep{carpenter2017stan} as it is a highly optimized and off-the-shelf probabilistic programming language, widely used by many practitioners.
\par
We set the lower and upper bounds of the BO search space, $\boldsymbol{\mathcal{X}}$, to be $\boldsymbol{\mathcal{X}}_{\text{lower}}=(1, 10^{-5}, \dots, 10^{-5})'$ and $\boldsymbol{\mathcal{X}}_{\text{upper}}=(1.5, \dots, 1.5)'$, respectively, except in settings 4 and 7, where we set $\boldsymbol{\mathcal{X}}_{\text{upper}}=(1.25, \dots, 1.25)'$ as we have more available data. Note that in any case, the search space contains WBB and WLB as potential solutions. To set the initial parameter values, we consider a pool of candidate values and choose the initialization that yields the largest posterior density. Further details on our initialization strategy can be found in the supplementary materials. It is worth noting that we initialize all the algorithms (i.e., BOB, WBB, NUTS, and ADVI) at the same point. Thus, if we see any difference in performance, it would be due to the sampling algorithm and not due to the initialization, as they all share the same initial values.
\par
We obtain $S=20000$ approximate posterior draws using 
BOB, WBB, and ADVI, while with NUTS, we run the algorithm for $2S=40000$ iterations and discard the first half as \textit{burn-in}. In the case of BOB, we use batches of size $S_{b}=4000$ to construct $\hat{\mathcal{L}}(\boldsymbol{x})$. We run all our simulations and data analyses on an Apple Silicon-based machine with 48 GB of memory and 16 CPU cores.

\subsection{Comparison Metrics}
\label{subsec:sims_metrics}
To assess the accuracy of posterior approximations, the Kolmogorov–Smirnov (KS) and the Total Variation (TV) distances between the Bayesian posterior and its approximation have been traditionally used. To ease computations, this assessment has been usually based on comparing the Bayesian marginal posteriors with the approximate marginal posteriors \citep{stringer2023fast, ng2022random}. However, in the context of mixture models, these comparisons are no longer viable because of the so-called label switching problem \citep{diebolt1994estimation}. One could use relabeling algorithms, as in \cite{stephens2000dealing}, but the problem would remain as the 
resulting ordering of the approximate marginals might be different to the ordering of the Bayesian marginals, even after employing a relabeling algorithm, making comparisons between the marginals virtually meaningless.
\par
We, therefore, base our comparisons on the posterior predictive distribution, as the posterior predictive is invariant to any permutation of $\boldsymbol{\theta}$ (i.e., it circumvents the label switching problem). More formally, let $p(\boldsymbol{\theta}|\mathbf{Y})$ be the Bayesian posterior and let $g(\boldsymbol{\theta}|\mathbf{Y})$ be its approximation. Then, the Bayesian posterior predictive distribution would be given by $p(\mathbf{y}_{\text{new}}|\mathbf{Y})=\int_{\boldsymbol{\Theta}}p(\mathbf{y}_{\text{new}}|\boldsymbol{\theta})p(\boldsymbol{\theta}|\mathbf{Y})d\boldsymbol{\theta}$, while the approximate posterior predictive would be $g(\mathbf{y}_{\text{new}}|\mathbf{Y})=\int_{\boldsymbol{\Theta}}p(\mathbf{y}_{\text{new}}|\boldsymbol{\theta})g(\boldsymbol{\theta}|\mathbf{Y})d\boldsymbol{\theta}$. 
Note, additionally, that the posterior predictive distribution reflects two kinds of uncertainty: (a) Sampling uncertainty about $\mathbf{y}$ conditional on $\boldsymbol{\theta}$ and (b) parametric uncertainty about $\boldsymbol{\theta}$ \citep{Lynch_posterior}. Following \cite{geisser2017predictive}, under a correctly specified model and a proper prior, the Bayes risk
\begin{equation*}
\int_{\boldsymbol{\Theta}}p(\boldsymbol{\theta})\left[\int_{\boldsymbol{\mathcal{Y}}}p(\mathbf{Y}|\boldsymbol{\theta})\text{D}_{\text{KL}}(p_{\text{true}}(\mathbf{y}_{\text{new}}) \rVert\, g(\mathbf{y}_{\text{new}}|\mathbf{Y})) d\mathbf{Y} \right] d\boldsymbol{\theta}
\end{equation*}
is minimized when $g(\mathbf{y}_{\text{new}}|\mathbf{Y}) = p(\mathbf{y}_{\text{new}}|\mathbf{Y})$, where $\text{D}_{\text{KL}}(p_{\text{true}}(\mathbf{y}_{\text{new}}) \rVert\, g(\mathbf{y}_{\text{new}}|\mathbf{Y}))$ denotes the KL divergence between the true data generating mechanism $p_{\text{true}}(\mathbf{y}_{\text{new}})$ and the approximate posterior predictive distribution $g(\mathbf{y}_{\text{new}}|\mathbf{Y})$, making the Bayesian posterior predictive distribution a natural benchmark when one is interested in quantifying uncertainty.
\par
Thus, we consider the TV and KS distances between $p(\mathbf{y}_{\text{new}}|\mathbf{Y})$ and $g(\mathbf{y}_{\text{new}}|\mathbf{Y})$ as comparison metrics, where smaller distances would suggest a better approximation to the Bayesian posterior and a better uncertainty quantification. To ease computations, we approximate the TV and KS distances between $p(\mathbf{y}_{\text{new}}|\mathbf{Y})$ and $g(\mathbf{y}_{\text{new}}|\mathbf{Y})$ as 
\begin{gather}
       \hat{\text{TV}} = \frac{1}{d}\sum_{j=1}^{d}\frac{1}{2}\sum_{y\in\mathcal{Y}}\left|\hat{P}^{\text{Bayes}}_{y_{\text{new},j}}(y)-\hat{G}^{(\cdot)}_{y_{\text{new},j}}(y)\right|,\\
       \hat{\text{KS}} = \frac{1}{d}\sum_{j=1}^{d}\sup_{y\in\mathcal{Y}}\left|\hat{P}^{\text{Bayes}}_{y_{\text{new},j}}(y)-\hat{G}^{(\cdot)}_{y_{\text{new},j}}(y)\right|,
\end{gather}
where $\hat{P}^{\text{Bayes}}_{y_{\text{new},j}}$ denotes the empirical CDF of the actual Bayesian posterior predictive distribution of $y_{\text{new},j}$ (i.e., the $j$-th element of $\mathbf{y}_{\text{new}}$) and $\hat{G}^{(\cdot)}_{y_{\text{new},j}}$ denotes the empirical CDF of the posterior predictive of $y_{\text{new},j}$ obtained by one of the approximations (i.e., BOB, WBB, NUTS, or ADVI). Details on how one can sample from the posterior predictive distributions and the actual Bayesian posterior (when the true latent variables, $\mathbf{Z}$, are known), can be found in the supplementary materials.

\subsection{Simulation Results}
\label{subsec:sims_results}

\begin{figure*}
\begin{center}
\includegraphics[width=\textwidth]{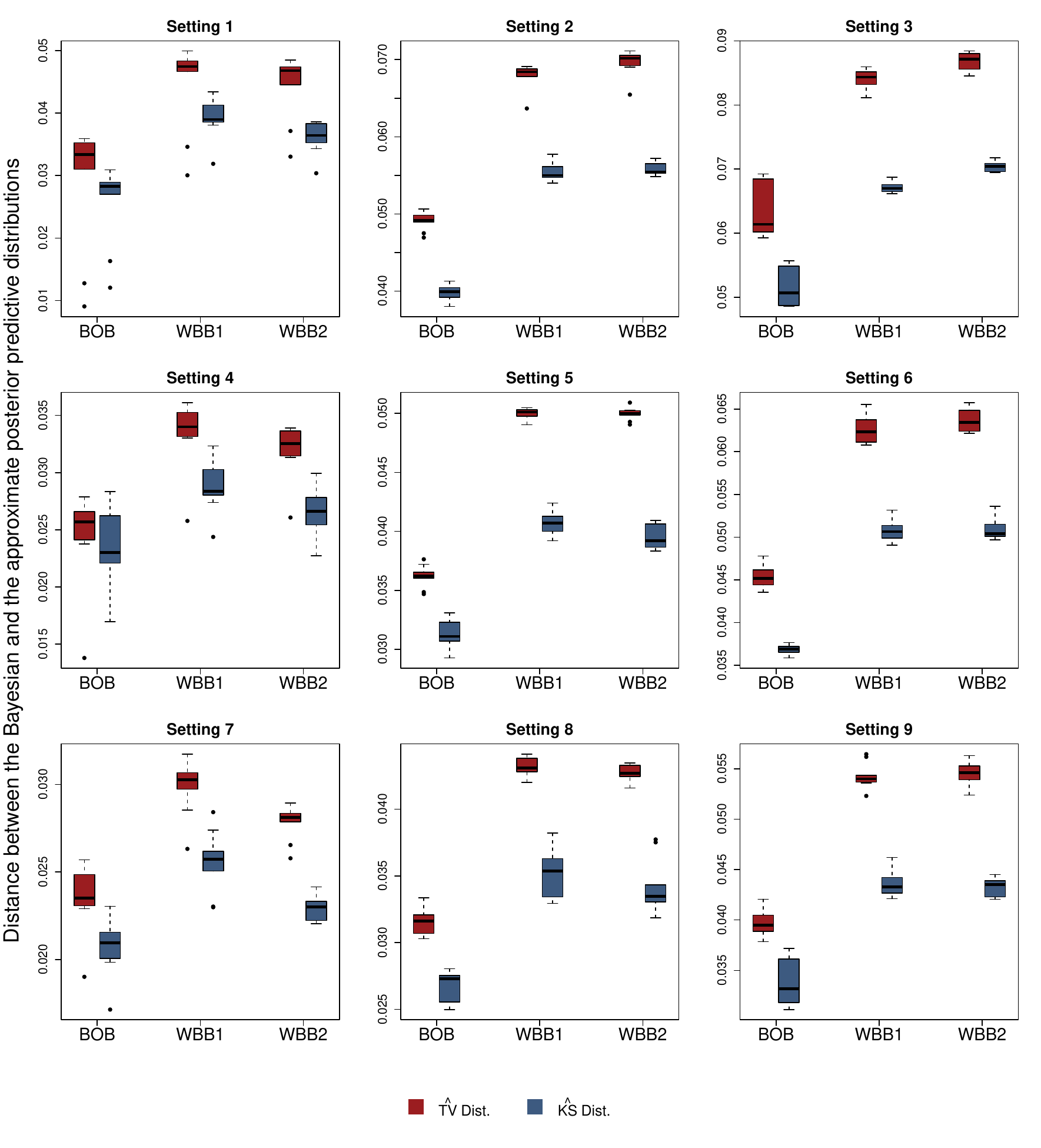}
\end{center}
\caption{Boxplots of the $\hat{\text{TV}}$ distances (in red) and $\hat{\text{KS}}$ distances (in blue)  between the Bayesian posterior predictive distribution and the posterior predictives obtained via BOB, WBB1, and WBB2, based on ten independent runs per setting.}
\label{fig:sim_results}
\end{figure*}

Figure \ref{fig:sim_results} and Table \ref{tab:sim_results} present the simulation results. Figure \ref{fig:sim_results} displays boxplots of the $\hat{\text{TV}}$ and $\hat{\text{KS}}$ distances between the Bayesian posterior predictive distribution and its approximations obtain via BOB, WBB1, and WBB2 based on ten independent runs per setting. Table \ref{tab:sim_results} shows the median (and the interquartile range) of the $\hat{\text{TV}}$ and $\hat{\text{KS}}$ distances, as well as of the elapsed (wall-clock) times, for all considered methods, based on the same ten independent runs. Figure \ref{fig:sim_results} clearly shows that, across the nine settings, BOB constantly outperforms both versions of WBB in recovering the Bayesian posterior predictive distribution. Interestingly, we can also observe that the performance of both versions of WBB is very similar, with no major differences between the two methods.

\begin{table}[h]
\caption{Median $\hat{\text{TV}}$ and $\hat{\text{KS}}$ distances between the Bayesian posterior predictive distribution and its approximations obtained via BOB, WBB1, WBB2, NUTS, and ADVI, as well as median elapsed (wall-clock) times in minutes, based on ten independent runs per setting.}
\label{tab:sim_results}
\begin{tabular*}{\textwidth}{@{\extracolsep\fill}llccccccccc}
\toprule%
& & \multicolumn{9}{@{}c@{}}{Setting} \\
\cmidrule{3-11}
Metric & Method & 1 & 2 & 3 & 4 & 5 & 6 & 7 & 8 & 9 \\
\midrule
$\hat{\text{TV}}$ & BOB & 0.033 & 0.049 & 0.061 & 0.026 & 0.036 & 0.045 & 0.023 & 0.032 & 0.039 \\ 
& & (0.003) & (0.001) & (0.007) & (0.002) & (0.000) & (0.001) & (0.001) & (0.001) & (0.001) \\ 
& WBB1 & 0.047 & 0.068 & 0.084 & 0.034 & 0.050 & 0.062 & 0.030 & 0.043 & 0.054 \\ 
& & (0.001) & (0.001) & (0.002) & (0.002) & (0.000) & (0.002) & (0.001) & (0.001) & (0.001) \\ 
& WBB2 & 0.047 & 0.070 & 0.087 & 0.033 & 0.050 & 0.063 & 0.028 & 0.043 & 0.055 \\ 
& & (0.002) & (0.001) & (0.002) & (0.002) & (0.000) & (0.002) & (0.000) & (0.001) & (0.001) \\ 
& NUTS & 0.006 & 0.006 & 0.056 & 0.039 & 0.070 & 0.057 & 0.072 & 0.058 & -- \\ 
& & (0.004) & (0.049) & (0.027) & (0.061) & (0.012) & (0.027) & (0.010) & (0.018) & -- \\ 
& ADVI & 0.083 & 0.107 & 0.284 & 0.074 & 0.094 & 0.159 & 0.071 & 0.095 & 0.113 \\ 
& & (0.018) & (0.040) & (0.077) & (0.016) & (0.058) & (0.032) & (0.015) & (0.042) & (0.076) \\ 
\midrule
$\hat{\text{KS}}$ & BOB & 0.028 & 0.040 & 0.051 & 0.023 & 0.031 & 0.037 & 0.021 & 0.027 & 0.033 \\ 
& & (0.002) & (0.001) & (0.005) & (0.004) & (0.001) & (0.001) & (0.001) & (0.002) & (0.004) \\ 
& WBB1 & 0.039 & 0.055 & 0.067 & 0.028 & 0.041 & 0.051 & 0.026 & 0.035 & 0.043 \\ 
& & (0.003) & (0.001) & (0.001) & (0.002) & (0.001) & (0.001) & (0.001) & (0.003) & (0.001) \\ 
& WBB2 & 0.036 & 0.055 & 0.070 & 0.027 & 0.039 & 0.050 & 0.023 & 0.033 & 0.044 \\ 
& & (0.003) & (0.001) & (0.001) & (0.002) & (0.002) & (0.001) & (0.001) & (0.001) & (0.001) \\ 
& NUTS & 0.008 & 0.009 & 0.058 & 0.048 & 0.063 & 0.065 & 0.049 & 0.047 & -- \\ 
& & (0.004) & (0.050) & (0.031) & (0.043) & (0.008) & (0.017) & (0.002) & (0.008) & -- \\ 
& ADVI & 0.078 & 0.081 & 0.151 & 0.062 & 0.075 & 0.102 & 0.052 & 0.066 & 0.081 \\ 
& & (0.013) & (0.010) & (0.034) & (0.012) & (0.015) & (0.019) & (0.002) & (0.019) & (0.040) \\ 
\midrule
Elapsed & BOB & 1.891 & 2.271 & 2.731 & 2.525 & 2.988 & 3.553 & 3.281 & 3.829 & 4.692 \\ 
& & (0.033) & (0.086) & (0.040) & (0.140) & (0.035) & (0.050) & (0.075) & (0.053) & (0.197) \\ 
& WBB1 & 0.002 & 0.003 & 0.004 & 0.003 & 0.005 & 0.008 & 0.004 & 0.007 & 0.011 \\ 
& & (0.000) & (0.001) & (0.001) & (0.000) & (0.002) & (0.001) & (0.001) & (0.001) & (0.001) \\ 
& WBB2 & 0.002 & 0.003 & 0.004 & 0.003 & 0.004 & 0.008 & 0.004 & 0.007 & 0.011 \\ 
& & (0.000) & (0.001) & (0.001) & (0.000) & (0.000) & (0.000) & (0.000) & (0.001) & (0.001) \\ 
& NUTS & 0.681 & 4.400 & 22.027 & 8.591 & 36.547 & 80.995 & 22.318 & 85.547 & -- \\ 
& & (0.073) & (6.519) & (8.028) & (5.367) & (3.141) & (2.646) & (5.641) & (21.072) & -- \\ 
& ADVI & 0.033 & 0.081 & 0.157 & 0.078 & 0.180 & 0.369 & 0.131 & 0.323 & 0.686 \\ 
& & (0.004) & (0.008) & (0.010) & (0.004) & (0.031) & (0.020) & (0.008) & (0.068) & (0.052) \\ 
\botrule
\end{tabular*}
\footnotetext{Note: Interquartile ranges are provided in parentheses.}
\end{table}

In Table \ref{tab:sim_results}, we can observe that both versions of WBB have the lowest running times. We can also observe that in simpler settings, the running times and accuracy of NUTS are comparable to those of BOB. However, as the simulation settings get more complicated, the median running times of BOB become significantly lower than those of NUTS. For instance, in setting 5, BOB takes a median running time of 2.99 minutes, while NUTS takes 36.55 minutes. This issue is more clear in settings 6 and 8, where BOB's median running times are 3.55 and 3.83 minutes, respectively, while NUTS takes median running times of 81 and 85.55 minutes, respectively. In other words, BOB can be 22.8 times faster than NUTS. Nonetheless, it is worth mentioning that in setting 9, NUTS was not able to run within a computing budget of 8 hours (480 minutes), while BOB took a median running time of 4.69 minutes, illustrating the poor scalability of MCMC algorithms to large problems. 
\par
More interestingly, though, we can observe that in more complicated settings, BOB is not only faster than NUTS, but also more accurate. For instance, in setting 7, the median $\hat{\text{TV}}$ distance obtained via NUTS is 3 times larger than the one of BOB, illustrating that BOB is, indeed, a reliable posterior sampler. It is also clear that, across all settings, ADVI produces the least accurate approximations of the Bayesian posterior predictive distribution. On the whole, even if WBB has the lowest running times, BOB constantly produces reliable approximations of the Bayesian posterior predictive distribution in a timely fashion, nicely balancing the trade-off between computational cost and accuracy. 

\begin{figure*}[!htp]
\centering
\begin{subfigure}{0.45\textwidth}
    \includegraphics[width=\textwidth]{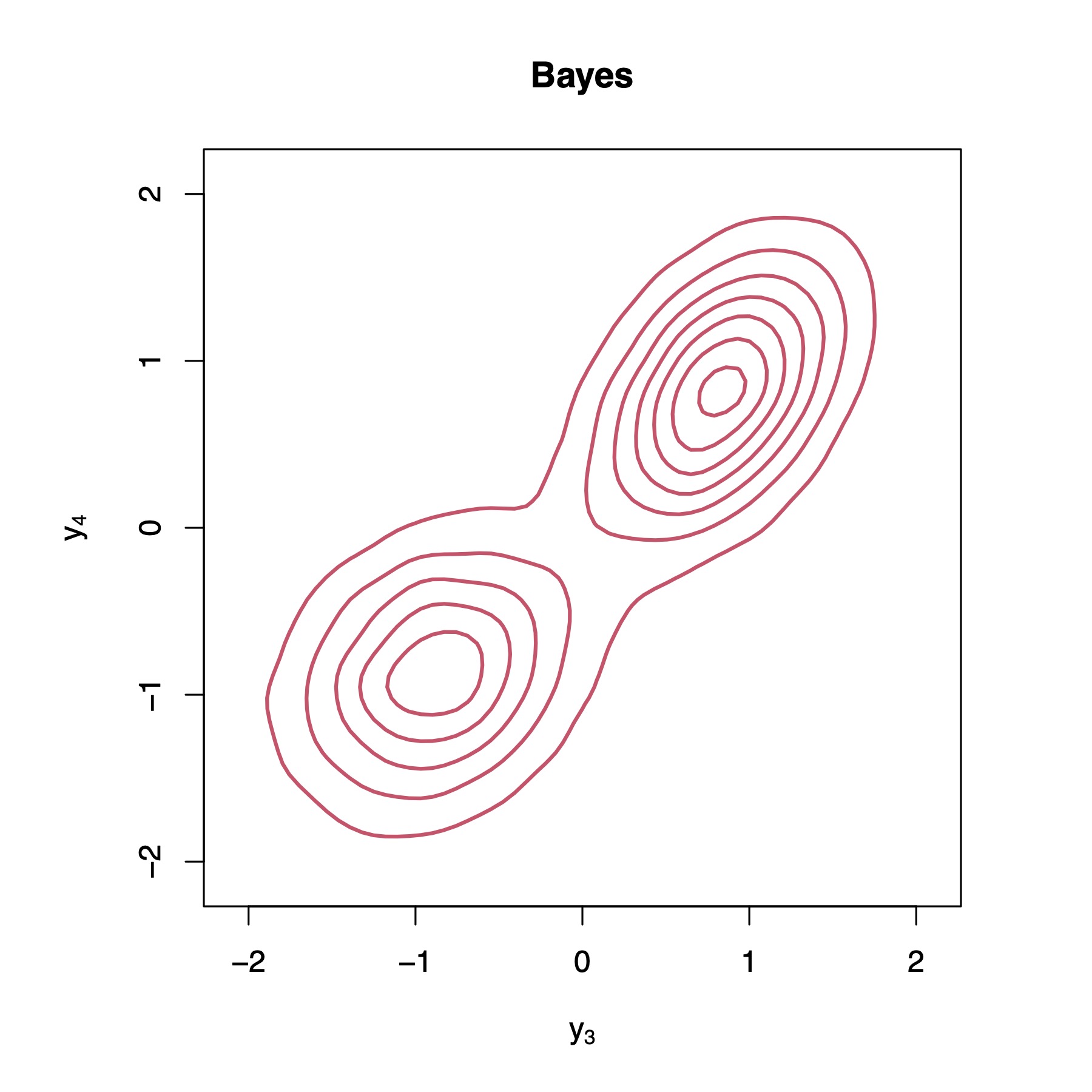}
\end{subfigure}
\hfill
\begin{subfigure}{0.45\textwidth}
    \includegraphics[width=\textwidth]{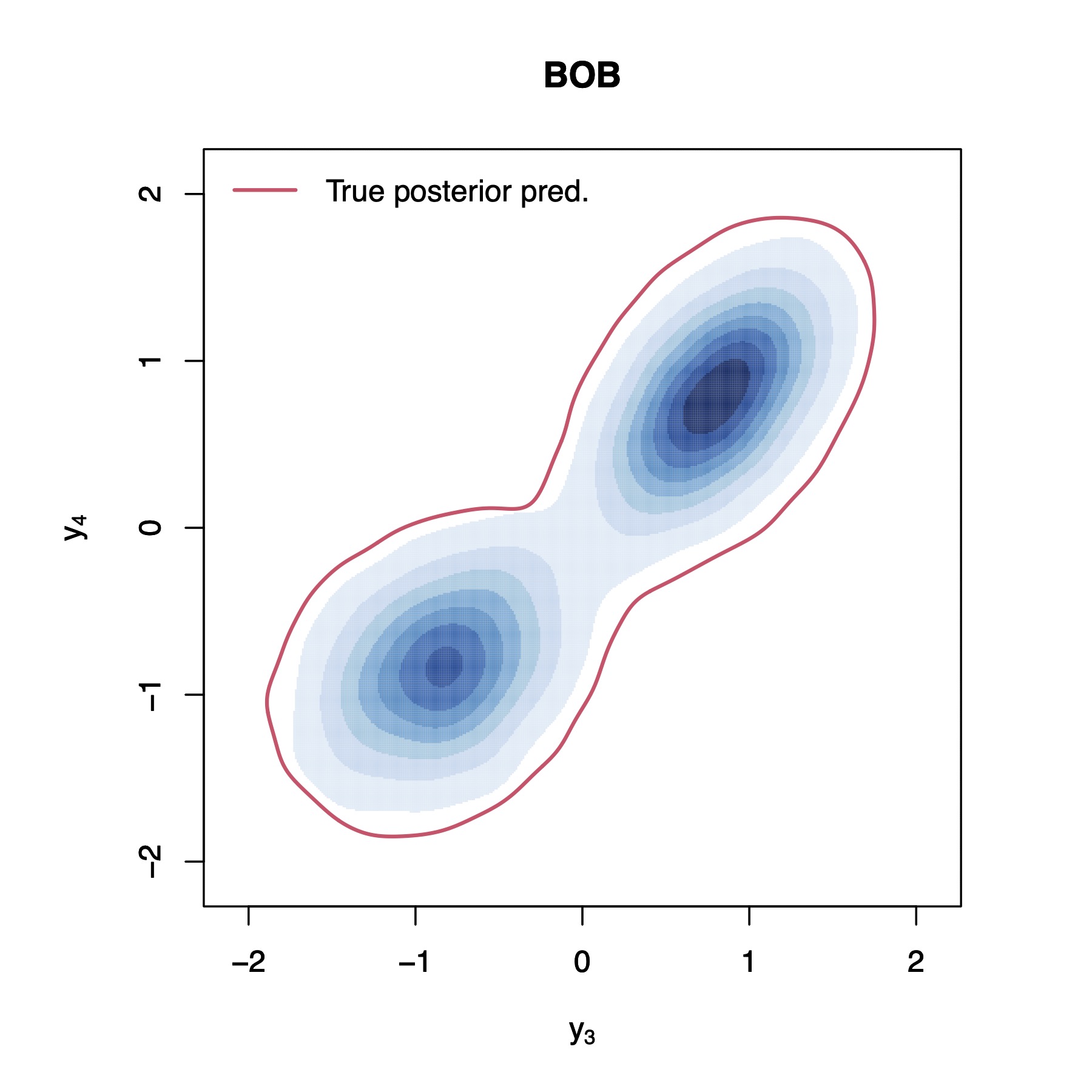}
\end{subfigure}
\hfill
\begin{subfigure}{0.45\textwidth}
    \includegraphics[width=\textwidth]{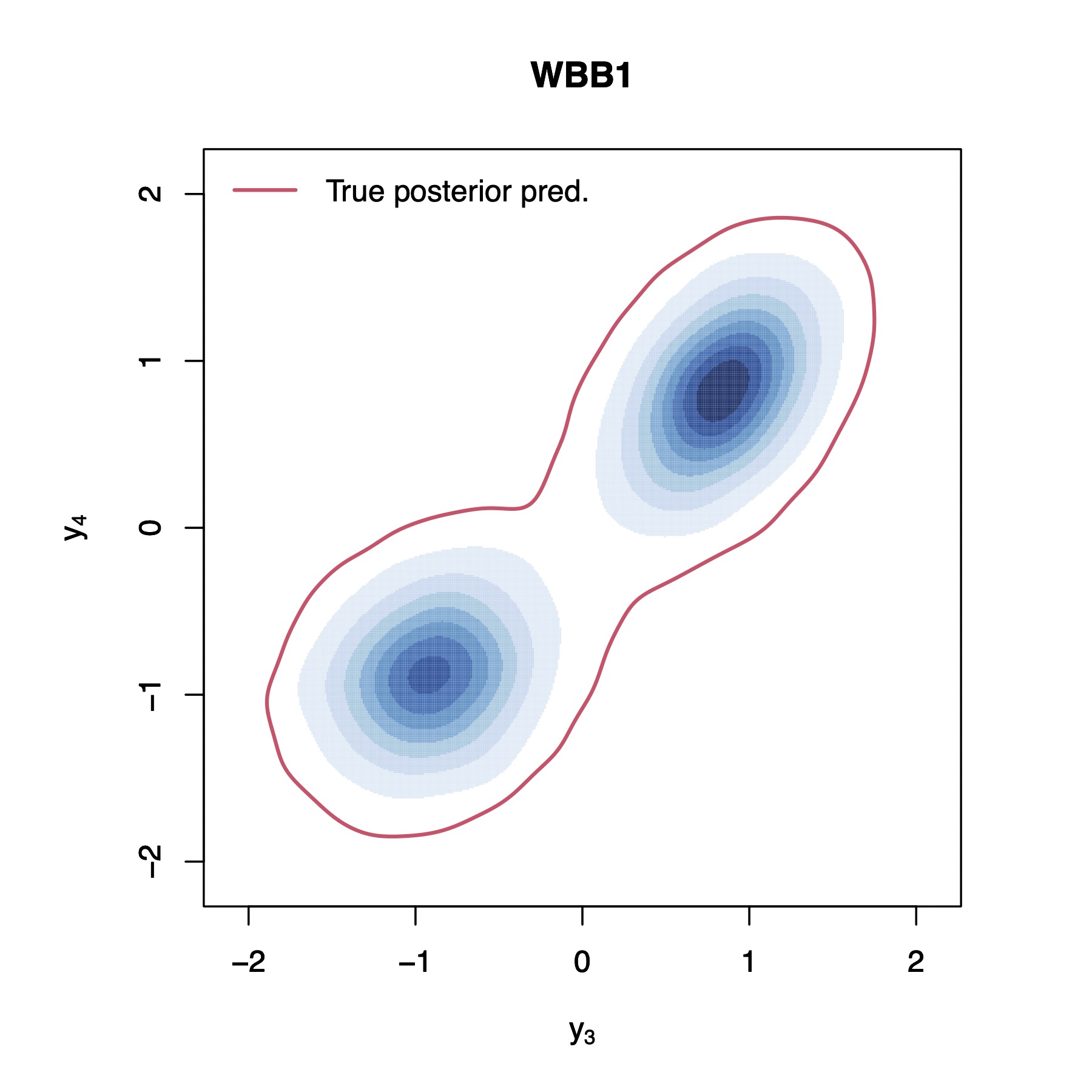}
\end{subfigure}
\hfill
\begin{subfigure}{0.45\textwidth}
    \includegraphics[width=\textwidth]{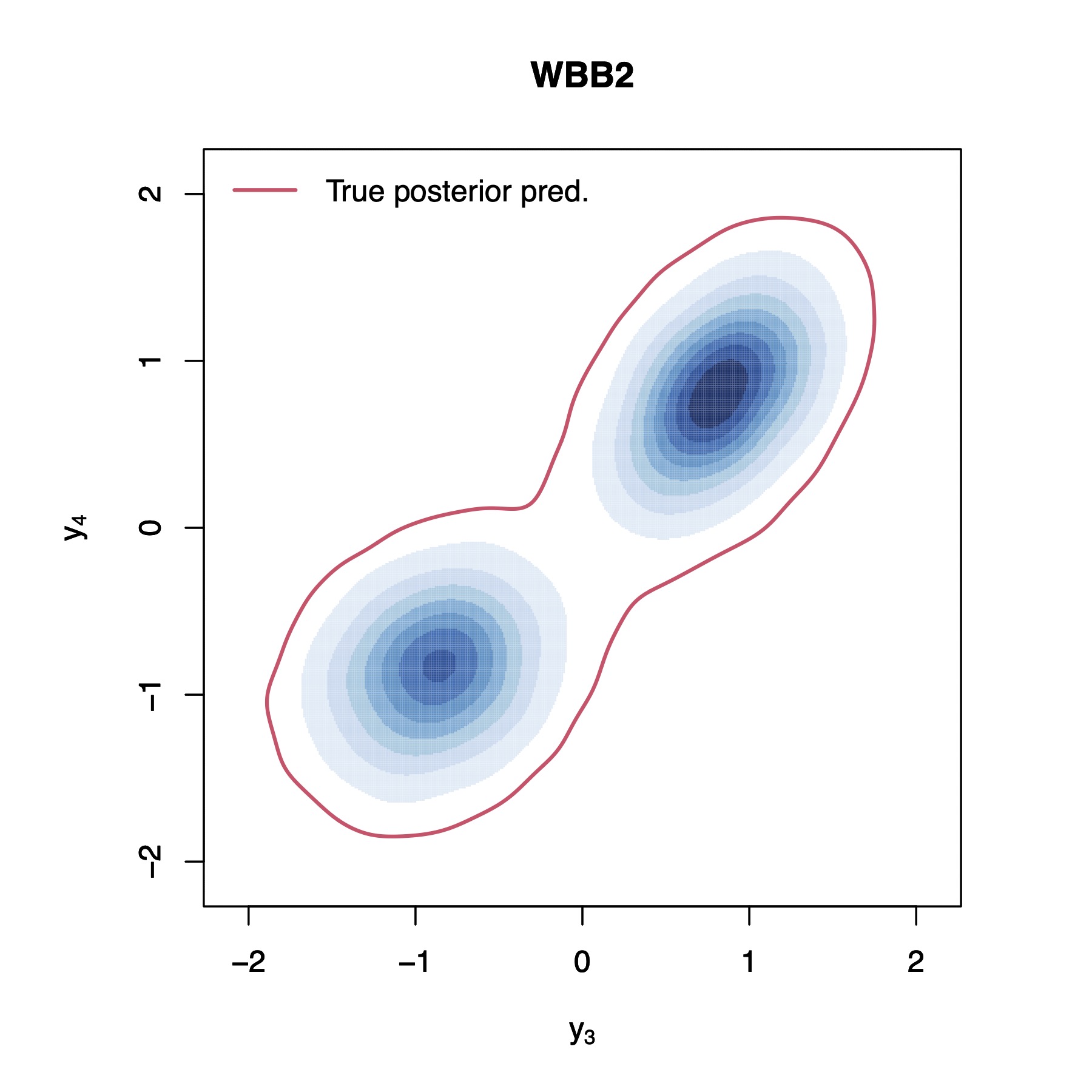}
\end{subfigure}
\hfill
\begin{subfigure}{0.45\textwidth}
    \includegraphics[width=\textwidth]{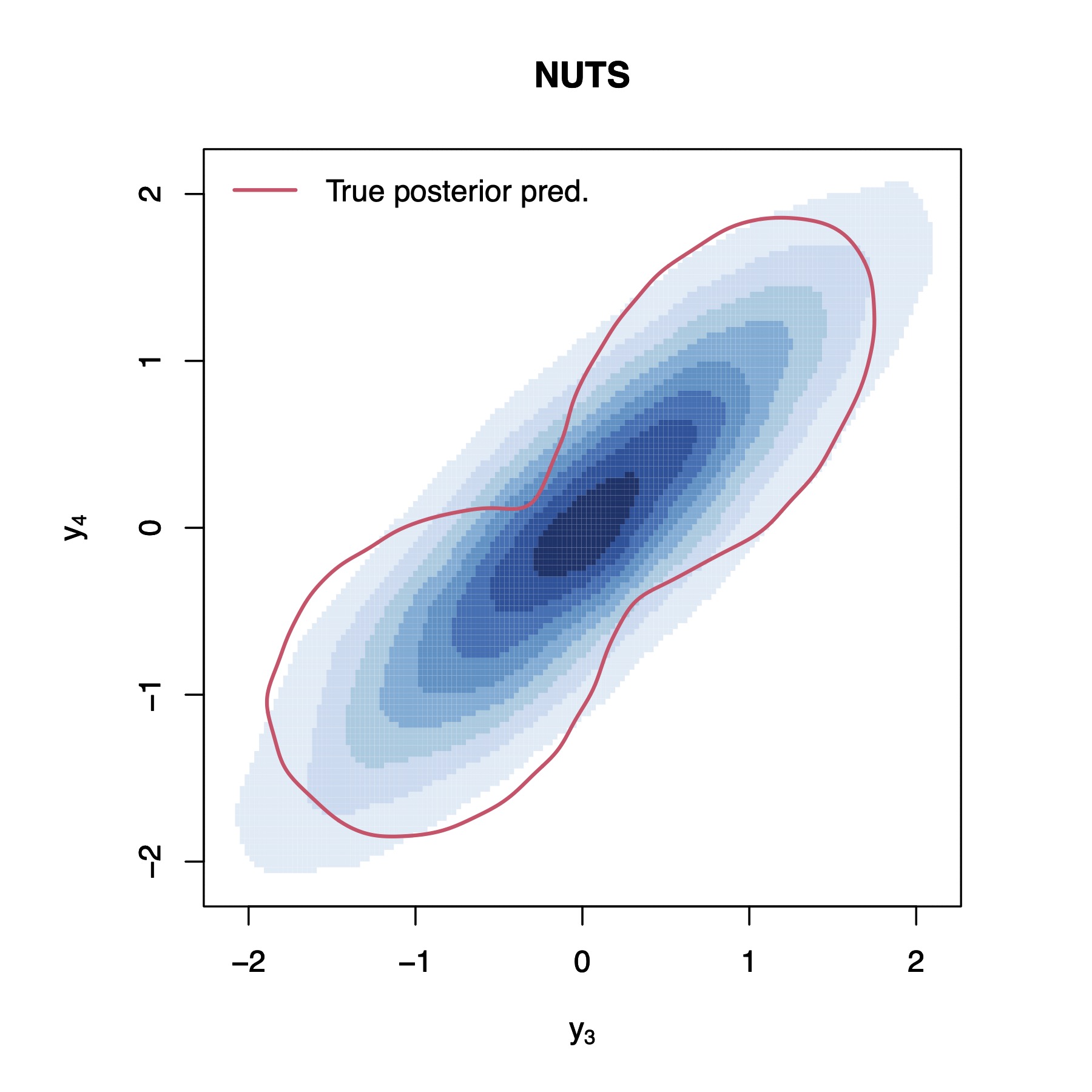}
\end{subfigure}
\hfill
\begin{subfigure}{0.45\textwidth}
    \includegraphics[width=\textwidth]{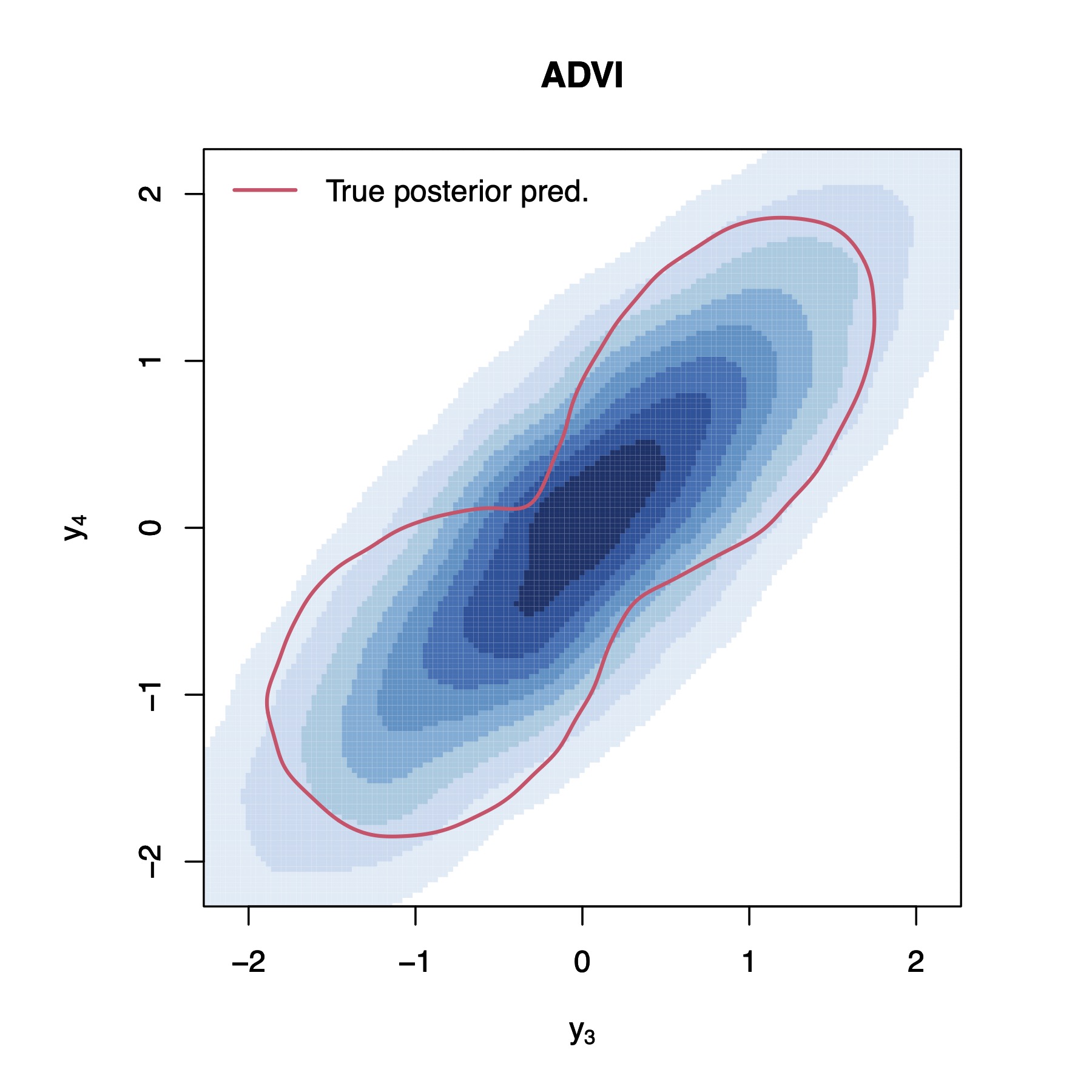}
\end{subfigure}
\caption{True Bayesian posterior predictive density (red contours) and its approximations (blue KDEs) obtained via BOB, WBB1, WBB2, NUTS, and ADVI, when $K=2$, $d = 10$, and $n = 50$.}
\label{fig:kdes_illustrative}
\end{figure*}

To provide a clearer demonstration of what is happening, let us bring up an illustrative example with $n = 50$, $d = 10$, and $K = 2$. Figure \ref{fig:kdes_illustrative} present KDEs of the true Bayesian posterior predictive distribution, as well as of the different approximations (details on how one can sample from the Bayesian posterior predictive distribution can be found in the supplementary materials). We can observe that the Bayesian posterior predictive density (i.e., the red contour plots) clearly indicates the existence of two clusters. We can also observe that WBB1 and WBB2 correctly capture the locations of such clusters, but do not correctly capture the dispersion of the Bayesian posterior. In fact, both versions of WBB produce overconfident posterior predictive distributions, which is not optimal for uncertainty quantification. Moreover, NUTS and ADVI fail to identify the two clusters in the data. Instead, they fit the data as one big cluster, which is not desirable either. These results are not surprising as the reparameterization used by ADVI forces the posterior to be unimodal \citep{morningstar2021automatic}. As a matter of fact, similar results for NUTS and ADVI have also been reported in \cite{fong2019scalable}, where the authors show that the samplers produce unimodal posteriors. Overall, it is clear that BOB produces the closest approximation to the Bayesian posterior predictive distribution and the best uncertainty quantification. Additional posterior predictive density plots are
provided in the supplementary materials, where we can observe similar results. 
\par
Table \ref{tab:illustration_results} presents the $\hat{\text{TV}}$ and $\hat{\text{KS}}$ distances between the true Bayesian posterior predictive distribution and its approximations, as well as the elapsed (wall-clock) times, for our illustrative example with  $n = 50$, $d = 10$, and $K = 2$. We can observe that BOB not only yields the smallest distances, but it is also 2.4 times faster than NUTS.

\begin{table}[h]
\caption{$\hat{\text{TV}}$ and $\hat{\text{KS}}$ distances between the true Bayesian posterior predictive distribution and its approximations obtained via BOB, WBB1, WBB2, NUTS, and ADVI, as well as elapsed (wall-clock) times in minutes for each method, when $K=2$, $d = 10$, and $n = 50$.}\label{tab:illustration_results}
\begin{tabular*}{\textwidth}{@{\extracolsep\fill}lccccc}
\toprule%
& \multicolumn{5}{@{}c@{}}{Method} \\
\cmidrule{2-6}%
Metric & BOB & WBB1 & WBB2 & NUTS & ADVI \\
\midrule
$\hat{\text{TV}}$ & 0.047 & 0.070 & 0.071 & 0.054 & 0.099 \\
$\hat{\text{KS}}$ & 0.040 & 0.055 & 0.055 & 0.055 & 0.059 \\
Elapsed (min)  & 2.136 & 0.005 & 0.005 & 5.144 & 0.083 \\
\botrule
\end{tabular*}
\end{table}

\begin{figure}
\begin{center}
\includegraphics[width=\textwidth]{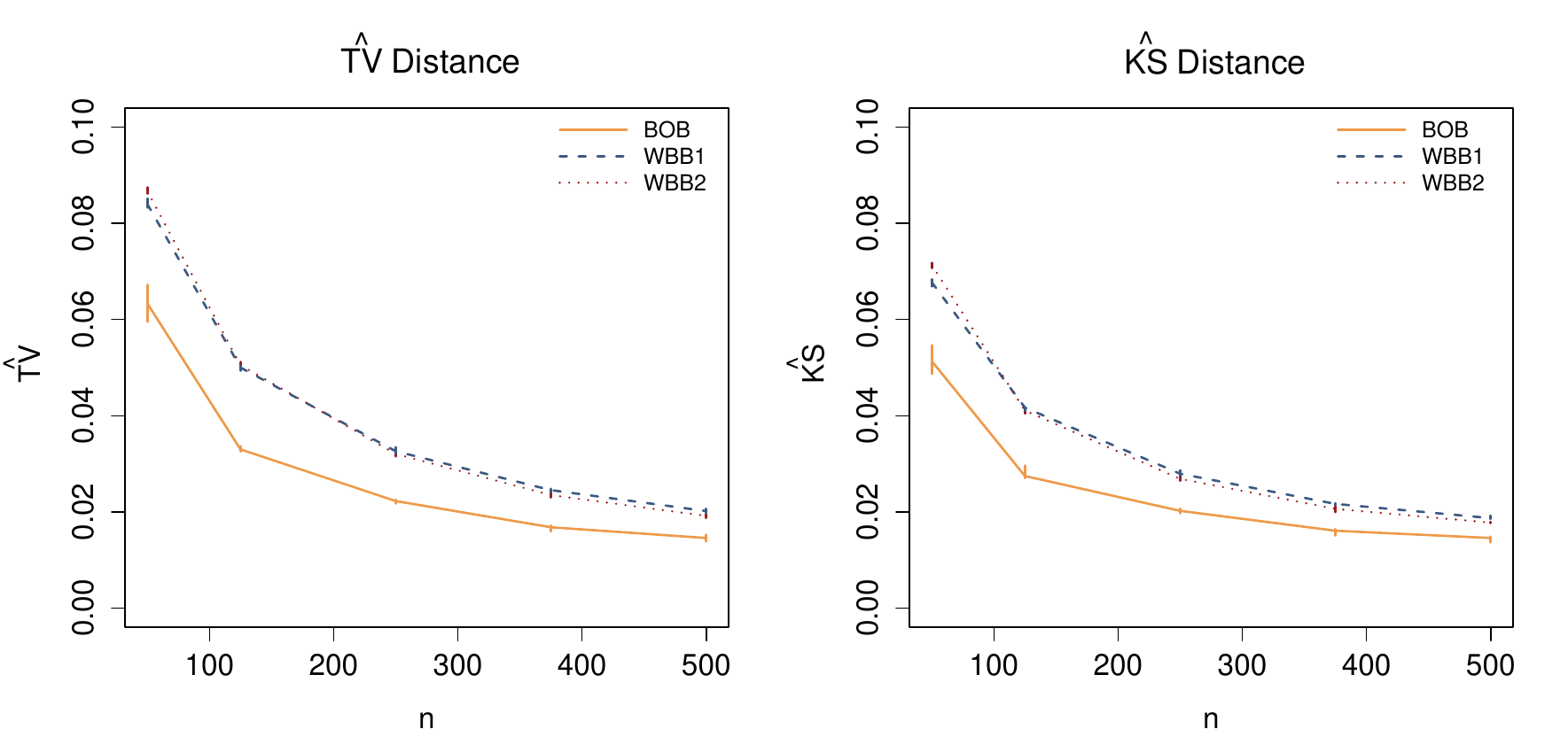}
\end{center}
\caption{$\hat{\text{TV}}$ and $\hat{\text{KS}}$ distances between the Bayesian posterior predictive and its approximations obtained via BOB (golden continuous line), WBB1 (blue dashed line), and WBB2 (red dotted line), as a function of $n$, with $d=15$ and $K=2$. The curves were produced by computing the median across ten independent runs. Error bars indicate the 25th and 75th percentiles, respectively.}
\label{fig:sims_varying_n}
\end{figure}

Lastly, we would like to investigate how the sample size affects both BOB and WBB. As discussed in section \ref{subsec:theory}, both BOB and WBB are first order correct for the Bayesian posterior. In other words, their approximations should get better with a growing sample size. Thus, we consider an additional experiment where we fix $d=15$ and $K=2$, and consider various sample sizes spanning from $n=50$ to $n=500$. Figure \ref{fig:sims_varying_n} displays the $\hat{\text{TV}}$ and $\hat{\text{KS}}$ distances between the Bayesian posterior predictive distribution and its approximations obtained via BOB, WBB1, and WBB2, as a function of $n$. We can observe that the $\hat{\text{TV}}$ and $\hat{\text{KS}}$ distances decrease as $n$ increases, which is expected. However, we can see that BOB constantly outperforms WBB, for all values of $n$. Asymptotically, though, we can observe that WBB and BOB tend to converge to the same limiting distribution, which is desirable as WBB possesses appealing asymptotic properties. Altogether, we have that BOB is a reliable method for approximate posterior sampling, which can be applied in a wide range of problems with different sample sizes.

\section{Analysis of Real-world Data}
\label{sec:data_analysis}
To further demonstrate the performance and practical utility of our proposed methods, we apply them to the widely analyzed \textit{Wine} \citep{AEBERHARD1994_wines} and \textit{Seeds} \citep{seeds2010complete} datasets. Both datasets are publicly available from the \href{https://archive.ics.uci.edu/}{\textcolor{DarkBlue}{UC Irvine Machine Learning Repository}}.

\subsection{Wine Data}
\label{subsec:wine}
The data consists of $d=13$ chemical properties for $178$ Italian wines, belonging to $K=3$ different types, namely Barbera, Barolo, and Grignolino. From the Barbera type we have 48 specimens, from Barolo we have 59 specimens, and from Grignolino we have 71 specimens. A detailed description of all the chemical properties from each wine is presented in the supplementary materials. The idea is to cluster types of wine based on these chemical features. To assess the validity of our model specification and model fitting, we randomly split our data into training and held-out sets, with $n=100$ observations in the training set. In all our analyses, we standardize the data so that each feature has mean 0 and variance 1. To set the prior hyperparameters, we follow the same approach as in section \ref{subsec:sim_setup}. We obtain $S=20000$ approximate posterior draws using BOB, WBB1, WBB2, and ADVI. In the case of NUTS, we run the algorithm for 40000 iterations and discard the first half as \textit{burn-in}. For BOB, we use batches of size $S_{b}=4000$ to construct $\hat{\mathcal{L}}(\boldsymbol{x})$. As before, we set the lower and upper bounds of the BO search space to be $\boldsymbol{\mathcal{X}}_{\text{lower}}=(1, 10^{-5}, \dots, 10^{-5})'$ and $\boldsymbol{\mathcal{X}}_{\text{upper}}=(1.5, \dots, 1.5)'$, respectively.
\subsection{Seeds Data}
\label{subsec:seeds}
Our second dataset is made up of $d=7$ observed measurements of geometrical properties for $210$ kernels belonging to $K=3$ different varieties of wheat, namely, Kama, Rosa and Canadian. There are 70 kernels of each variety and the idea is to cluster varieties of wheat based on these observed measurements. Further details on these observed measurements are presented in the supplementary materials. Again, we randomly split our data into training and held-out sets, with $n=110$ observations in the training set. All other configurations are set as in subsection \ref{subsec:wine}.
\subsection{Results}
\label{subsec:data_results}

\begin{figure*}[!htp]
\centering
\begin{subfigure}{0.45\textwidth}
    \includegraphics[width=\textwidth]{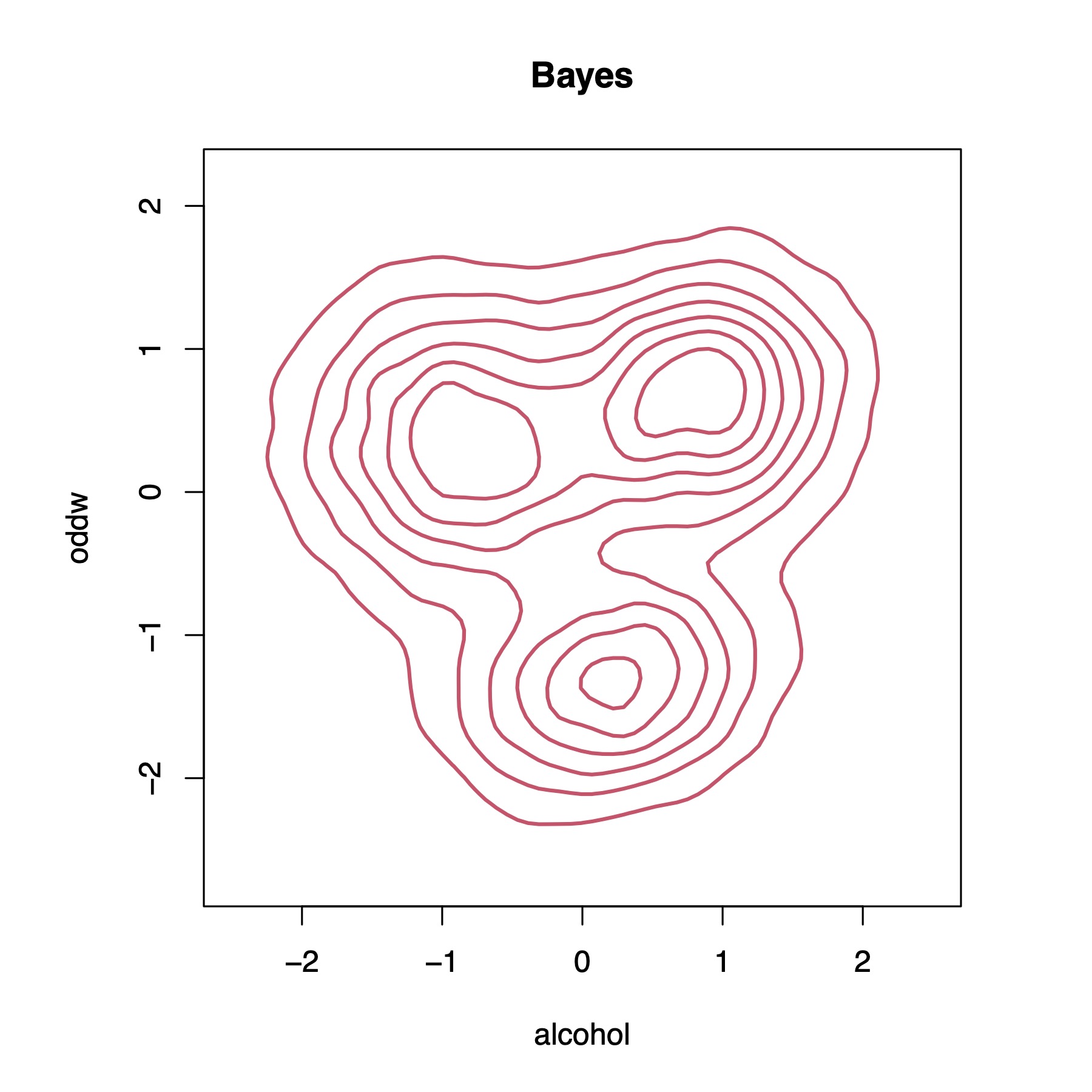}
\end{subfigure}
\hfill
\begin{subfigure}{0.45\textwidth}
    \includegraphics[width=\textwidth]{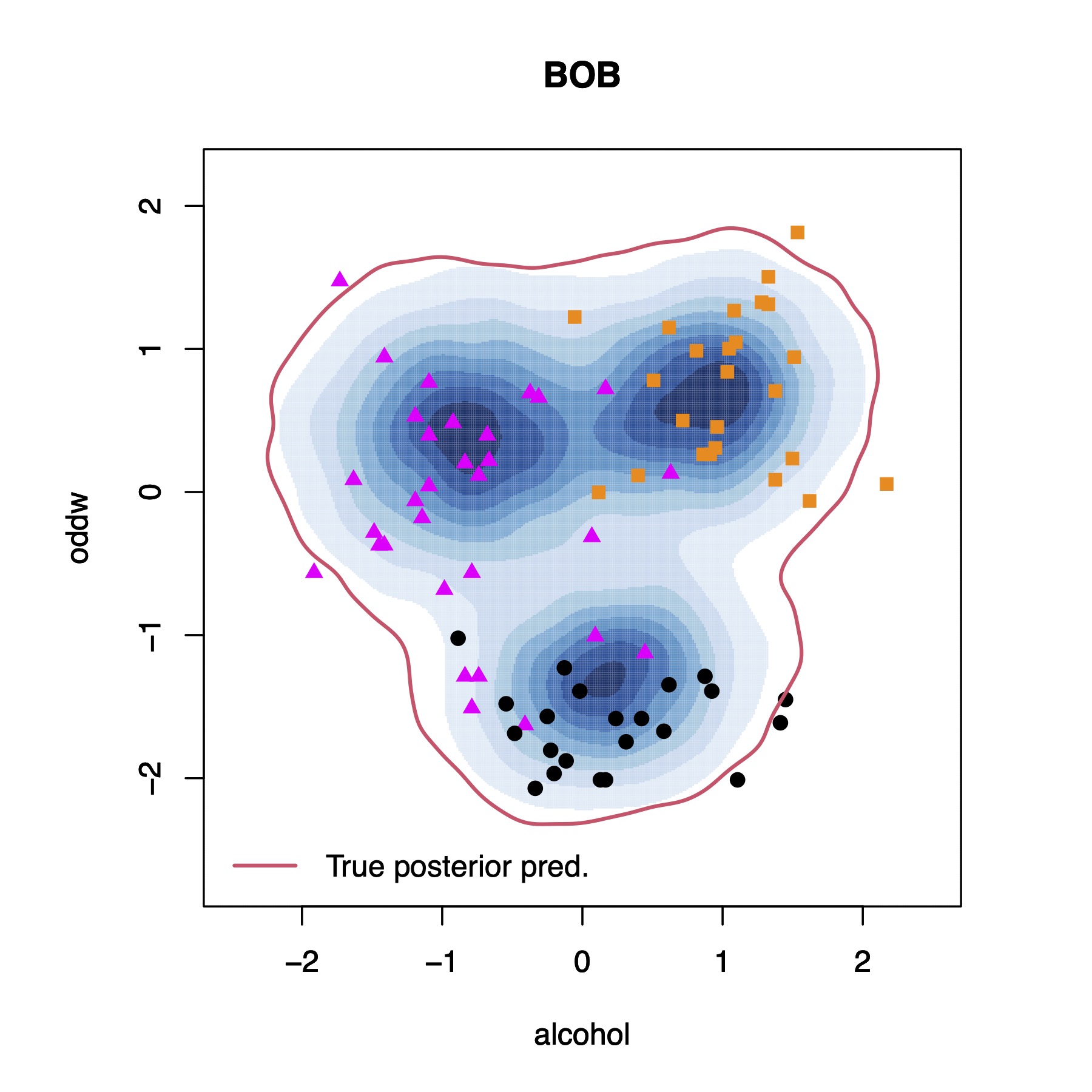}
\end{subfigure}
\hfill
\begin{subfigure}{0.45\textwidth}
    \includegraphics[width=\textwidth]{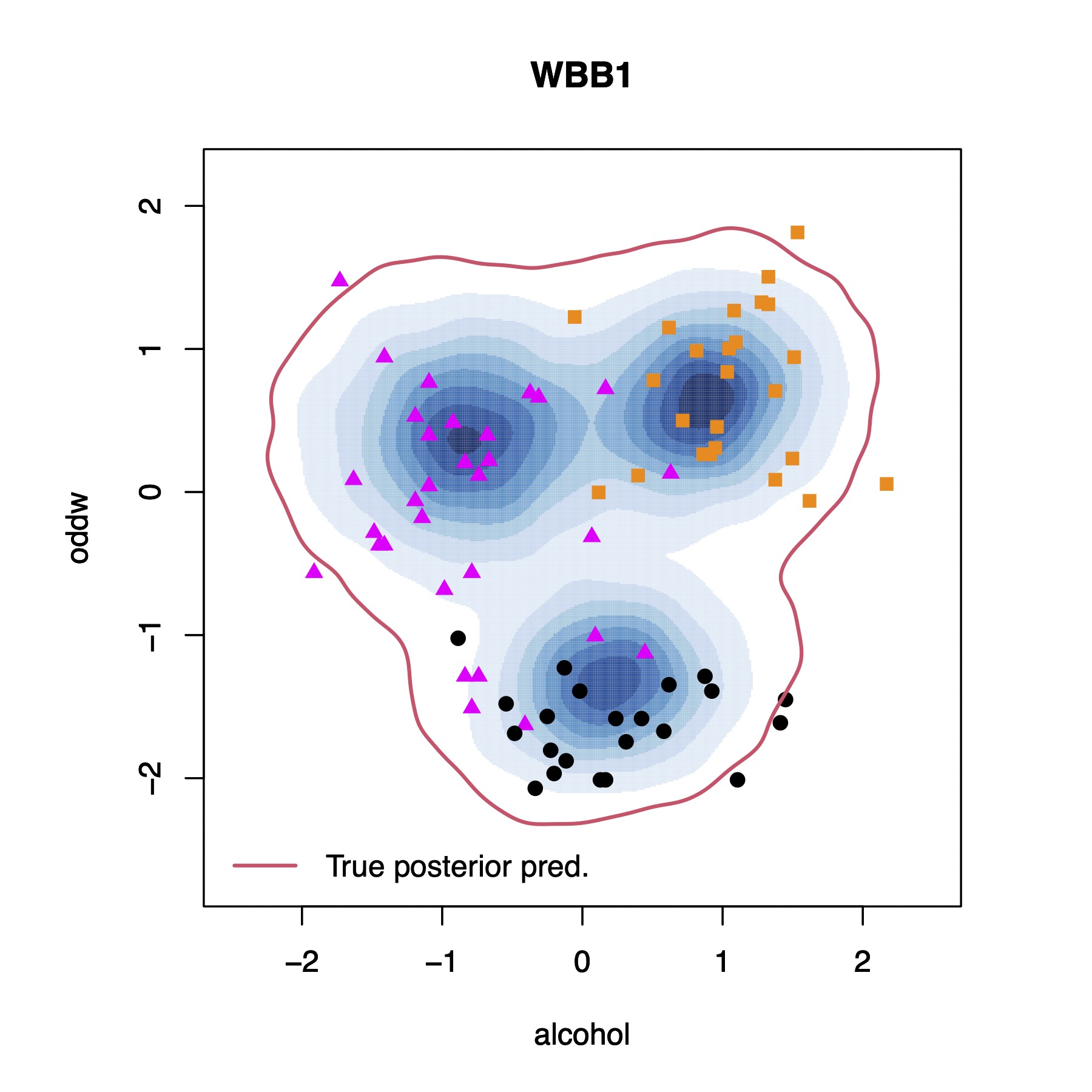}
\end{subfigure}
\hfill
\begin{subfigure}{0.45\textwidth}
    \includegraphics[width=\textwidth]{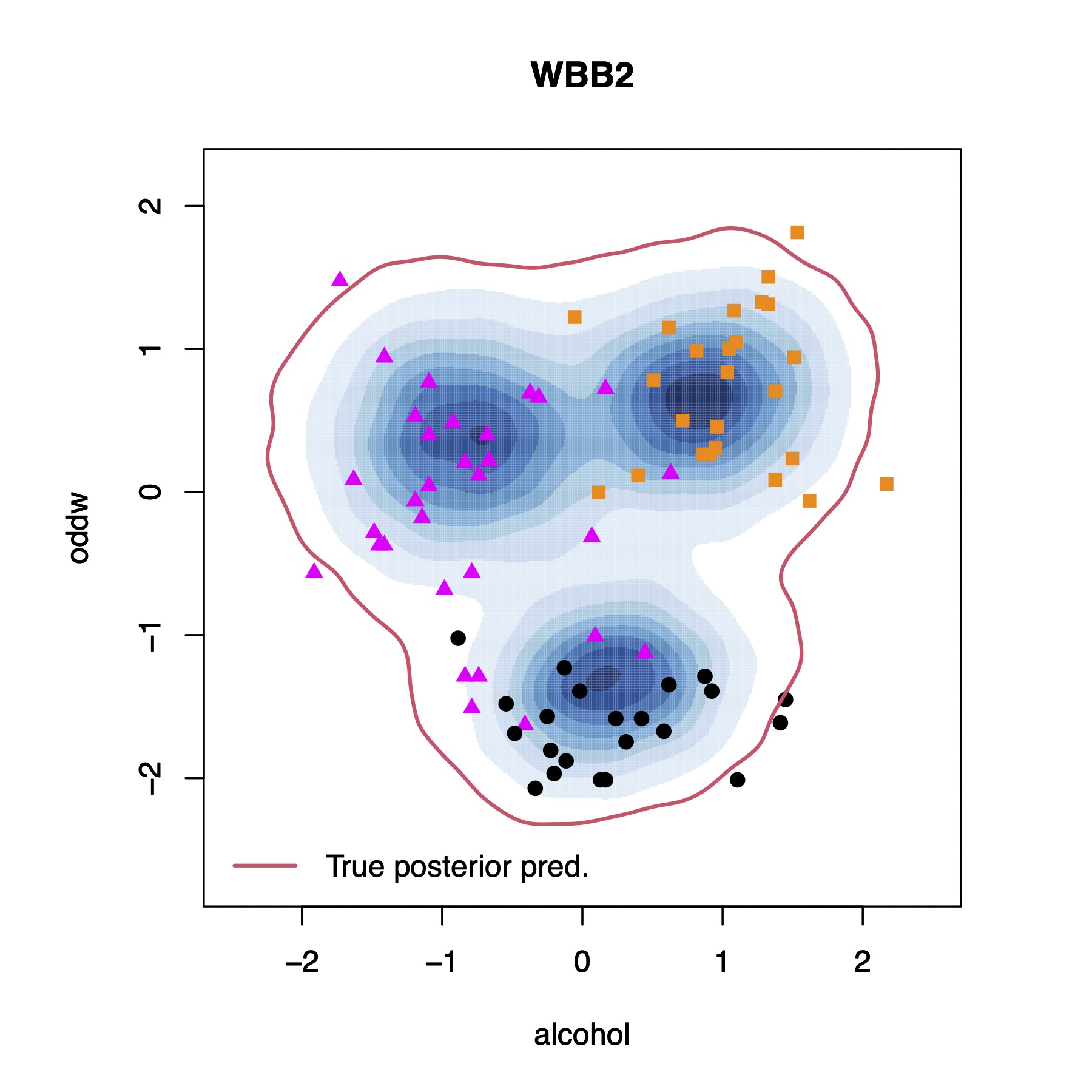}
\end{subfigure}
\hfill
\begin{subfigure}{0.45\textwidth}
    \includegraphics[width=\textwidth]{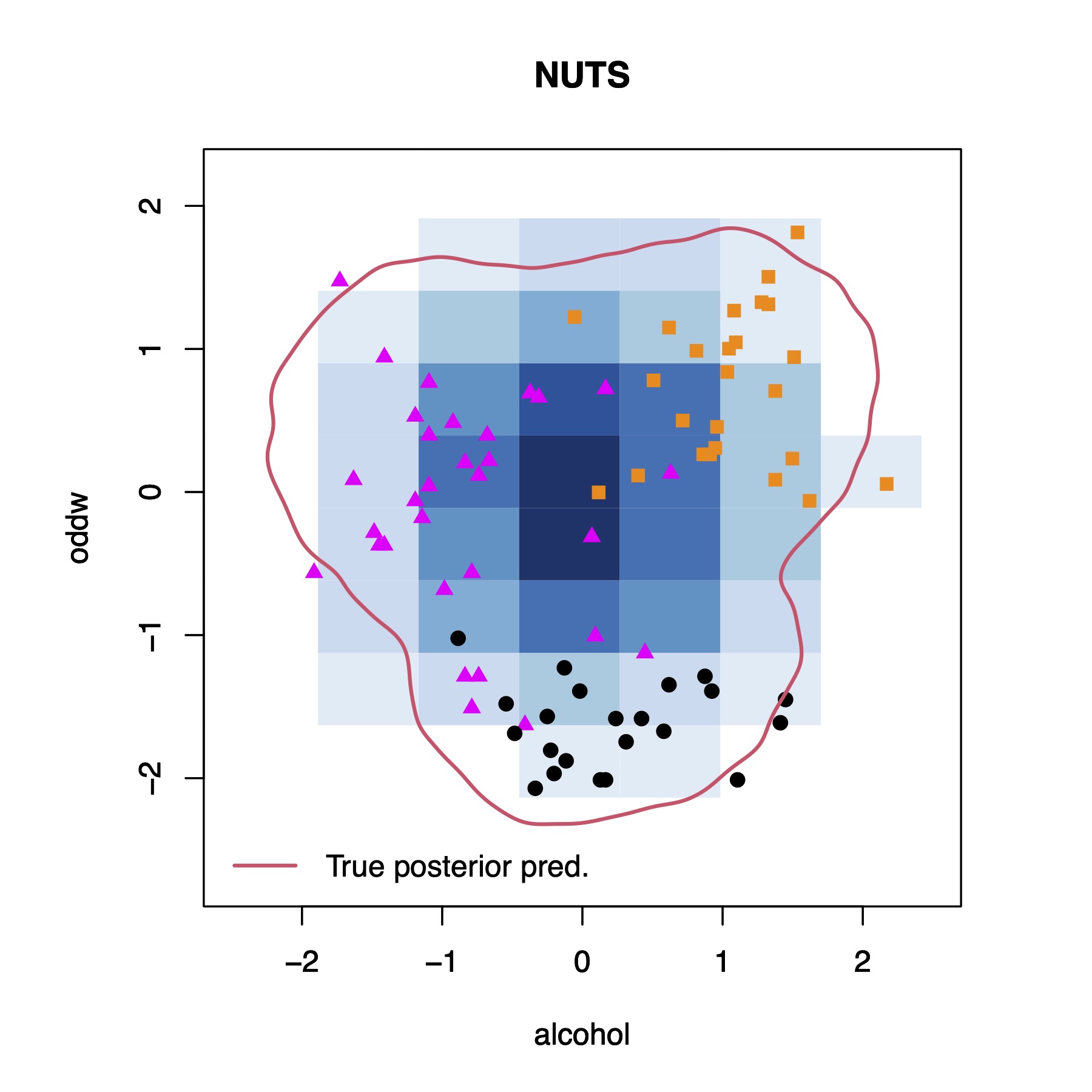}
\end{subfigure}
\hfill
\begin{subfigure}{0.45\textwidth}
    \includegraphics[width=\textwidth]{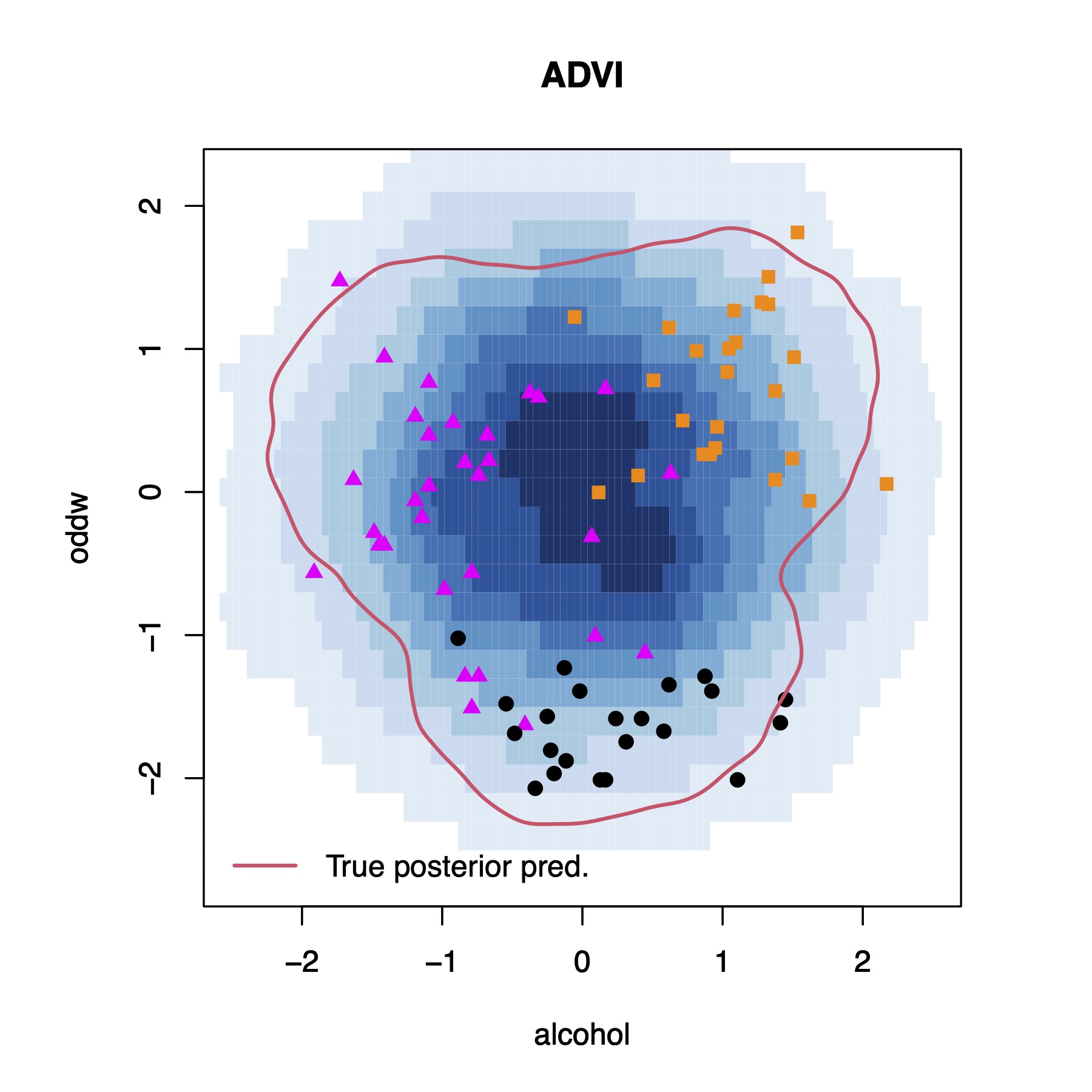}
\end{subfigure}
\caption{True Bayesian posterior predictive density (red contours) and its approximations (blue KDEs) obtained via BOB, WBB1, WBB2, NUTS, and ADVI, for selected variables from the wine data. Contours and KDEs were computed using only the training data. Scatter plots depict the held-out data. Orange squares, pink triangles and dark circles represent Barolo, Grignolino, and Barbera wines, respectively.}
\label{fig:densities_wine}
\end{figure*}

\begin{figure*}[!htp]
\centering
\begin{subfigure}{0.45\textwidth}
    \includegraphics[width=\textwidth]{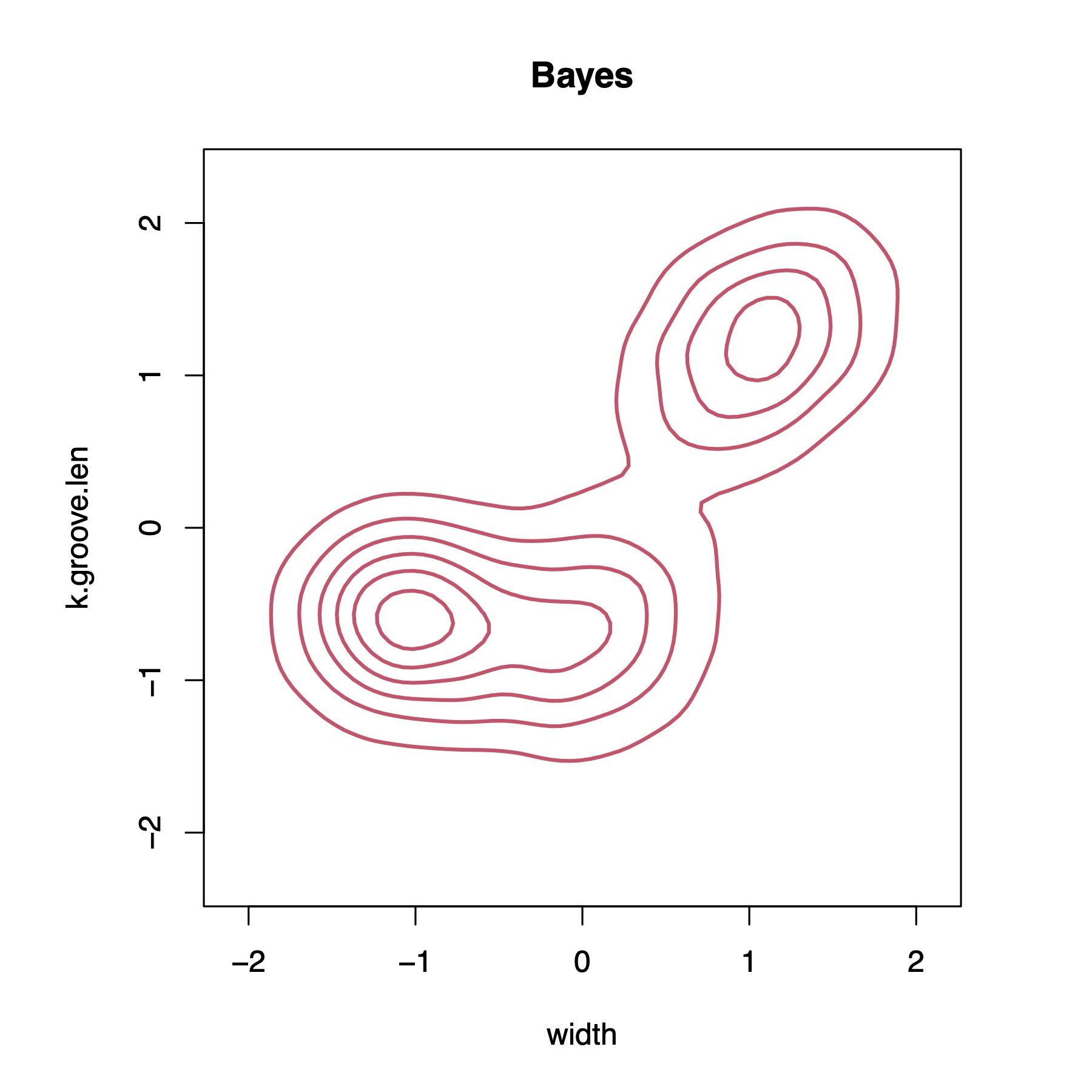}
\end{subfigure}
\hfill
\begin{subfigure}{0.45\textwidth}
    \includegraphics[width=\textwidth]{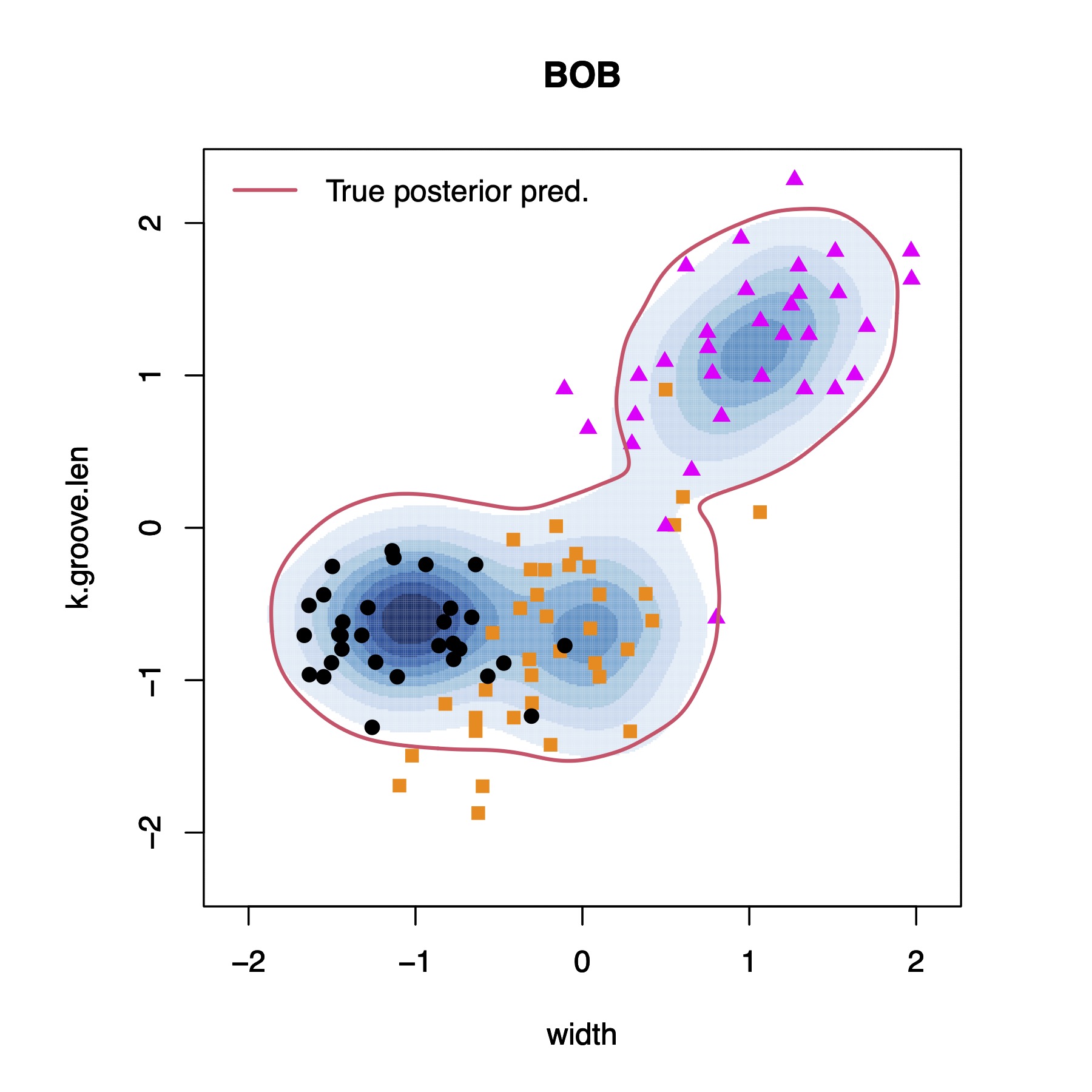}
\end{subfigure}
\hfill
\begin{subfigure}{0.45\textwidth}
    \includegraphics[width=\textwidth]{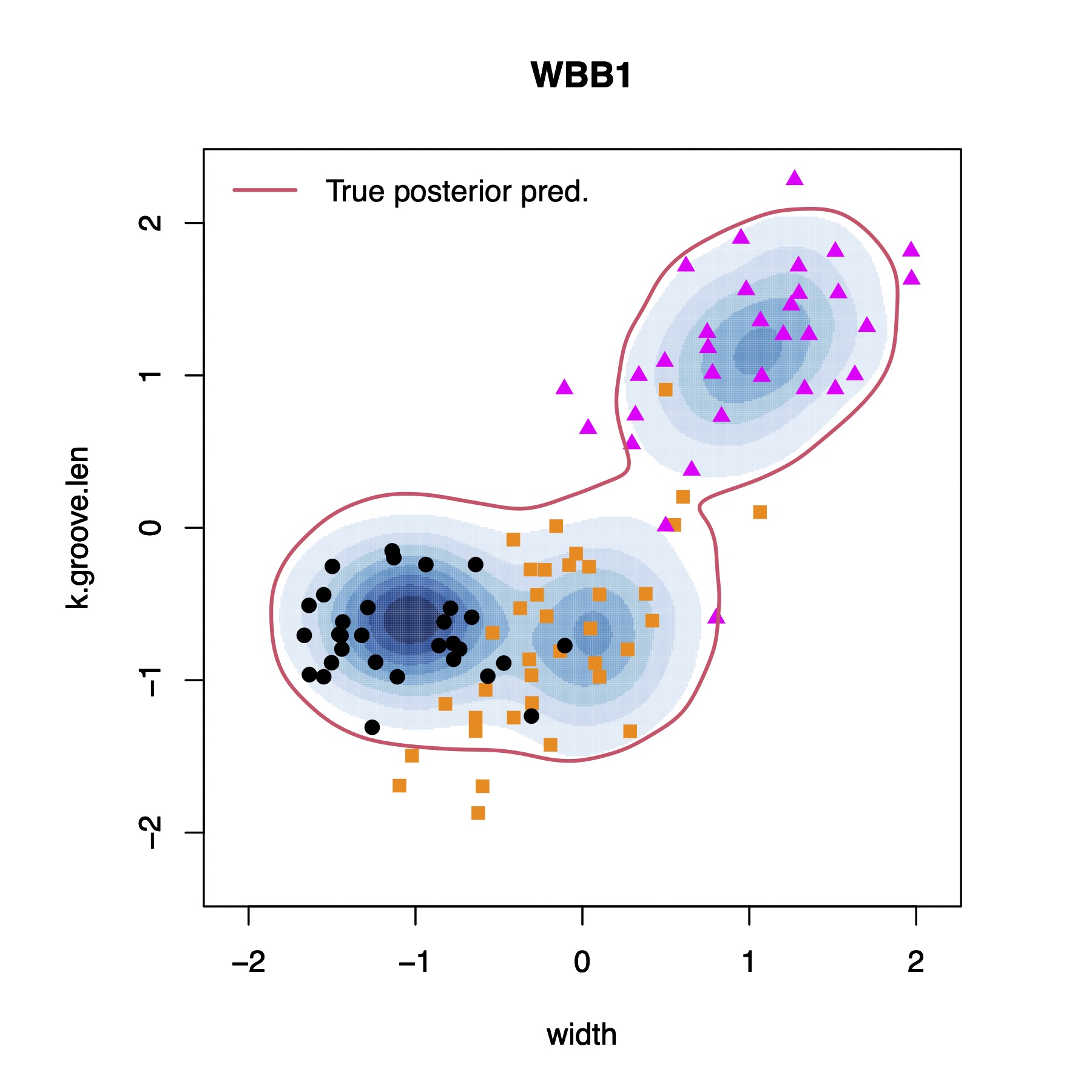}
\end{subfigure}
\hfill
\begin{subfigure}{0.45\textwidth}
    \includegraphics[width=\textwidth]{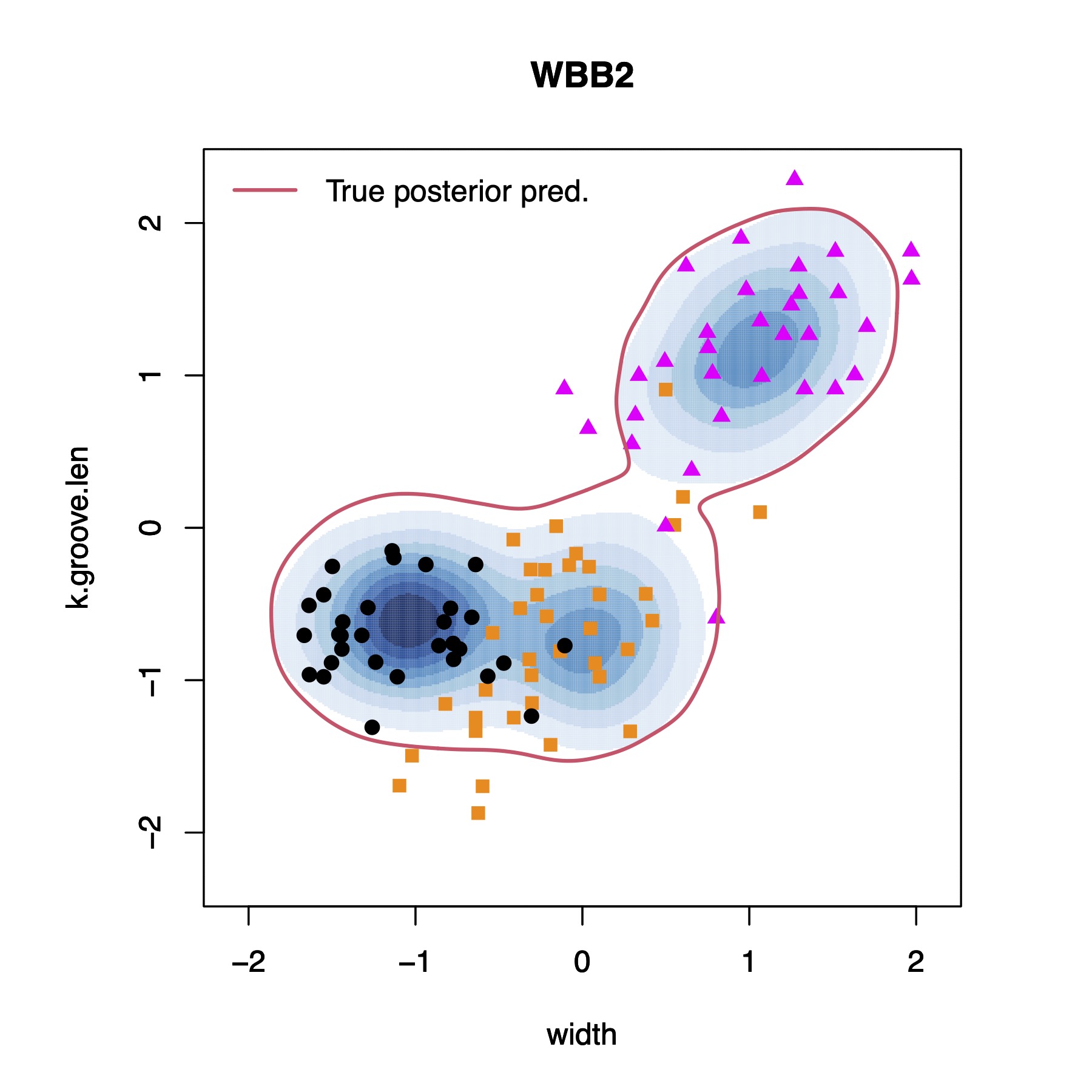}
\end{subfigure}
\hfill
\begin{subfigure}{0.45\textwidth}
    \includegraphics[width=\textwidth]{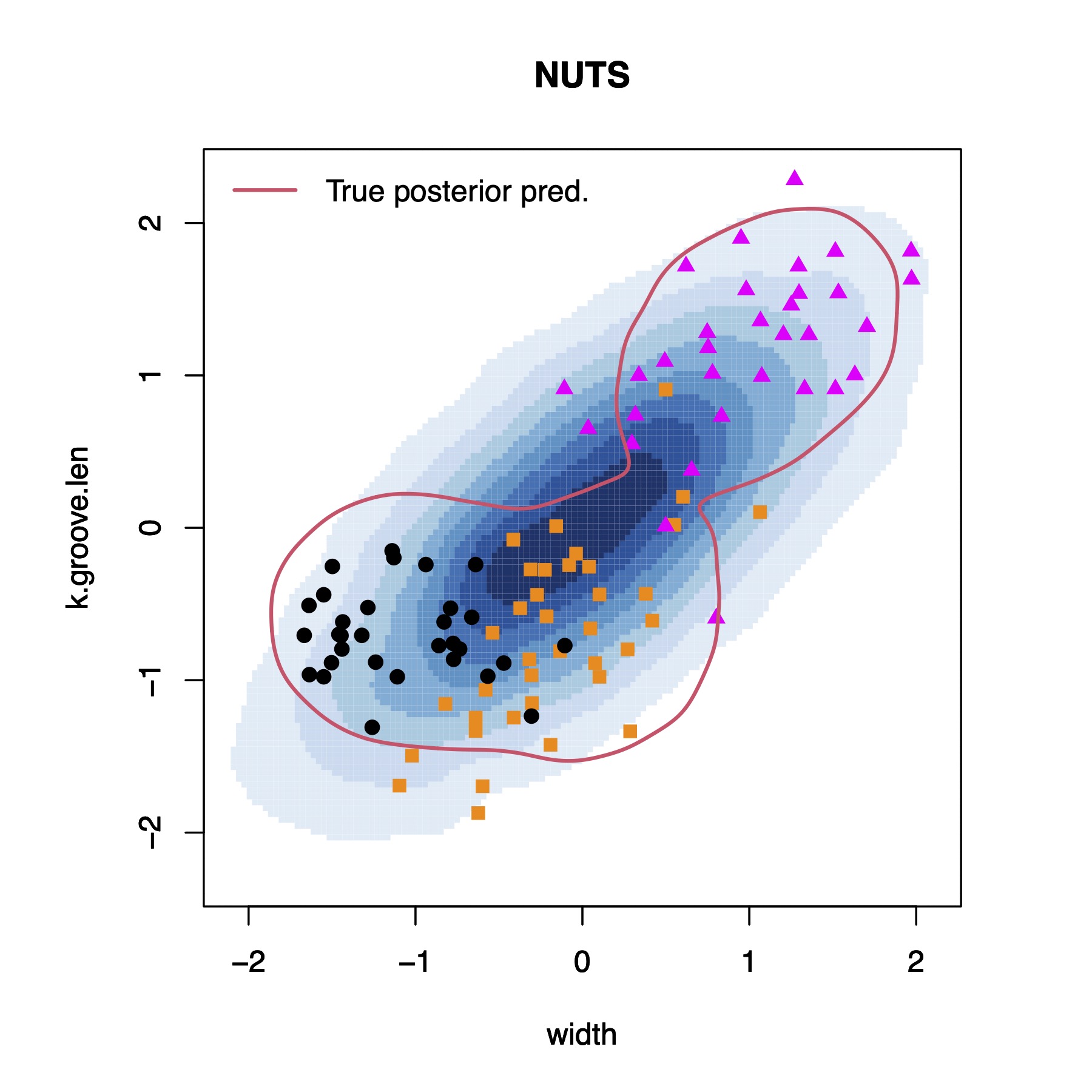}
\end{subfigure}
\hfill
\begin{subfigure}{0.45\textwidth}
    \includegraphics[width=\textwidth]{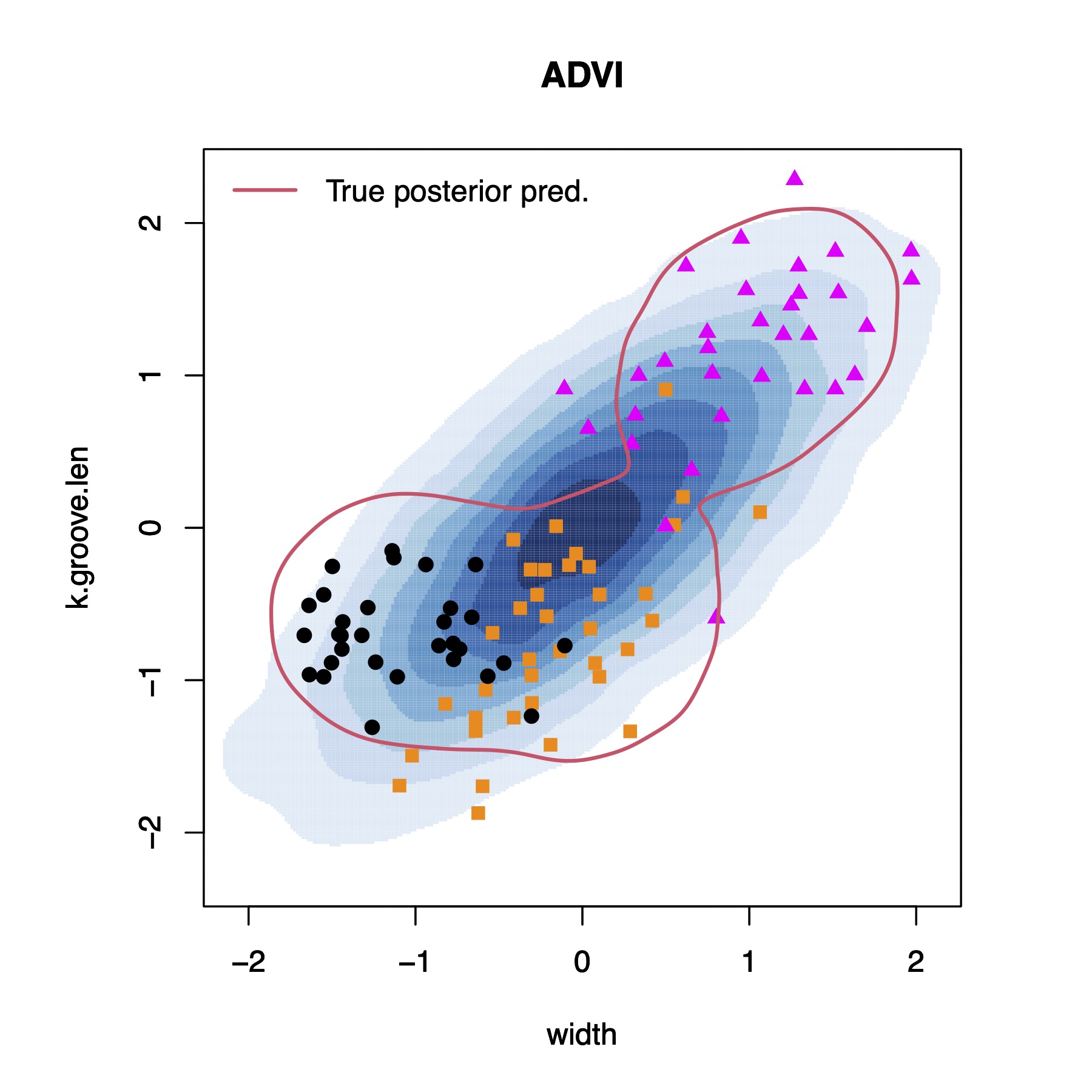}
\end{subfigure}
\caption{True Bayesian posterior predictive density (red contours) and its approximations (blue KDEs) obtained via BOB, WBB1, WBB2, NUTS, and ADVI, for selected variables from the seeds data. Contours and KDEs were computed using only the training data. Scatter plots depict the held-out data. Orange squares, pink triangles and dark circles represent  Kama, Rosa, and Canadian wheat kernels, respectively.}
\label{fig:densities_seeds}
\end{figure*}

Figures \ref{fig:densities_wine} and \ref{fig:densities_seeds} present posterior predictive density plots for selected variables from the wine and seeds datasets, respectively. As before, the red contours represent the true Bayesian posterior predictive density, while the blue KDEs represent the different approximations. These contours and KDEs were obtained using only the training data. On the other hand, the scatter plots represent the held-out data, which can help us evaluate how well the different methods generalize to unseen data and how well the different approximations recover the true underlying data generating mechanism. Different symbols and colors in the scatter plots represent the actual labels of the held-out data. For instance, in Figure \ref{fig:densities_wine}, orange squares, pink triangles and dark circles represent Barolo, Grignolino, and Barbera wines, respectively. In Figure \ref{fig:densities_seeds}, orange squares, pink triangles and dark circles represent Kama, Rosa, and Canadian wheat kernels, respectively.

In Figures \ref{fig:densities_wine} and \ref{fig:densities_seeds}, the Bayesian posterior predictive density clearly indicates the existence of three clusters, which align with the held-out data, suggesting that the model assumptions are reasonable. We can also observe that both versions of WBB correctly capture the location of these clusters. However, both versions of WBB produce overconfident posterior predictive distributions and do not capture the dispersion of the Bayesian posterior. NUTS and ADVI are unable to identify the three clusters in the data. Having said that, it is clear that in the wine and seeds datasets, BOB produces the closest approximation to the Bayesian posterior predictive distribution and the best uncertainty quantification. Note that these results are consistent with the results from section \ref{sec:sims}. Additional posterior density plots are provided in the supplementary materials, where we can observe similar patterns across different variables from the wine and seeds datasets.

Moreover, Table \ref{tab:real_data_results} presents elapsed (wall-clock) times in minutes, as well as the $\hat{\text{TV}}$ and $\hat{\text{KS}}$ distances between the Bayesian posterior predictive distribution and its approximations obtained via BOB, WBB1, WBB2, NUTS, and ADVI, for the wine and seeds datasets. We can observe that BOB returns the smallest $\hat{\text{TV}}$ and $\hat{\text{KS}}$ distances in both datasets. Additionally, note that in the wine dataset, NUTS takes 69.4 minutes to run. BOB, on the other hand, takes only 3.16 minutes. This is a 22-fold reduction in the running time, suggesting that BOB not only outperforms competing approaches in recovering the Bayesian posterior predictive distribution, but it can also be substantially faster than MCMC samplers. On the whole, this illustrates BOB's practical utility in real-world problems.

\begin{table}[h]
\caption{$\hat{\text{TV}}$ and $\hat{\text{KS}}$ distances between the Bayesian posterior predictive distribution and its approximations obtained via BOB, WBB1, WBB2, NUTS, and ADVI, as well as elapsed (wall-clock) times in minutes, for the wine and seeds datasets.}\label{tab:real_data_results}
\begin{tabular*}{\textwidth}{@{\extracolsep\fill}llccccc}
\toprule%
& & \multicolumn{5}{@{}c@{}}{Method} \\
\cmidrule{3-7}%
Dataset & Metric & BOB & WBB1 & WBB2 & NUTS & ADVI \\
\midrule
Wine&$\hat{\text{TV}}$ & 0.039 & 0.056 & 0.058 & 0.065 & 0.117 \\
    &$\hat{\text{KS}}$ & 0.032 & 0.048 & 0.049 & 0.047 & 0.071 \\
    & Elapsed (min)   & 3.158 & 0.010 & 0.007 & 69.399 & 0.256 \\
\midrule
Seeds&$\hat{\text{TV}}$& 0.019 & 0.025 & 0.023 & 0.072 & 0.071 \\
     &$\hat{\text{KS}}$& 0.020 & 0.027 & 0.023 & 0.079 & 0.079 \\
& Elapsed (min)& 2.869 & 0.005 & 0.005 & 13.657 & 0.131 \\
\botrule
\end{tabular*}
\end{table}

\section{Discussion}
\label{sec:conc}

In this article, we have investigated alternatives to MCMC algorithms in order to sample from the posterior distribution of GMMs. More precisely, we developed a randomly weighted EM algorithm and introduced BOB, a novel computational methodology for approximate posterior sampling. We build on WLB and WBB ideas, which are based on the premise that optimizing randomly weighted posterior densities can be faster and less computationally intensive than sampling from an intractable posterior. BOB, however, tackles the problem of automatically selecting the random weights under arbitrary sample sizes, which has not been addressed before. This is done by minimizing, through Bayesian Optimization, a black-box and noisy version of the reverse KL divergence between the Bayesian posterior and an approximate posterior induced by random weighting. We have demonstrated that BOB constantly outperforms competing approaches in recovering the Bayesian posterior, it provides a better uncertainty quantification, and it retains key asymptotic properties from existing methods. Moreover, we showed that BOB is not only more accurate than all the other existing methods, but it can also be substantially faster than MCMC algorithms, making BOB a competitive approximate posterior sampler. 

\textit{Limitations}: As with many other methodological developments, BOB presents both strengths and limitations, which can make it either highly suitable, or impractical, depending on the specific context of its application. That said, we can identify two main limitations: (1) Somehow ironic, the first limitation involves the use of Bayesian Optimization. Despite being one of the most efficient approaches to optimize a black-box objective, BO does not scale well to high-dimensional search spaces \citep{moriconi2020high}. Recall that our search space, $\boldsymbol{\mathcal{X}}$, is $2(K+1)$-dimensional, and thus, we should expect that BOB's applicability and performance would be significantly better under small-medium values of $K$ (i.e., the number of components in the mixture). Fortunately, high-dimensional black-box optimization is an active and vibrant area of research within the scientific community. Incorporating such developments and comparing different black-box optimizers could be an avenue for future research. (2) The second limitation is with respect to the sensitivity of our algorithms to the initial parameter values. Despite the incorporation of a tempering profile, both WBB and BOB are still sensitive to the initial parameter values. Poorly chosen initial values might lead the algorithm to saddle at a sub-optimal solution, and no amount of tempering would help us escape such a local optima. Fortunately, recent developments in clustering methods can help us obtain adequate starting points. Lastly, throughout this article, we have assumed that the number of clusters, $K$, is fixed and known. The problem of selecting $K$ within a Bayesian paradigm has been widely discussed in the literature (see e.g. \cite{richardson1997bayesian}, \cite{Stephens2000components}, \cite{nobile2007bayesian}, or \cite{baudry2010combining}), and it is outside the scope of this article.

On the whole, BOB joins a growing body of literature, which aims to draw practitioner's attention toward posterior samplers beyond traditional MCMC algorithms \citep{deep_boots, newtonBoots}. Thus, extending BOB (and BOB-like methodologies) from GMMs with conjugate priors to arbitrary posterior distributions presents an exciting research opportunity, which we intend to pursue in the future.

\section*{Acknowledgements}
We would like to thank Michael Martin for many constructive comments and helpful suggestions. Any remaining errors are, of course, our own.

\bibliography{sn-bibliography}

\begin{thebibliography}{}
\providecommand{\doi}[1]{\url{https://doi.org/#1}}
\bibcommenthead

\bibitem[\protect\citeauthoryear{Gelman, Carlin, Stern, Dunson, Vehtari, and Rubin}{Gelman et~al.}{2014}]{gelman2013bayesian}
Gelman, A., J.B. Carlin, H.S. Stern, D.B. Dunson, A.~Vehtari, and D.B. Rubin. 2014.
\newblock {\em Bayesian Data Analysis\textup{, 3rd edn.}}
\newblock CRC Press, Boca Raton, FL.

\bibitem[\protect\citeauthoryear{MacQueen}{MacQueen}{1967}]{macqueen1967some}
MacQueen, J. 1967.
\newblock Some methods for classification and analysis of multivariate observations.
\newblock {\em Proceedings of the Fifth Berkeley Symposium on Mathematical Statistics and Probability\/}~{\em 1\/}(14): 281--297 .

\bibitem[\protect\citeauthoryear{Raymaekers and Zamar}{Raymaekers and Zamar}{2022}]{HT_k_means_raymaekers2022}
Raymaekers, J. and R.H. Zamar. 2022.
\newblock Regularized k-means through hard-thresholding.
\newblock {\em Journal of Machine Learning Research\/}~{\em 23\/}(93): 1--48 .

\bibitem[\protect\citeauthoryear{Scrucca, Fraley, Murphy, and Raftery}{Scrucca et~al.}{2023}]{mclustScrucca}
Scrucca, L., C.~Fraley, T.B. Murphy, and A.E. Raftery. 2023.
\newblock {\em Model-Based Clustering, Classification, and Density Estimation Using {mclust} in {R}}.
\newblock Chapman and Hall/CRC.

\bibitem[\protect\citeauthoryear{Witten and Tibshirani}{Witten and Tibshirani}{2010}]{witten2010framework}
Witten, D.M. and R.~Tibshirani. 2010.
\newblock A framework for feature selection in clustering.
\newblock {\em Journal of the American Statistical Association\/}~{\em 105\/}(490): 713--726 .

\end{thebibliography}


\begin{thebibliography}{}
\providecommand{\doi}[1]{\url{https://doi.org/#1}}
\bibcommenthead

\bibitem[\protect\citeauthoryear{Aeberhard, Coomans, and {de Vel}}{Aeberhard et~al.}{1994}]{AEBERHARD1994_wines}
Aeberhard, S., D.~Coomans, and O.~{de Vel}. 1994.
\newblock Comparative analysis of statistical pattern recognition methods in high dimensional settings.
\newblock {\em Pattern Recognition\/}~{\em 27\/}(8): 1065--1077 .

\bibitem[\protect\citeauthoryear{Allassonni{\`e}re and Chevallier}{Allassonni{\`e}re and Chevallier}{2021}]{allassonniere2021temp}
Allassonni{\`e}re, S. and J.~Chevallier. 2021.
\newblock A new class of stochastic em algorithms. escaping local maxima and handling intractable sampling.
\newblock {\em Computational Statistics \& Data Analysis\/}~159: 107159 .

\bibitem[\protect\citeauthoryear{Baudry, Raftery, Celeux, Lo, and Gottardo}{Baudry et~al.}{2010}]{baudry2010combining}
Baudry, J.P., A.E. Raftery, G.~Celeux, K.~Lo, and R.~Gottardo. 2010.
\newblock Combining mixture components for clustering.
\newblock {\em Journal of Computational and Graphical Statistics\/}~{\em 19\/}(2): 332--353 .

\bibitem[\protect\citeauthoryear{Bishop}{Bishop}{2006}]{bishop2006pattern}
Bishop, C.M. 2006.
\newblock {\em Pattern recognition and machine learning}.
\newblock Springer New York, NY.

\bibitem[\protect\citeauthoryear{Blei, Kucukelbir, and McAuliffe}{Blei et~al.}{2017}]{blei2017variational}
Blei, D.M., A.~Kucukelbir, and J.D. McAuliffe. 2017.
\newblock Variational inference: A review for statisticians.
\newblock {\em Journal of the American Statistical Association\/}~{\em 112\/}(518): 859--877 .

\bibitem[\protect\citeauthoryear{Bull}{Bull}{2011}]{bull2011convergence}
Bull, A.D. 2011.
\newblock Convergence rates of efficient global optimization algorithms.
\newblock {\em Journal of Machine Learning Research\/}~{\em 12\/}(88): 2879--2904 .

\bibitem[\protect\citeauthoryear{Carpenter, Gelman, Hoffman, Lee, Goodrich, Betancourt, Brubaker, Guo, Li, and Riddell}{Carpenter et~al.}{2017}]{carpenter2017stan}
Carpenter, B., A.~Gelman, M.D. Hoffman, D.~Lee, B.~Goodrich, M.~Betancourt, M.~Brubaker, J.~Guo, P.~Li, and A.~Riddell. 2017.
\newblock Stan: A probabilistic programming language.
\newblock {\em Journal of Statistical Software\/}~{\em 76\/}(1): 1–32 .

\bibitem[\protect\citeauthoryear{Celeux, Hurn, and Robert}{Celeux et~al.}{2000}]{celeux2000computational}
Celeux, G., M.~Hurn, and C.P. Robert. 2000.
\newblock Computational and inferential difficulties with mixture posterior distributions.
\newblock {\em Journal of the American Statistical Association\/}~{\em 95\/}(451): 957--970 .

\bibitem[\protect\citeauthoryear{Charytanowicz, Niewczas, Kulczycki, Kowalski, {\L}ukasik, and {\.Z}ak}{Charytanowicz et~al.}{2010}]{seeds2010complete}
Charytanowicz, M., J.~Niewczas, P.~Kulczycki, P.A. Kowalski, S.~{\L}ukasik, and S.~{\.Z}ak. 2010.
\newblock Complete gradient clustering algorithm for features analysis of x-ray images.
\newblock {\em Information Technologies in Biomedicine\/}~2: 15--24 .

\bibitem[\protect\citeauthoryear{Dempster, Laird, and Rubin}{Dempster et~al.}{1977}]{dempster1977EM}
Dempster, A.P., N.M. Laird, and D.B. Rubin. 1977.
\newblock Maximum likelihood from incomplete data via the em algorithm.
\newblock {\em Journal of the Royal Statistical Society Series B: Statistical Methodology\/}~{\em 39\/}(1): 1--22 .

\bibitem[\protect\citeauthoryear{Diebolt and Robert}{Diebolt and Robert}{1994}]{diebolt1994estimation}
Diebolt, J. and C.P. Robert. 1994.
\newblock Estimation of finite mixture distributions through bayesian sampling.
\newblock {\em Journal of the Royal Statistical Society Series B: Statistical Methodology\/}~{\em 56\/}(2): 363--375 .

\bibitem[\protect\citeauthoryear{Fong, Lyddon, and Holmes}{Fong et~al.}{2019}]{fong2019scalable}
Fong, E., S.~Lyddon, and C.~Holmes. 2019.
\newblock Scalable nonparametric sampling from multimodal posteriors with the posterior bootstrap.
\newblock {\em Proceedings of the 36th International Conference on Machine Learning\/}~97: 1952--1962 .

\bibitem[\protect\citeauthoryear{Fraley and Raftery}{Fraley and Raftery}{2002}]{reftery_model_clust}
Fraley, C. and A.E. Raftery. 2002.
\newblock Model-based clustering, discriminant analysis, and density estimation.
\newblock {\em Journal of the American Statistical Association\/}~{\em 97\/}(458): 611--631 .

\bibitem[\protect\citeauthoryear{Friedman, Hastie, and Tibshirani}{Friedman et~al.}{2008}]{friedman_glasso}
Friedman, J., T.~Hastie, and R.~Tibshirani. 2008, 12.
\newblock {Sparse inverse covariance estimation with the graphical lasso}.
\newblock {\em Biostatistics\/}~{\em 9\/}(3): 432--441 .

\bibitem[\protect\citeauthoryear{Geisser}{Geisser}{2017}]{geisser2017predictive}
Geisser, S. 2017.
\newblock {\em Predictive Inference}.
\newblock Chapman and Hall/CRC.

\bibitem[\protect\citeauthoryear{Gelman and Rubin}{Gelman and Rubin}{1992}]{gelman1992inference}
Gelman, A. and D.B. Rubin. 1992.
\newblock Inference from iterative simulation using multiple sequences.
\newblock {\em Statistical Science\/}~{\em 7\/}(4): 457--472 .

\bibitem[\protect\citeauthoryear{Hoffman and Gelman}{Hoffman and Gelman}{2014}]{hoffman2014no}
Hoffman, M.D. and A.~Gelman. 2014.
\newblock The no-u-turn sampler: Adaptively setting path lengths in hamiltonian monte carlo.
\newblock {\em Journal of Machine Learning Research\/}~{\em 15\/}(47): 1593--1623 .

\bibitem[\protect\citeauthoryear{Hofmeyr}{Hofmeyr}{2021}]{FKSUM_2021}
Hofmeyr, D.P. 2021.
\newblock Fast exact evaluation of univariate kernel sums.
\newblock {\em IEEE Transactions on Pattern Analysis and Machine Intelligence\/}~{\em 43\/}(2): 447--458 .

\bibitem[\protect\citeauthoryear{Hofmeyr}{Hofmeyr}{2022}]{FKSUM_2022}
Hofmeyr, D.P. 2022.
\newblock Fast kernel smoothing in r with applications to projection pursuit.
\newblock {\em Journal of Statistical Software\/}~{\em 101\/}(3): 1–33 .

\bibitem[\protect\citeauthoryear{Izenman and Sommer}{Izenman and Sommer}{1988}]{izenman1988}
Izenman, A.J. and C.J. Sommer. 1988.
\newblock Philatelic mixtures and multimodal densities.
\newblock {\em Journal of the American Statistical Association\/}~{\em 83\/}(404): 941--953 .

\bibitem[\protect\citeauthoryear{Jones}{Jones}{2001}]{jones2001taxonomy}
Jones, D.R. 2001.
\newblock A taxonomy of global optimization methods based on response surfaces.
\newblock {\em Journal of Global Optimization\/}~21: 345--383 .

\bibitem[\protect\citeauthoryear{Jones, Schonlau, and Welch}{Jones et~al.}{1998}]{jones1998efficient}
Jones, D.R., M.~Schonlau, and W.J. Welch. 1998.
\newblock Efficient global optimization of expensive black-box functions.
\newblock {\em Journal of Global Optimization\/}~13: 455--492 .

\bibitem[\protect\citeauthoryear{Kandasamy, Vysyaraju, Neiswanger, Paria, Collins, Schneider, Poczos, and Xing}{Kandasamy et~al.}{2020}]{dragonfly_2020}
Kandasamy, K., K.R. Vysyaraju, W.~Neiswanger, B.~Paria, C.R. Collins, J.~Schneider, B.~Poczos, and E.P. Xing. 2020.
\newblock Tuning hyperparameters without grad students: Scalable and robust bayesian optimisation with dragonfly.
\newblock {\em Journal of Machine Learning Research\/}~{\em 21\/}(81): 1--27 .

\bibitem[\protect\citeauthoryear{Kirkpatrick, Gelatt~Jr, and Vecchi}{Kirkpatrick et~al.}{1983}]{kirkpatrick1983optimization}
Kirkpatrick, S., C.D. Gelatt~Jr, and M.P. Vecchi. 1983.
\newblock Optimization by simulated annealing.
\newblock {\em Science\/}~{\em 220\/}(4598): 671--680 .

\bibitem[\protect\citeauthoryear{Kucukelbir, Tran, Ranganath, Gelman, and Blei}{Kucukelbir et~al.}{2017}]{kucukelbir2017automatic}
Kucukelbir, A., D.~Tran, R.~Ranganath, A.~Gelman, and D.M. Blei. 2017.
\newblock Automatic differentiation variational inference.
\newblock {\em Journal of Machine Learning Research\/}~{\em 18\/}(14): 1--45 .

\bibitem[\protect\citeauthoryear{Lartigue, Durrleman, and Allassonni{\`e}re}{Lartigue et~al.}{2022}]{lartigue2022deterministic}
Lartigue, T., S.~Durrleman, and S.~Allassonni{\`e}re. 2022.
\newblock Deterministic approximate em algorithm; application to the riemann approximation em and the tempered em.
\newblock {\em Algorithms\/}~{\em 15\/}(3): 78 .

\bibitem[\protect\citeauthoryear{Le~Riche and Picheny}{Le~Riche and Picheny}{2021}]{DiceOptim2021}
Le~Riche, R. and V.~Picheny. 2021.
\newblock Revisiting bayesian optimization in the light of the coco benchmark.
\newblock {\em Structural and Multidisciplinary Optimization\/}~{\em 64\/}(5): 3063--3087 .

\bibitem[\protect\citeauthoryear{Lyddon, Holmes, and Walker}{Lyddon et~al.}{2019}]{lyddon2019general}
Lyddon, S.P., C.~Holmes, and S.~Walker. 2019.
\newblock General bayesian updating and the loss-likelihood bootstrap.
\newblock {\em Biometrika\/}~{\em 106\/}(2): 465--478 .

\bibitem[\protect\citeauthoryear{Lynch and Western}{Lynch and Western}{2004}]{Lynch_posterior}
Lynch, S.M. and B.~Western. 2004.
\newblock Bayesian posterior predictive checks for complex models.
\newblock {\em Sociological Methods \& Research\/}~{\em 32\/}(3): 301--335 .

\bibitem[\protect\citeauthoryear{Moriconi, Deisenroth, and Sesh~Kumar}{Moriconi et~al.}{2020}]{moriconi2020high}
Moriconi, R., M.P. Deisenroth, and K.~Sesh~Kumar. 2020.
\newblock High-dimensional bayesian optimization using low-dimensional feature spaces.
\newblock {\em Machine Learning\/}~109: 1925--1943 .

\bibitem[\protect\citeauthoryear{Morningstar, Vikram, Ham, Gallagher, and Dillon}{Morningstar et~al.}{2021}]{morningstar2021automatic}
Morningstar, W., S.~Vikram, C.~Ham, A.~Gallagher, and J.~Dillon. 2021.
\newblock Automatic differentiation variational inference with mixtures.
\newblock {\em International Conference on Artificial Intelligence and Statistics\/}~130: 3250--3258 .

\bibitem[\protect\citeauthoryear{Newton, Polson, and Xu}{Newton et~al.}{2021}]{newtonBoots}
Newton, M.A., N.G. Polson, and J.~Xu. 2021.
\newblock Weighted bayesian bootstrap for scalable posterior distributions.
\newblock {\em Canadian Journal of Statistics\/}~{\em 49\/}(2): 421--437 .

\bibitem[\protect\citeauthoryear{Newton and Raftery}{Newton and Raftery}{1994}]{raftery_boostrap_likelihood}
Newton, M.A. and A.E. Raftery. 1994.
\newblock Approximate bayesian inference with the weighted likelihood bootstrap.
\newblock {\em Journal of the Royal Statistical Society Series B: Statistical Methodology\/}~{\em 56\/}(1): 3--48 .

\bibitem[\protect\citeauthoryear{Ng and Newton}{Ng and Newton}{2022}]{ng2022random}
Ng, T.L. and M.A. Newton. 2022.
\newblock Random weighting in lasso regression.
\newblock {\em Electronic Journal of Statistics\/}~{\em 16\/}(1): 3430--3481 .

\bibitem[\protect\citeauthoryear{Ni, M{\"u}ller, Diesendruck, Williamson, Zhu, and Ji}{Ni et~al.}{2020}]{ni2020scalable}
Ni, Y., P.~M{\"u}ller, M.~Diesendruck, S.~Williamson, Y.~Zhu, and Y.~Ji. 2020.
\newblock Scalable bayesian nonparametric clustering and classification.
\newblock {\em Journal of Computational and Graphical Statistics\/}~{\em 29\/}(1): 53--65 .

\bibitem[\protect\citeauthoryear{Nie and Ročková}{Nie and Ročková}{2023a}]{rockova2022}
Nie, L. and V.~Ročková. 2023a.
\newblock Bayesian bootstrap spike-and-slab lasso.
\newblock {\em Journal of the American Statistical Association\/}~{\em 118\/}(543): 2013--2028 .

\bibitem[\protect\citeauthoryear{Nie and Ročková}{Nie and Ročková}{2023b}]{deep_boots}
Nie, L. and V.~Ročková. 2023b.
\newblock Deep bootstrap for bayesian inference.
\newblock {\em Philosophical Transactions of the Royal Society A\/}~{\em 381\/}(2247): 20220154 .

\bibitem[\protect\citeauthoryear{Nobile and Fearnside}{Nobile and Fearnside}{2007}]{nobile2007bayesian}
Nobile, A. and A.T. Fearnside. 2007.
\newblock Bayesian finite mixtures with an unknown number of components: The allocation sampler.
\newblock {\em Statistics and Computing\/}~17: 147--162 .

\bibitem[\protect\citeauthoryear{Omori, Chib, Shephard, and Nakajima}{Omori et~al.}{2007}]{OMORI2007425}
Omori, Y., S.~Chib, N.~Shephard, and J.~Nakajima. 2007.
\newblock Stochastic volatility with leverage: Fast and efficient likelihood inference.
\newblock {\em Journal of Econometrics\/}~{\em 140\/}(2): 425--449 .

\bibitem[\protect\citeauthoryear{Pompe}{Pompe}{2021}]{pompe2021introducing}
Pompe, E. 2021.
\newblock Introducing prior information in weighted likelihood bootstrap with applications to model misspecification.
\newblock {\em arXiv preprint arXiv:2103.14445\/} .

\bibitem[\protect\citeauthoryear{Provost and Zang}{Provost and Zang}{2024}]{provost2024nonparametric}
Provost, S.B. and Y.~Zang. 2024.
\newblock Nonparametric copula density estimation methodologies.
\newblock {\em Mathematics\/}~{\em 12\/}(3): 398 .

\bibitem[\protect\citeauthoryear{Rasmussen and Williams}{Rasmussen and Williams}{2005}]{rasmussen2006gaussian}
Rasmussen, C.E. and C.K.I. Williams. 2005, 11.
\newblock {\em Gaussian Processes for Machine Learning}.
\newblock The MIT Press.

\bibitem[\protect\citeauthoryear{Richardson and Green}{Richardson and Green}{1997}]{richardson1997bayesian}
Richardson, S. and P.J. Green. 1997.
\newblock On bayesian analysis of mixtures with an unknown number of components (with discussion).
\newblock {\em Journal of the Royal Statistical Society Series B: Statistical Methodology\/}~{\em 59\/}(4): 731--792 .

\bibitem[\protect\citeauthoryear{Roustant, Ginsbourger, and Deville}{Roustant et~al.}{2012}]{DiceOptim2012}
Roustant, O., D.~Ginsbourger, and Y.~Deville. 2012.
\newblock Dicekriging, diceoptim: Two r packages for the analysis of computer experiments by kriging-based metamodeling and optimization.
\newblock {\em Journal of Statistical Software\/}~{\em 51\/}(1): 1–55 .

\bibitem[\protect\citeauthoryear{Sambridge}{Sambridge}{2014}]{Sambridge_2014_Parallel}
Sambridge, M. 2014.
\newblock {A Parallel Tempering algorithm for probabilistic sampling and multimodal optimization}.
\newblock {\em Geophysical Journal International\/}~{\em 196\/}(1): 357--374 .

\bibitem[\protect\citeauthoryear{Schilling}{Schilling}{2017}]{schilling2017measures}
Schilling, R.L. 2017.
\newblock {\em Measures, Integrals and Martingales}.
\newblock Cambridge University Press.

\bibitem[\protect\citeauthoryear{Sklar}{Sklar}{1959}]{sklar1959fonctions}
Sklar, M. 1959.
\newblock {Fonctions de r{\'e}partition {\`a} N dimensions et leurs marges}.
\newblock {\em {Annales de l'ISUP}\/}~{\em 8\/}(3): 229--231 .

\bibitem[\protect\citeauthoryear{Snoek, Larochelle, and Adams}{Snoek et~al.}{2012}]{snoek2012practical}
Snoek, J., H.~Larochelle, and R.P. Adams. 2012.
\newblock Practical bayesian optimization of machine learning algorithms.
\newblock {\em Advances in Neural Information Processing Systems\/}~25: 2951–2959 .

\bibitem[\protect\citeauthoryear{Stephens}{Stephens}{2000a}]{Stephens2000components}
Stephens, M. 2000a.
\newblock {Bayesian analysis of mixture models with an unknown number of components—an alternative to reversible jump methods}.
\newblock {\em The Annals of Statistics\/}~{\em 28\/}(1): 40 -- 74 .

\bibitem[\protect\citeauthoryear{Stephens}{Stephens}{2000b}]{stephens2000dealing}
Stephens, M. 2000b.
\newblock Dealing with label switching in mixture models.
\newblock {\em Journal of the Royal Statistical Society Series B: Statistical Methodology\/}~{\em 62\/}(4): 795--809 .

\bibitem[\protect\citeauthoryear{Stringer, Brown, and Stafford}{Stringer et~al.}{2023}]{stringer2023fast}
Stringer, A., P.~Brown, and J.~Stafford. 2023.
\newblock Fast, scalable approximations to posterior distributions in extended latent gaussian models.
\newblock {\em Journal of Computational and Graphical Statistics\/}~{\em 32\/}(1): 84--98 .

\bibitem[\protect\citeauthoryear{Van~Havre, White, Rousseau, and Mengersen}{Van~Havre et~al.}{2015}]{van2015overfitting}
Van~Havre, Z., N.~White, J.~Rousseau, and K.~Mengersen. 2015.
\newblock Overfitting bayesian mixture models with an unknown number of components.
\newblock {\em PloS one\/}~{\em 10\/}(7): e0131739 .

\bibitem[\protect\citeauthoryear{Wade and Ghahramani}{Wade and Ghahramani}{2018}]{s_wade_cluster_inference}
Wade, S. and Z.~Ghahramani. 2018.
\newblock {Bayesian Cluster Analysis: Point Estimation and Credible Balls (with Discussion)}.
\newblock {\em Bayesian Analysis\/}~{\em 13\/}(2): 559 -- 626 .

\end{thebibliography}


\newpage

\begin{center}
    \Titlefont{Supplementary Materials for ``BOB: Bayesian Optimized Bootstrap for Uncertainty Quantification in Gaussian Mixture Models"}
    \\
    \vspace{25px}
    \Authorfont{Santiago Marin, \quad Bronwyn Loong, \quad Anton H. Westveld}
    \\
    \vspace{20px}
\end{center}

\section*{S.1 Proof of Proposition \ref{Prop_joint_mu_Omega}}
\label{sec:Proof_Prop_joint_mu_Omega}

From \eqref{surrogate_objective}, we have that the update of $\boldsymbol{\mu}_{k}$ and $\boldsymbol{\Sigma}_{k}$ is given by
\begin{equation*}
    \bigl(\boldsymbol{\mu}_{k}^{(t+1)}, \boldsymbol{\Sigma}_{k}^{(t+1)}\bigr) = \argmax_{\boldsymbol{\mu}_{k},\boldsymbol{\Sigma}_{k}}\bigl\{ h(\boldsymbol{\mu}_{k}, \boldsymbol{\Sigma}_{k})\bigr\},
\end{equation*}
with
\begin{equation*}
    \begin{split}
        h(\boldsymbol{\mu}_{k}, \boldsymbol{\Sigma}_{k}) & = 
        -\frac{1}{2}\sum_{i=1}^{n}\left(u_{i}q_{ik}\left[\log|\boldsymbol{\Sigma}_{k}| + \left(\mathbf{y}_{i} - \boldsymbol{\mu}_{k}\right)'\boldsymbol{\Sigma}_{k}^{-1}\left(\mathbf{y}_{i} - \boldsymbol{\mu}_{k}\right)\right]\right)- \frac{\Tilde{u}_{\Sigma_{k}}\tr(\boldsymbol{\Psi}_{k}\boldsymbol{\Sigma}_{k}^{-1})}{2}\\
        & 
        \quad -\frac{\Tilde{u}_{\mu_{k}}\lambda_{k}}{2}(\boldsymbol{\mu}_{k} - \boldsymbol{\beta}_{k})'\boldsymbol{\Sigma}_{k}^{-1}(\boldsymbol{\mu}_{k}- \boldsymbol{\beta}_{k}) 
        -\Tilde{u}_{\Sigma_{k}}\left(\frac{\nu_{k}+d}{2}+1\right)\log|\boldsymbol{\Sigma}_{k}|\\
        & \overset{(*)}{=} -\frac{\Tilde{n}_{k}}{2}\log|\boldsymbol{\Sigma}_{k}| - \frac{1}{2}\tr\left[\left(\Tilde{n}_{k}(\Tilde{\Bar{\mathbf{y}}}_{k} - \boldsymbol{\mu}_{k})(\Tilde{\Bar{\mathbf{y}}}_{k} - \boldsymbol{\mu}_{k})' + \Tilde{\mathbf{S}}_{k}\right)\boldsymbol{\Sigma}_{k}^{-1} \right]\\
         &  \quad -\left(\frac{\Tilde{\nu}_{k}+d}{2} + 1\right)\log|\boldsymbol{\Sigma}_{k}| - \frac{1}{2}\tr\left(\Tilde{\boldsymbol{\Psi}}_{k}\boldsymbol{\Sigma}_{k}^{-1}\right)
         - \frac{\Tilde{\lambda}_{k}}{2}\tr\left[(\boldsymbol{\mu}_{k} - \boldsymbol{\beta}_{k})(\boldsymbol{\mu}_{k}-\boldsymbol{\beta}_{k})'\boldsymbol{\Sigma}_{k}^{-1}\right] \\
         & =  -\left(\frac{\Bar{\nu}_{k} + d}{2} + 1\right)\log|\boldsymbol{\Sigma}_{k}| - \frac{1}{2}\tr\left(\Bar{\boldsymbol{\Psi}}_{k}\boldsymbol{\Sigma}_{k}^{-1}\right) -   \frac{\Bar{\lambda}_{k}}{2}(\boldsymbol{\mu}_{k} - \Bar{\boldsymbol{\beta}}_{k})'\boldsymbol{\Sigma}_{k}^{-1}(\boldsymbol{\mu}_{k} - \Bar{\boldsymbol{\beta}}_{k}),
    \end{split}
\end{equation*}
where $(*)$ follows from the fact that $\sum_{i=1}^{n}\mathbf{b}_{i}'\mathbf{A}\mathbf{b}_{i} = \tr(\mathbf{B}'\mathbf{BA})$, where $\mathbf{B}\in\mathbb{R}^{n\times d}$ denotes the matrix whose $i$-th row is $\mathbf{b}_{i}'\in\mathbb{R}^{1\times d}$, and by writing $(\mathbf{y}_{i} - \boldsymbol{\mu}_{k})$ as $(\mathbf{y}_{i} - \Tilde{\Bar{\mathbf{y}}}_{k} + \Tilde{\Bar{\mathbf{y}}}_{k} - \boldsymbol{\mu}_{k})$. \hfill\qed

\section*{S.2 Sampling from the Posterior Predictive Distribution}
\label{sec:post_pred}

Given (approximate or actual) posterior draws $\left\{\boldsymbol{\theta}_{(s)}\right\}_{s=1}^{S}$, one can easily obtain posterior predictive draws as described in algorithm \ref{alg:post_pred_draws}. 

\renewcommand{\thealgorithm}{S.1}
\begin{algorithm}
\caption{Posterior Predictive Sampler}\label{alg:post_pred_draws}
\textbf{Input:} \\
\hspace*{\algorithmicindent} Posterior draws: $\left\{\boldsymbol{\theta}_{(s)}\right\}_{s=1}^{S}$ \\
\hspace*{\algorithmicindent} Model: $p(\mathbf{y}|\,\boldsymbol{\theta})$\\
\textbf{Output:} \\
\hspace*{\algorithmicindent} Posterior predictive draws: $\left\{\mathbf{y}_{\text{new},(s)}\right\}_{s=1}^{S}$
\begin{algorithmic}[1]
\For{$s\in\{1\,\dots,S\}$} \Comment{in parallel}
    \State Sample $\mathbf{y}_{\text{new},(s)}\sim p(\mathbf{y}|\,\boldsymbol{\theta}_{(s)})$
\EndFor
\end{algorithmic}
\end{algorithm}

\section*{S.3 Sampling from the Actual Bayesian Posterior}
\label{sec:bayes_post}

Following \citeSupp{gelman2013bayesian}, under the likelihood and priors specified in section \ref{subsec:Model_setup}, the Bayesian posteriors for $\left\{\boldsymbol{\mu}_{k}|\mathbf{Y},\mathbf{Z}\right\}$, $\left\{\boldsymbol{\Sigma}_{k}|\mathbf{Y},\mathbf{Z}\right\}$, and $\left\{\boldsymbol{\pi}|\mathbf{Y},\mathbf{Z}\right\}$ are given by
\begin{gather*}
     \boldsymbol{\mu}_{k}|\mathbf{Y}, {\mathbf{Z}}\sim\textit{t}_{(\hat{\nu}_{k}-d+1)}\left(\hat{\boldsymbol{\beta}}_{k},{{\hat{\boldsymbol{\Psi}}_{k}}/{(\hat{\lambda}_{k}(\hat{\nu}_{k}-d+1))}}\right),\\
    \boldsymbol{\Sigma}_{k}|\mathbf{Y}, {\mathbf{Z}}\sim\mathcal{IW}\left(\hat{\nu}_{k}, \hat{\boldsymbol{\Psi}}_{k}^{-1} \right),\\
    \boldsymbol{\pi}|\mathbf{Y}, {\mathbf{Z}}\sim\mathcal{D}\Bigl(\hat{a}_{1},\dots,\,\hat{a}_{K}\Bigr),
\end{gather*}
where $\hat{\boldsymbol{\beta}}_{k} = (\hat{\beta}_{k1},\dots,\hat{\beta}_{kd})' = \lambda_{k}/(\lambda_{k} + n_{z_{k}})\boldsymbol{\beta}_{k} + n_{z_{k}}/(\lambda_{k} + n_{z_{k}})\Bar{\mathbf{y}}_{z_{k}}$, 
$\hat{\boldsymbol{\Psi}}_{k} = \left(\hat{\psi}_{kjj'}\right) = \boldsymbol{\Psi}_{k} + \mathbf{S}_{z_{k}} + (\lambda_{k}n_{z_{k}})/(\lambda_{k} + n_{z_{k}})(\Bar{\mathbf{y}}_{z_{k}} - \boldsymbol{\beta}_{k})(\Bar{\mathbf{y}}_{z_{k}}-\boldsymbol{\beta}_{k})'$, $\hat{\lambda}_{k} = \lambda_{k} + n_{z_{k}}$, $\hat{\nu}_{k} = \nu_{k} + n_{z_{k}}$, and $\hat{a}_{k} = a_{k} + n_{z_{k}}$, with $n_{z_{k}} = \sum_{i=1}^{n}z_{ik}$, $\Bar{\mathbf{y}}_{z_{k}} = (n_{z_{k}})^{-1}\sum_{i:z_{ik}=1}\mathbf{y}_{i}$, and $\mathbf{S}_{z_{k}} = \sum_{i:z_{ik}=1}(\mathbf{y}_{i} - \Bar{\mathbf{y}}_{z_{k}})(\mathbf{y}_{i} - \Bar{\mathbf{y}}_{z_{k}})'$. When the true latent indicator variables, $\mathbf{Z}$, are known (e.g., under simulated data or when we know the true cluster labels), one can easily compute, evaluate or obtain draws from the joint Bayesian posterior distribution. More precisely, algorithm \ref{alg:bayes_post} generates draws from the joint Bayesian posterior, namely $\left\{\boldsymbol{\theta}_{\text{Bayes},(s)}\right\}_{s=1}^{S}$. Then, with the posterior draws $\left\{\boldsymbol{\theta}_{\text{Bayes},(s)}\right\}_{s=1}^{S}$, we can proceed to obtain posterior predictive draws as in algorithm \ref{alg:post_pred_draws}.

\renewcommand{\thealgorithm}{S.2}
\begin{algorithm}
\caption{Bayesian Posterior Sampler}\label{alg:bayes_post}
\textbf{Input:} \\
\hspace*{\algorithmicindent} Data: $\mathbf{Y}$ \\
\hspace*{\algorithmicindent} True latent indicator variables: $\mathbf{Z}$ \\
\hspace*{\algorithmicindent} Total number of posterior draws: $S$ \\
\hspace*{\algorithmicindent} Posterior hyper-parameters: $\bigl\{\hat{\boldsymbol{\beta}}_{k}, \hat{\boldsymbol{\Psi}}_{k}, \hat{\lambda}_{k}, \hat{\nu}_{k}, \hat{a}_{k}   \bigr\}_{k=1}^{K}$ \\
\textbf{Output:} \\
\hspace*{\algorithmicindent} Bayesian posterior draws: $\left\{\boldsymbol{\theta}_{\text{Bayes},(s)}\right\}_{s=1}^{S}$
\begin{algorithmic}[1]
\For{$s\in\{1\,\dots,S\}$} \Comment{in parallel}
    \State Sample $\boldsymbol{\Sigma}_{k,(s)}|\mathbf{Y}, {\mathbf{Z}} \sim \mathcal{IW}\bigl(\hat{\nu}_{k}, \hat{\boldsymbol{\Psi}}_{k}^{-1}\bigr)$, $\forall k\in[K]$ 
    \State Sample  $\,\boldsymbol{\mu}_{k,(s)}|\mathbf{Y}, {\mathbf{Z}}, \boldsymbol{\Sigma}_{k,(s)} \sim \mathcal{N}\bigl(\hat{\boldsymbol{\beta}}_{k}, \boldsymbol{\Sigma}_{k,(s)}/\hat{\lambda}_{k}\bigr)$, $\forall k\in[K]$
    \State Sample $\boldsymbol{\pi}_{(s)}|\mathbf{Y}, {\mathbf{Z}}\sim\mathcal{D}\bigl(\hat{a}_{1},\dots,\,\hat{a}_{K}\bigr)$
    \State Set $\boldsymbol{\theta}_{\text{Bayes},(s)} \leftarrow \bigl\{\pi_{k,(s)}, \boldsymbol{\mu}_{k,(s)}, \boldsymbol{\Sigma}_{k,(s)}  \bigr\}_{k=1}^{K}$
\EndFor
\end{algorithmic}
\end{algorithm}

\section*{S.4 Warm Initialization Strategy}

To initialize our algorithms, we consider a pool of candidate initial values. These candidate values are obtained via (1) Hard-thresholded $K$-means \citepSupp{HT_k_means_raymaekers2022}, where the HTK-means penalty term is selected using AIC, BIC and a regularization path plot; (2) Sparse $K$-means \citepSupp{witten2010framework}, where the sparse $K$-means shrinkage parameter is selected using a permutation approach; (3) Model-based clustering using the  ``\texttt{mclust}" \texttt{R} package \citepSupp{mclustScrucca}, which itself is initialized by hierarchical model-based agglomerative clustering; and (4) $K$-means clustering \citepSupp{macqueen1967some}. Then, we choose as a starting point the values that yield the largest posterior density. To facilitate comparisons, we initialize all algorithms at this starting point. 

\section*{S.5 Features in Benchmark Data}

Tables \ref{tab:variables_wine} and \ref{tab:variables_seeds} present details and descriptions of the variables in the Wine and Seeds datasets, respectively.

\begin{table}[h]
\renewcommand\thetable{S.1} 
\caption{Features in the Wine Data}\label{tab:variables_wine}
\begin{tabular*}{\textwidth}{@{\extracolsep\fill}llll}
\toprule%
\multicolumn{2}{@{}c@{}}{Wine Feature} & \multicolumn{2}{@{}c@{}}{Wine Feature} \\\cmidrule{1-2} \cmidrule{3-4}%
\multicolumn{1}{@{}r@{}}{Abbreviation} &  \multicolumn{1}{@{}c@{}}{Description} & \multicolumn{1}{@{}r@{}}{Abbreviation} &  \multicolumn{1}{@{}c@{}}{Description} \\
\midrule
\texttt{alco} & Alcohol percentage & \texttt{nflav} & Nonflavanoid phenols \\
\texttt{mal} & Malic acid & \texttt{proa} & Proanthocyanins \\
\texttt{ash} & Ash & \texttt{col} & Colour intensity \\
\texttt{akash} & Alkalinity of ash & \texttt{hue} & Hue \\
\texttt{mag} & Magnesium & \texttt{ODdw}  & $\frac{OD_{280}}{OD_{315}}$ of diluted wines \\
\texttt{phen} & Total phenols & \texttt{prol} &  Proline \\
\texttt{flav} & Flavanoids  \\
\botrule
\end{tabular*}
\end{table}

\begin{table}[h]
\renewcommand\thetable{S.2} 
\caption{Features in the Seeds Data}\label{tab:variables_seeds}
\begin{tabular*}{\textwidth}{@{\extracolsep\fill}llll}
\toprule%
\multicolumn{2}{@{}c@{}}{Seed Feature} & \multicolumn{2}{@{}c@{}}{Seed Feature} \\\cmidrule{1-2} \cmidrule{3-4}%
\multicolumn{1}{@{}r@{}}{Abbreviation} &  \multicolumn{1}{@{}c@{}}{Description} & \multicolumn{1}{@{}r@{}}{Abbreviation} &  \multicolumn{1}{@{}c@{}}{Description} \\
\midrule
\texttt{area} & Area of kernel & \texttt{k.width} & Width of kernel \\
\texttt{perimeter} & Perimeter of kernel & \texttt{asymmetry}   & Asymmetry coefficient \\
\texttt{compactness} & Compactness of kernel & \texttt{k.gr.len} & Length of kernel groove \\
\texttt{k.len} & Length of kernel \\ 
\botrule
\end{tabular*}
\end{table}

\section*{S.6 Additional Posterior Density Plots}
\label{sec:additional_density_plots}

\begin{figure*}[!htp]
\centering
\begin{subfigure}{0.45\textwidth}
    \includegraphics[width=\textwidth]{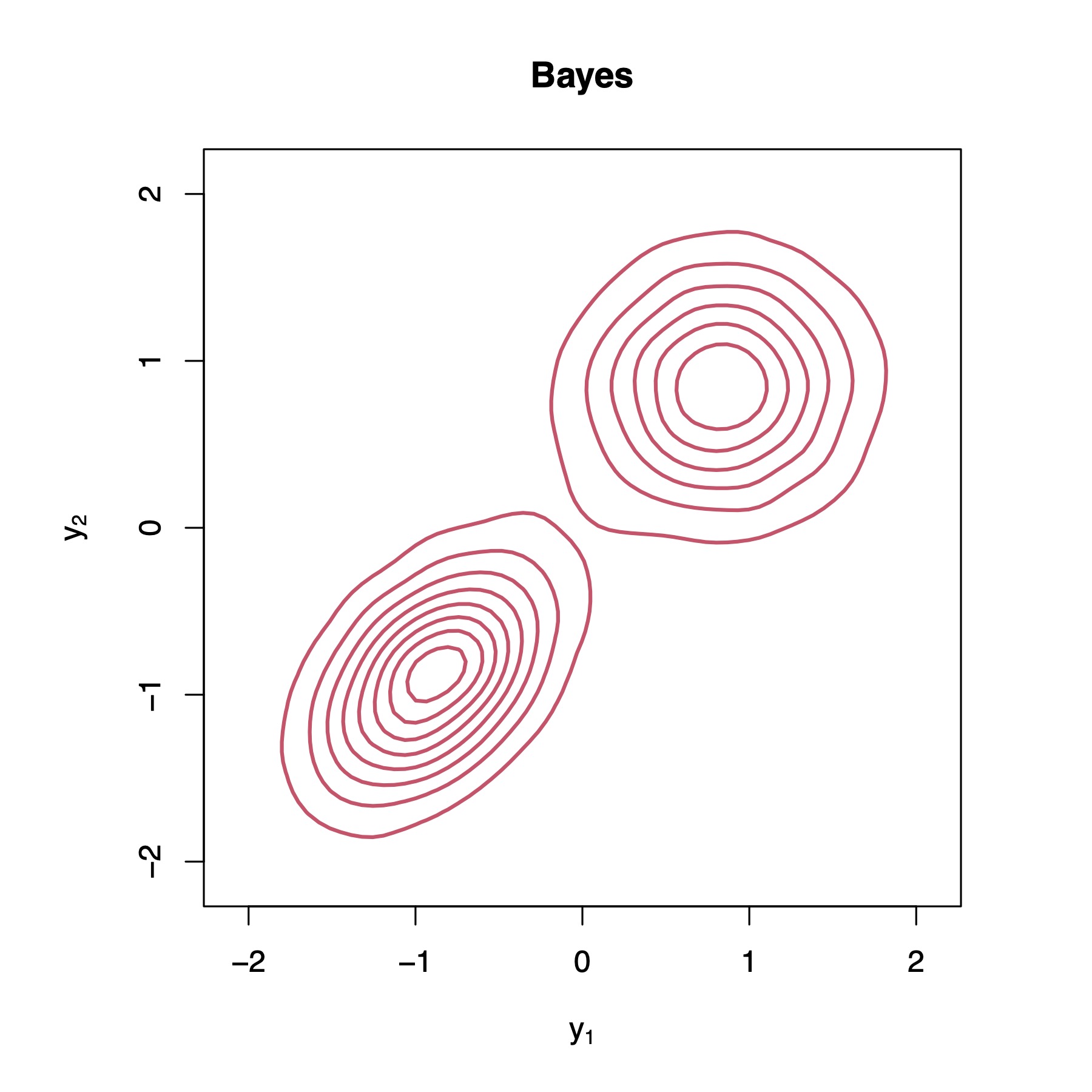}
\end{subfigure}
\hfill
\begin{subfigure}{0.45\textwidth}
    \includegraphics[width=\textwidth]{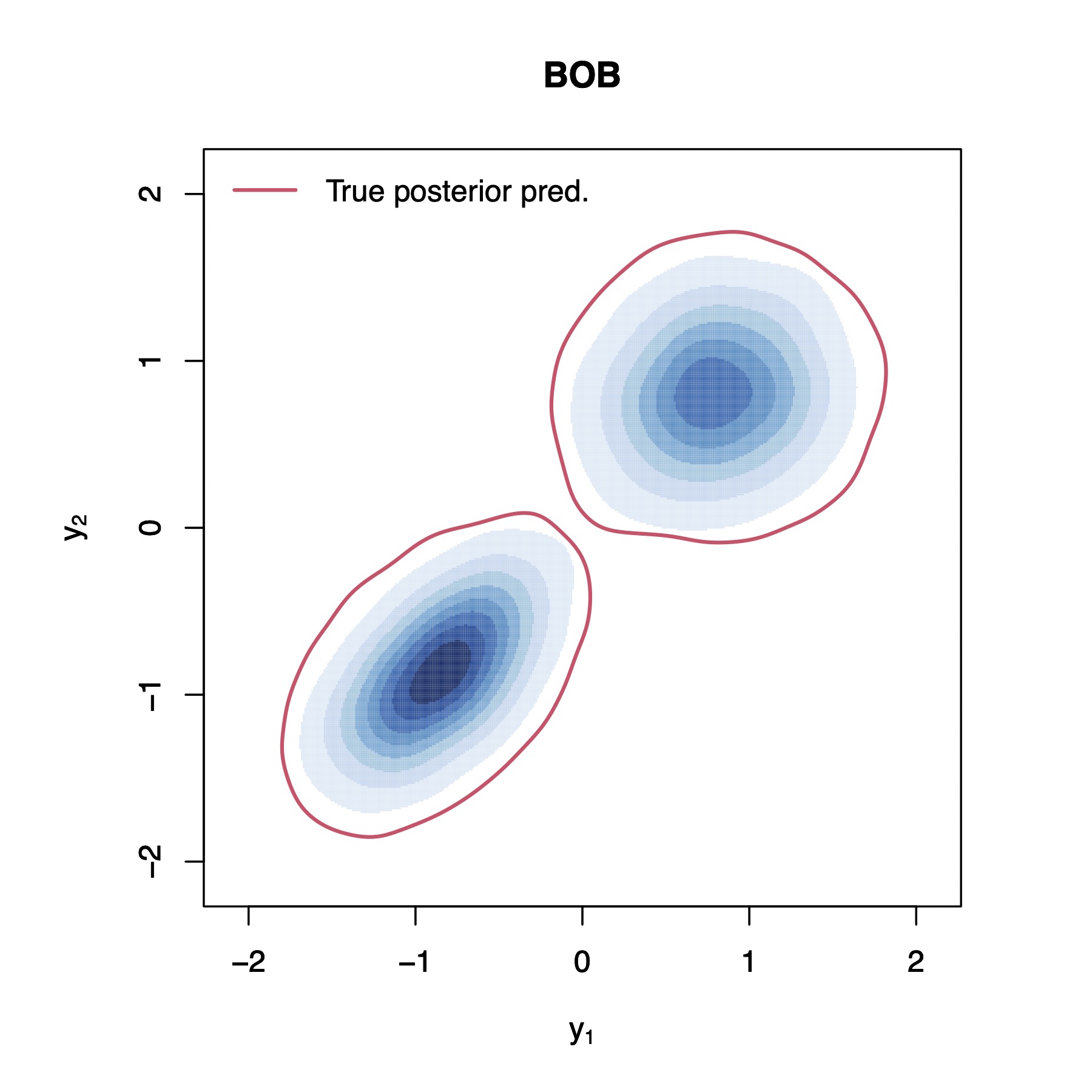}
\end{subfigure}
\hfill
\begin{subfigure}{0.45\textwidth}
    \includegraphics[width=\textwidth]{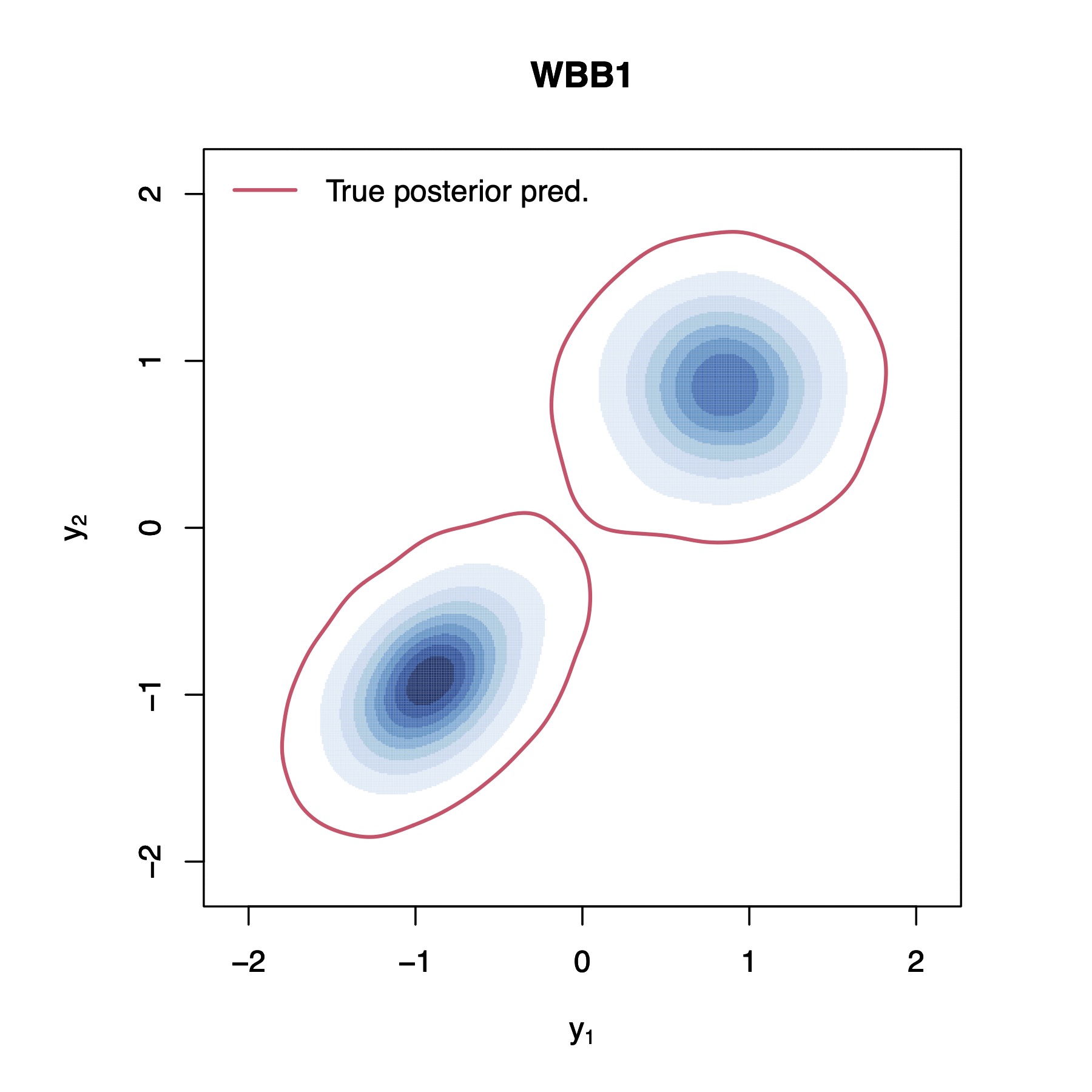}
\end{subfigure}
\hfill
\begin{subfigure}{0.45\textwidth}
    \includegraphics[width=\textwidth]{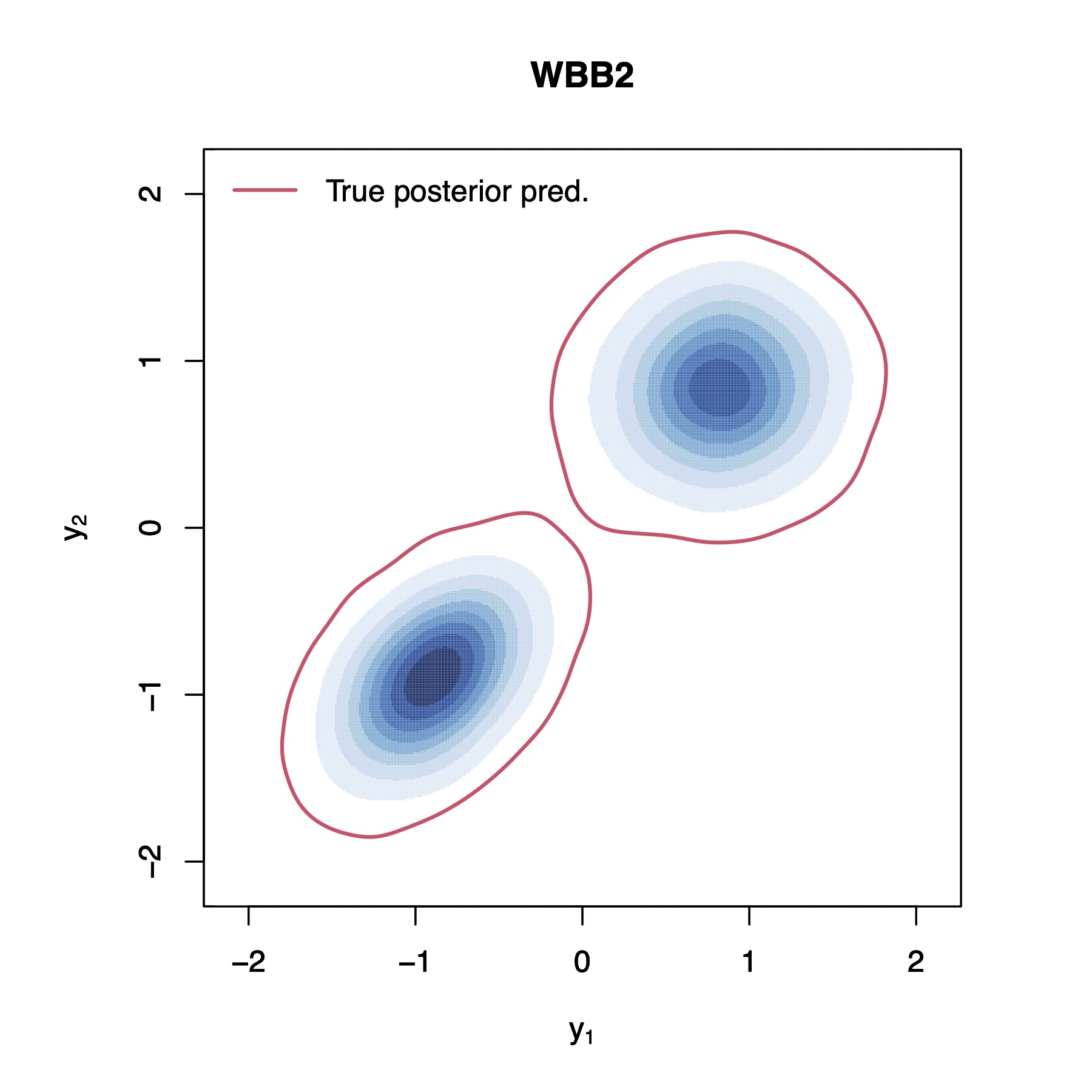}
\end{subfigure}
\hfill
\begin{subfigure}{0.45\textwidth}
    \includegraphics[width=\textwidth]{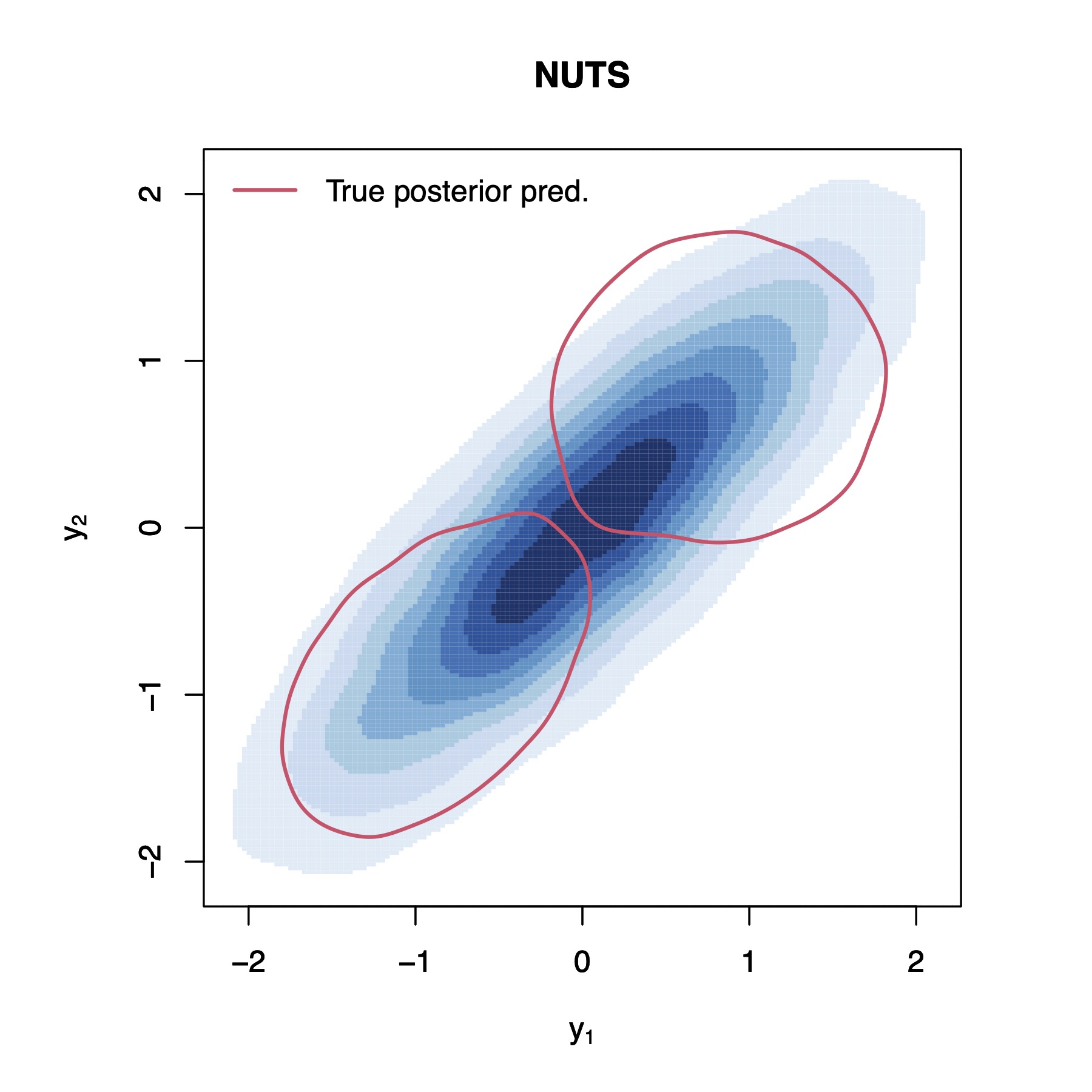}
\end{subfigure}
\hfill
\begin{subfigure}{0.45\textwidth}
    \includegraphics[width=\textwidth]{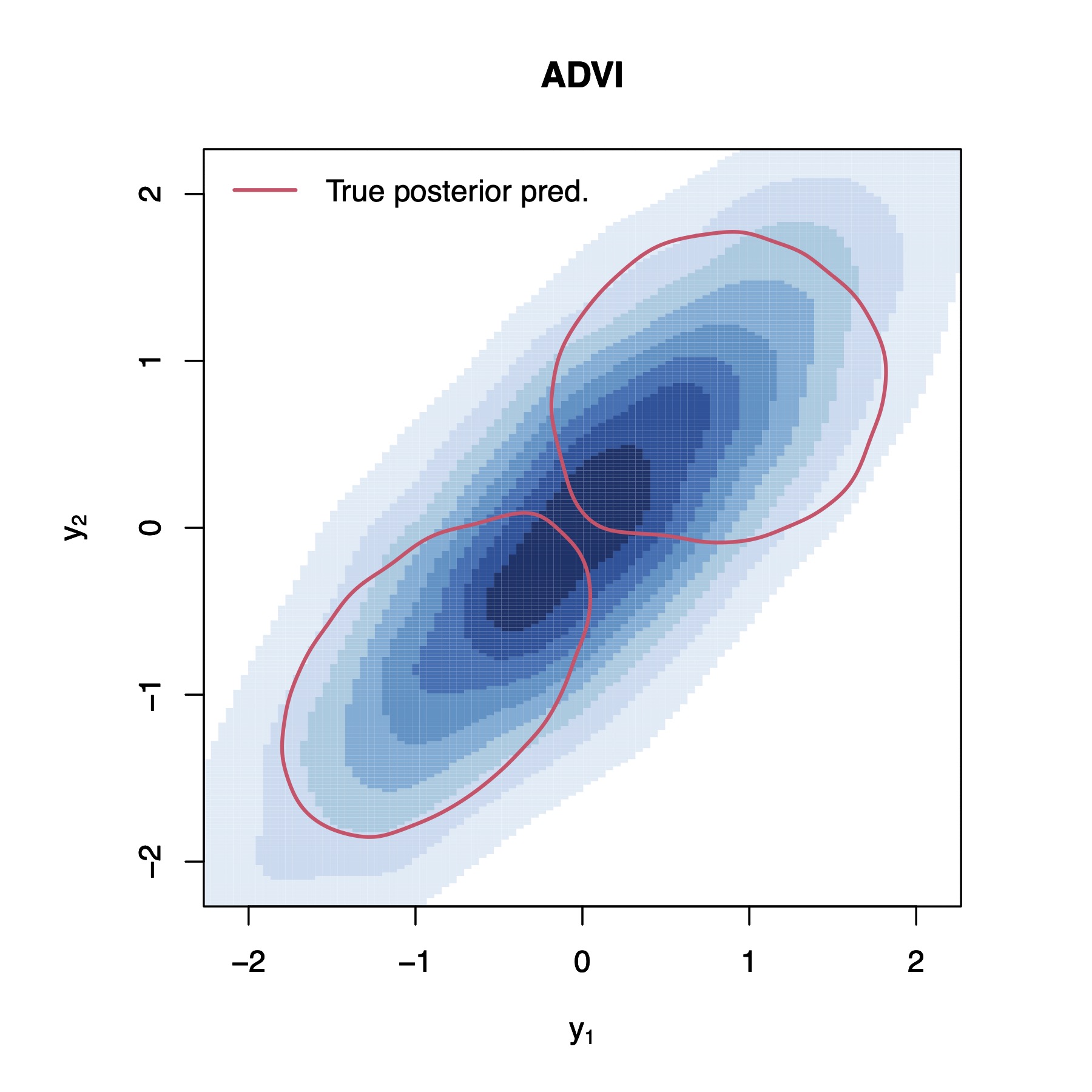}
\end{subfigure}
\caption{True Bayesian posterior predictive density (red contours) and its approximations (blue KDEs) obtained via BOB, WBB1, WBB2, NUTS, and ADVI, when $K=2$, $d = 10$, and $n = 50$.}
\label{fig:kdes_illustrative_supp}
\end{figure*}

Figure \ref{fig:kdes_illustrative_supp} presents additional posterior predictive density plots for our illustrative example, with $n = 50$, $d = 10$, and $K = 2$. We can observe similar patterns as in the main article. More precisely, both versions of WBB fail to capture the variance from the Bayesian posterior, while NUTS and ADVI produce unimodal posterior predictive distributions, failing to identify the clusters in the data. It is clear that BOB offers the best approximation to the Bayesian posterior predictive distribution and the best uncertainty quantification. 

\begin{figure*}[!htp]
\centering
\begin{subfigure}{0.45\textwidth}
    \includegraphics[width=\textwidth]{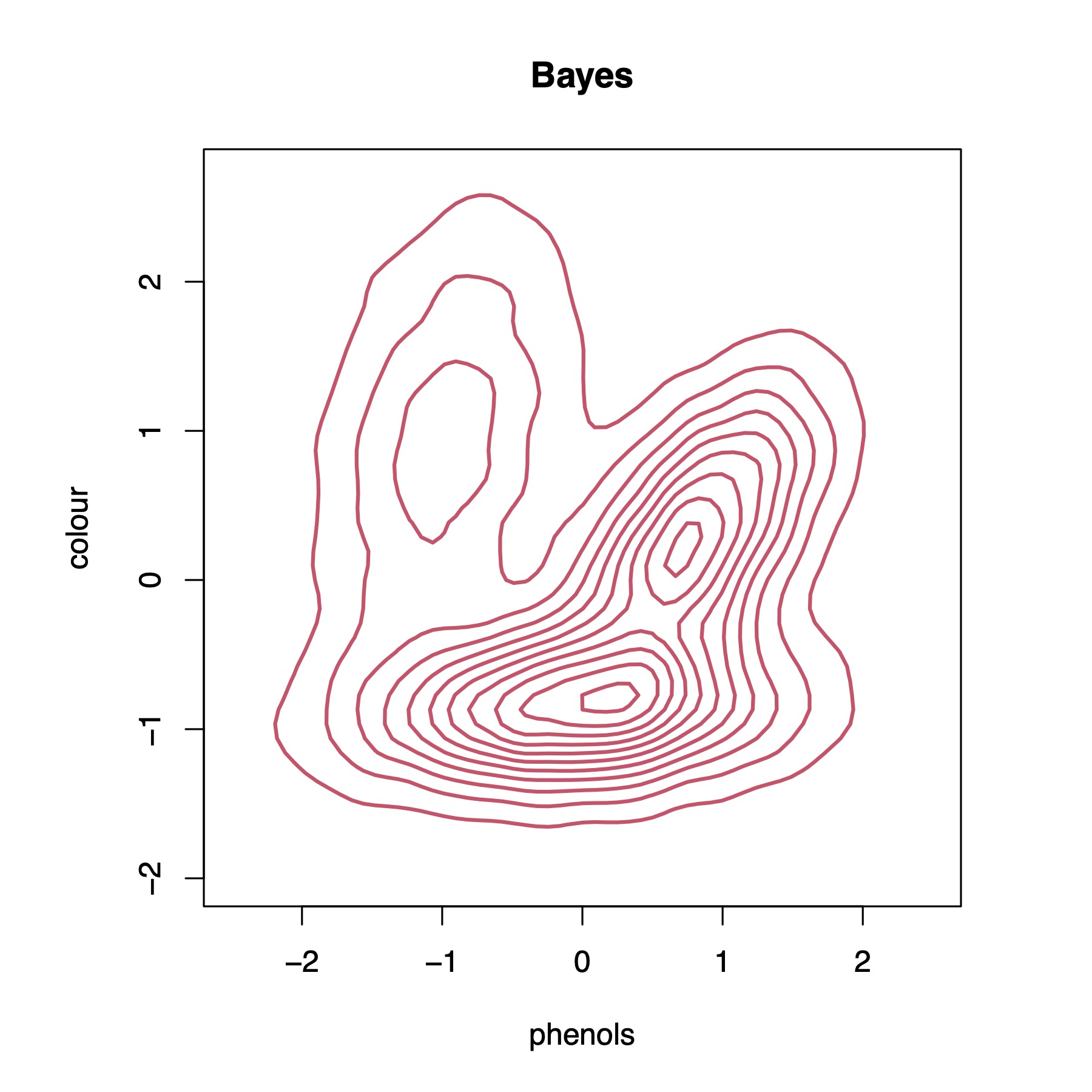}
\end{subfigure}
\hfill
\begin{subfigure}{0.45\textwidth}
    \includegraphics[width=\textwidth]{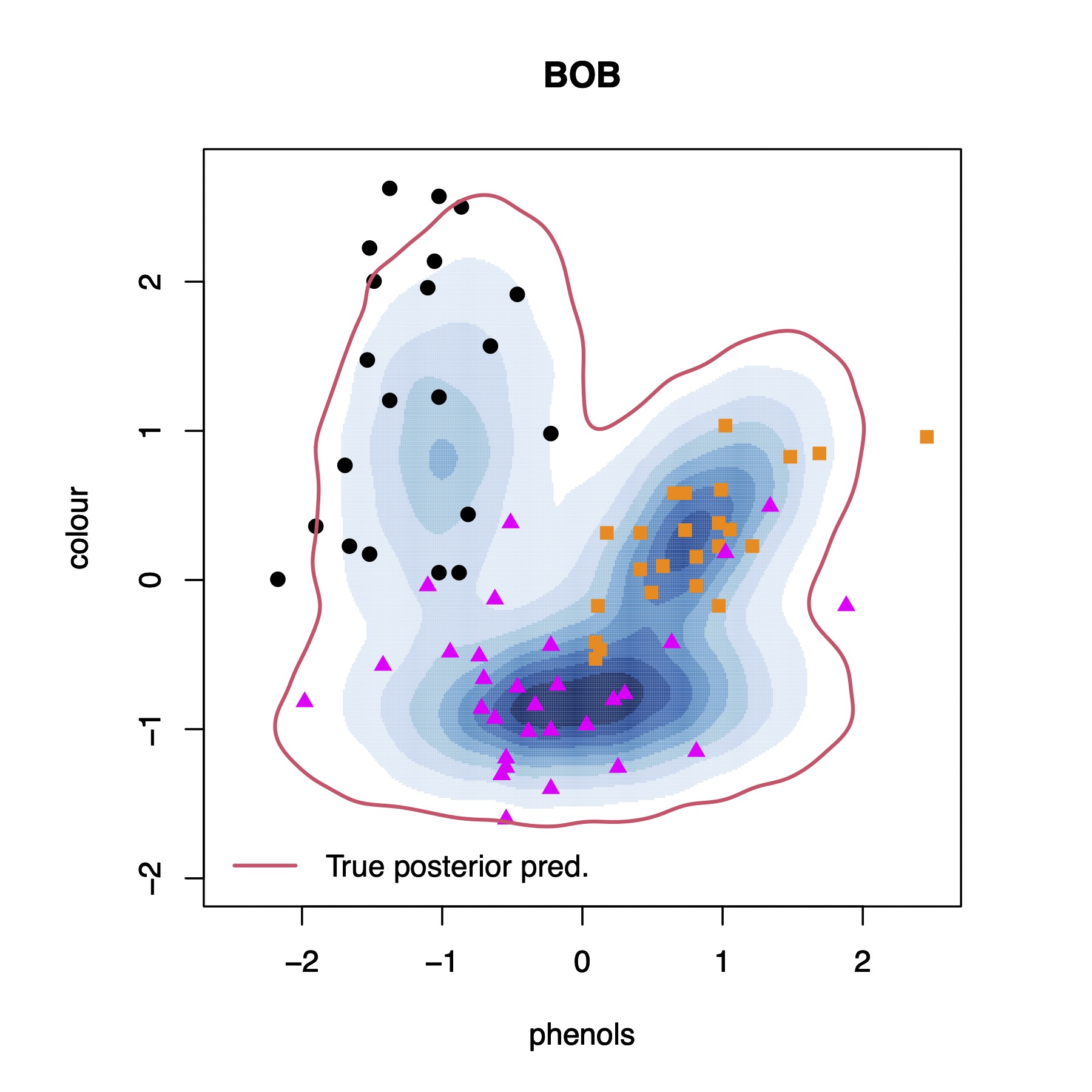}
\end{subfigure}
\hfill
\begin{subfigure}{0.45\textwidth}
    \includegraphics[width=\textwidth]{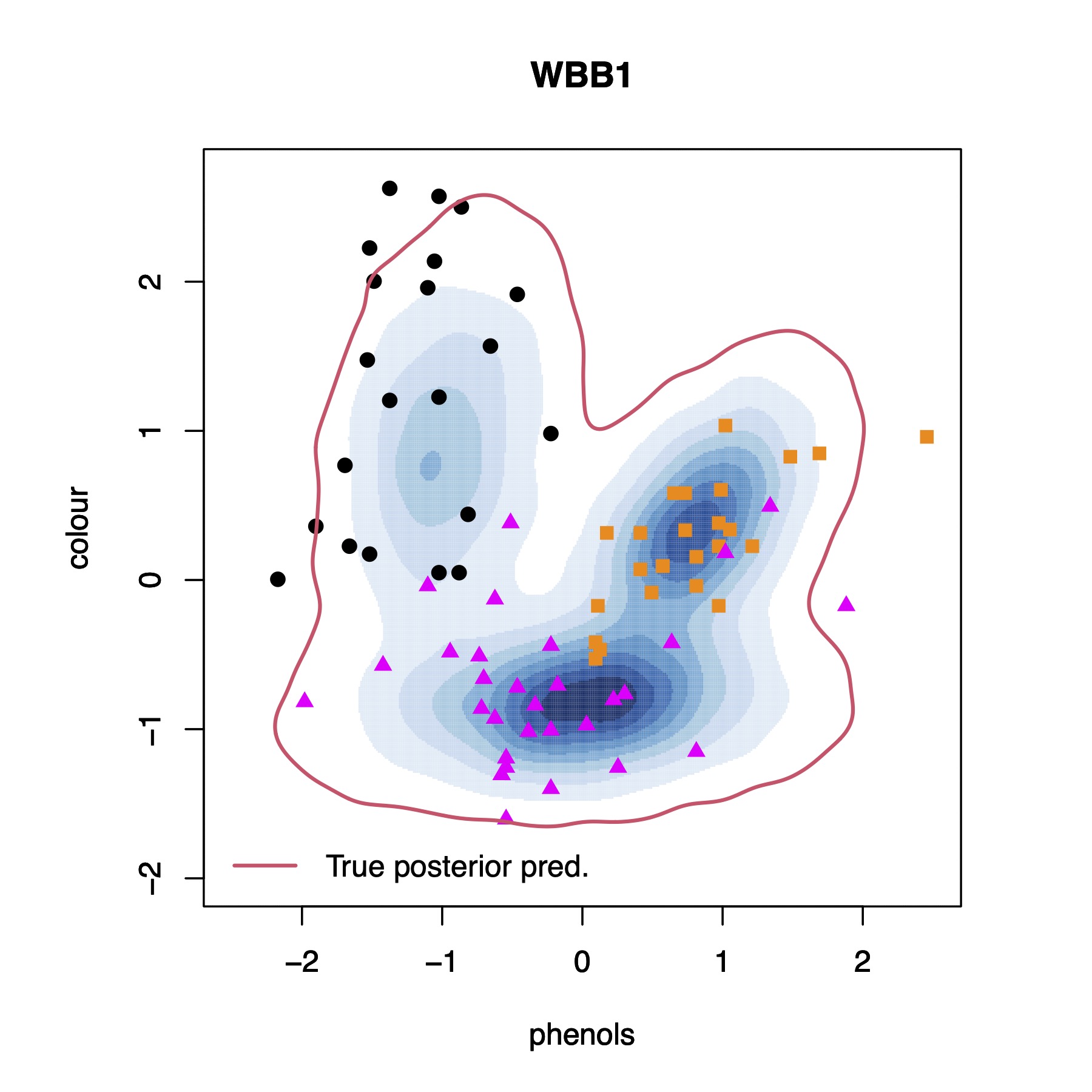}
\end{subfigure}
\hfill
\begin{subfigure}{0.45\textwidth}
    \includegraphics[width=\textwidth]{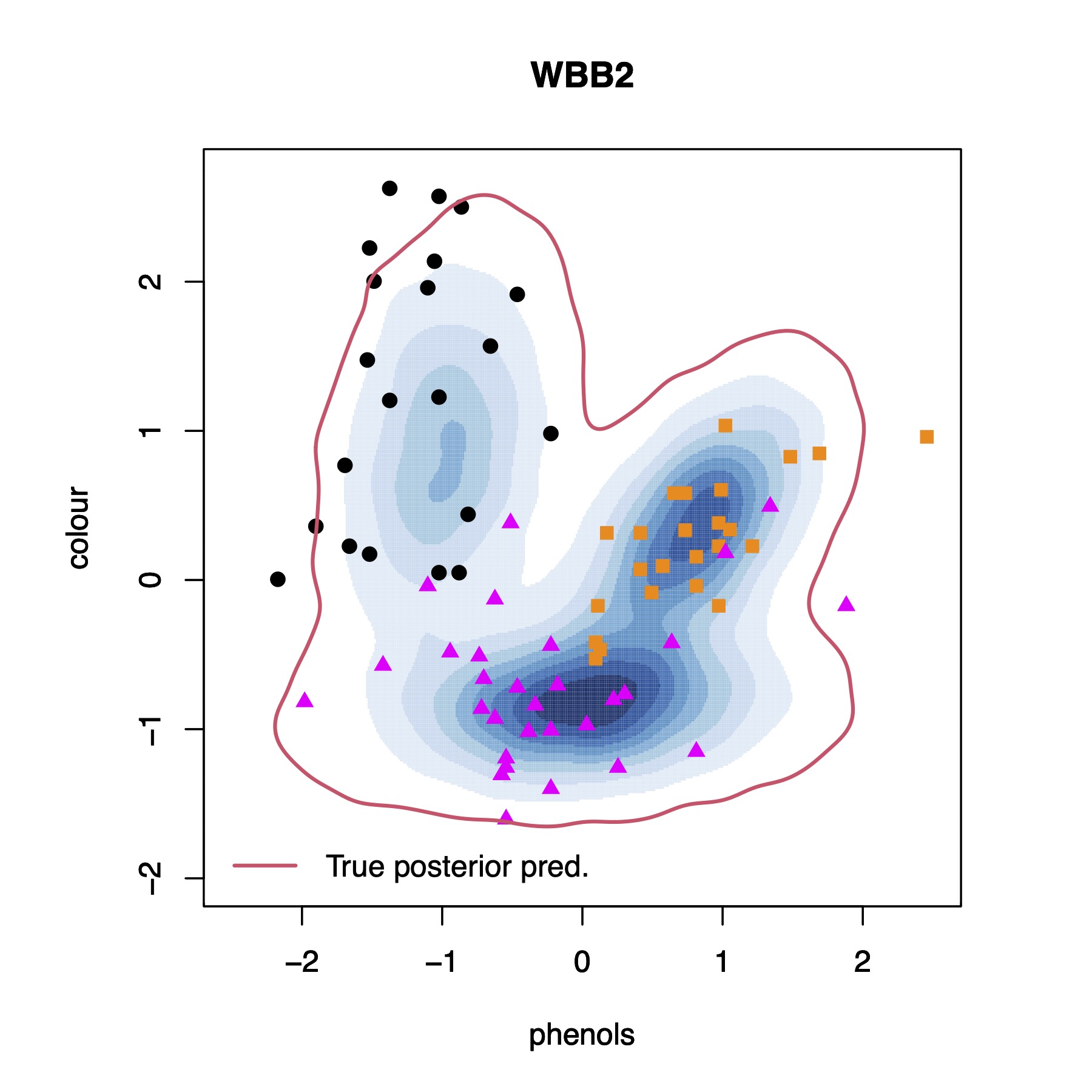}
\end{subfigure}
\hfill
\begin{subfigure}{0.45\textwidth}
    \includegraphics[width=\textwidth]{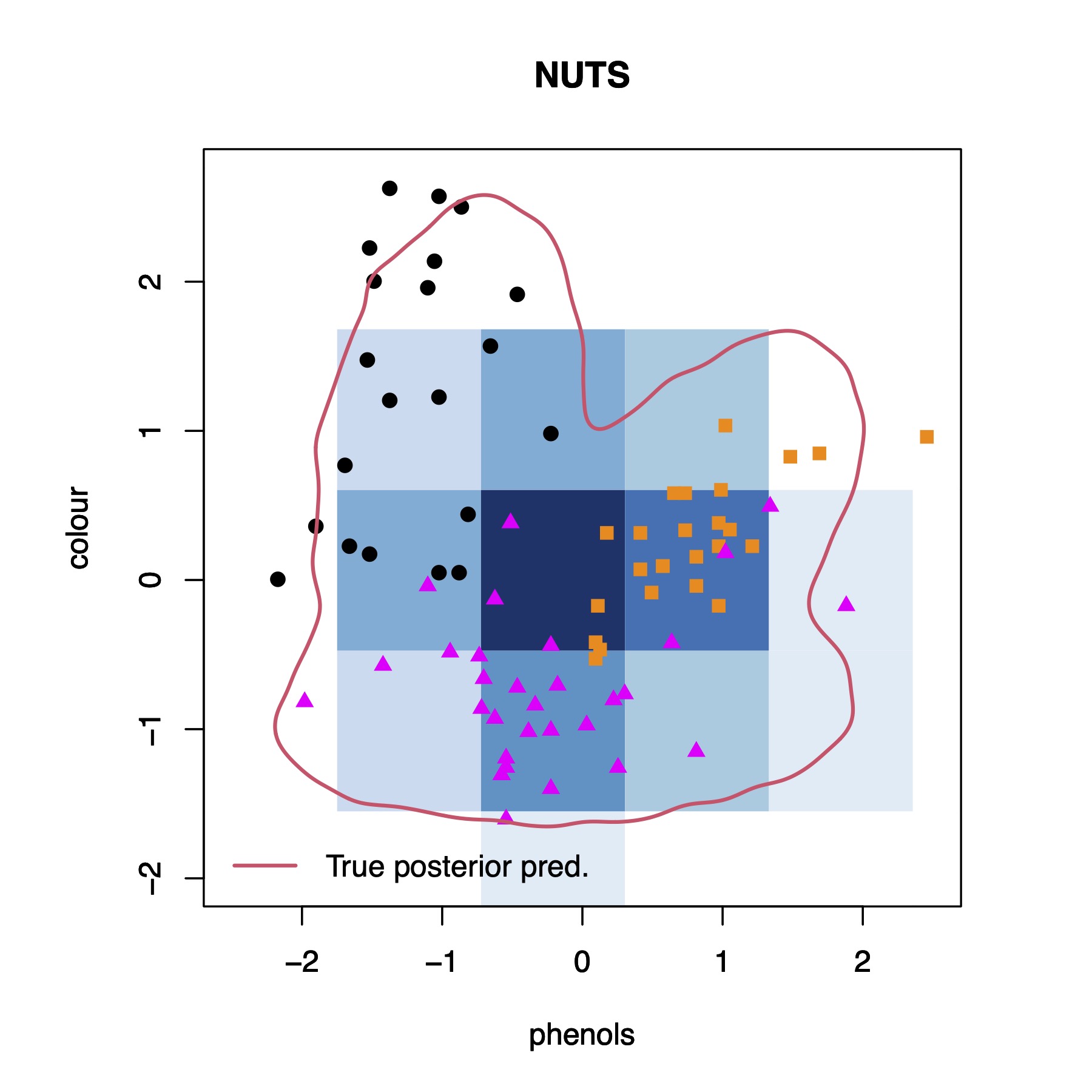}
\end{subfigure}
\hfill
\begin{subfigure}{0.45\textwidth}
    \includegraphics[width=\textwidth]{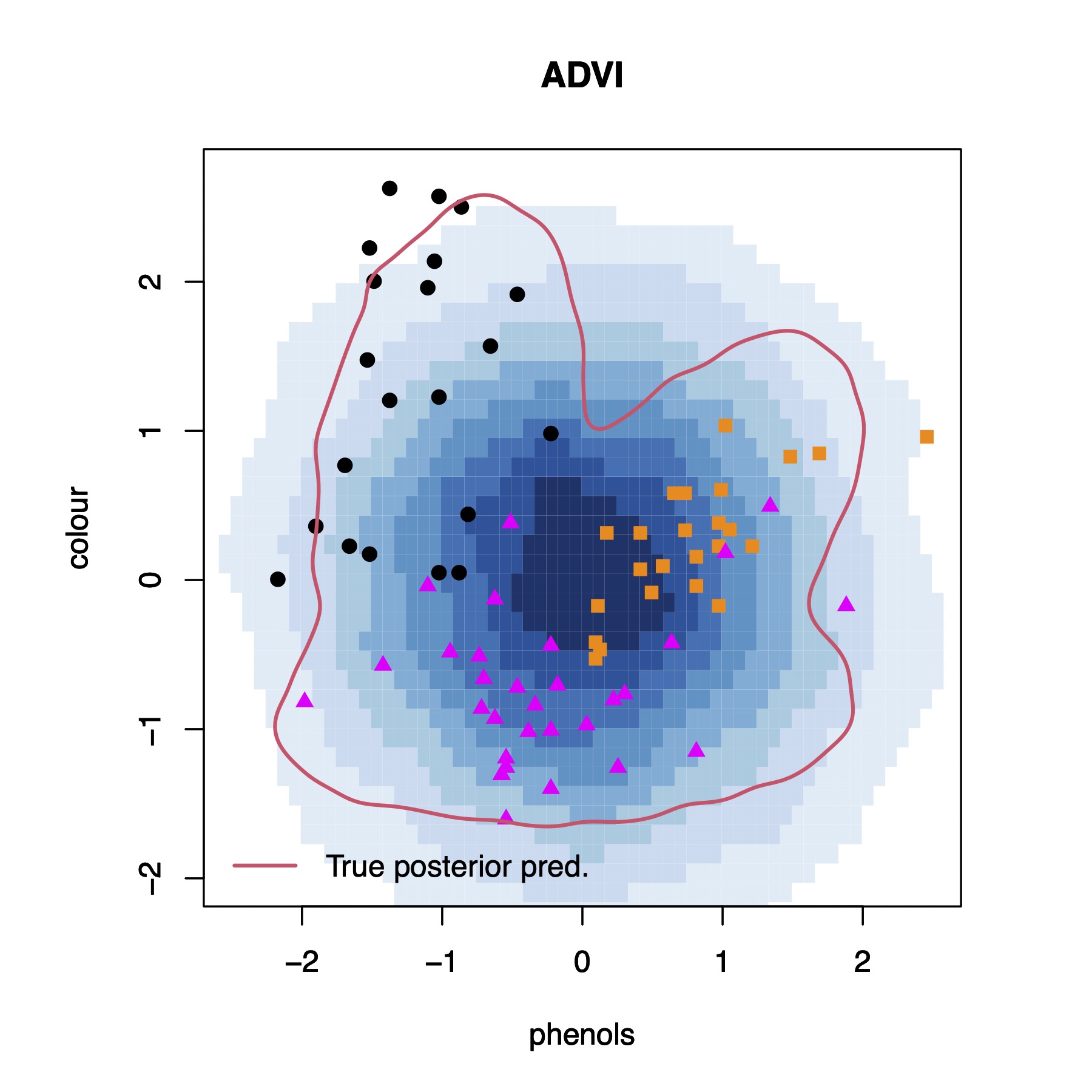}
\end{subfigure}
\caption{True Bayesian posterior predictive density (red contours) and its approximations (blue KDEs) obtained via BOB, WBB1, WBB2, NUTS, and ADVI, for additional variables from the wine data. Contours and KDEs were computed using only the training data. Scatter plots depict the held-out data. Orange squares, pink triangles and dark circles represent Barolo, Grignolino, and Barbera wines, respectively.}
\label{fig:supp_densities_wine}
\end{figure*}

\begin{figure*}[!htp]
\centering
\begin{subfigure}{0.45\textwidth}
    \includegraphics[width=\textwidth]{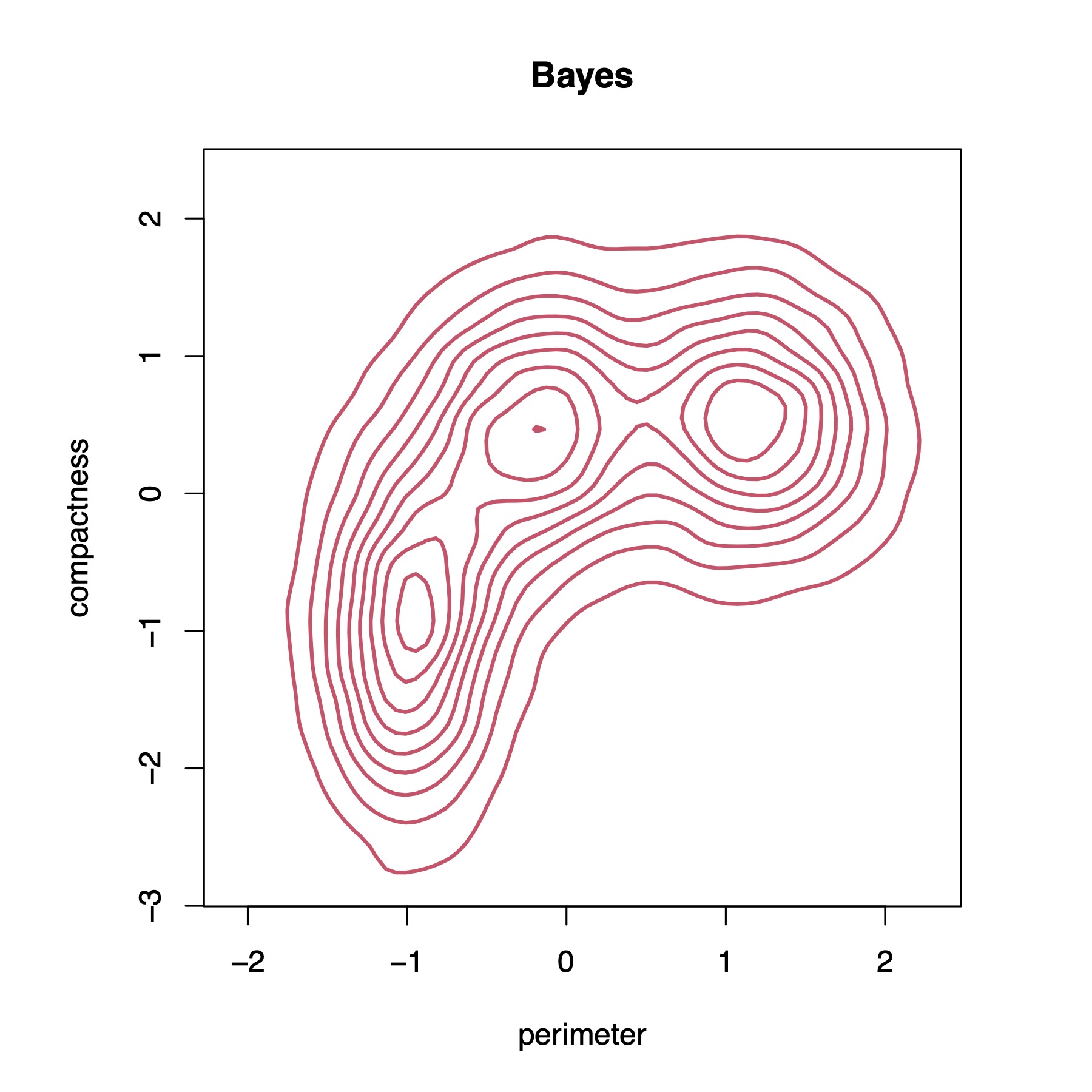}
\end{subfigure}
\hfill
\begin{subfigure}{0.45\textwidth}
    \includegraphics[width=\textwidth]{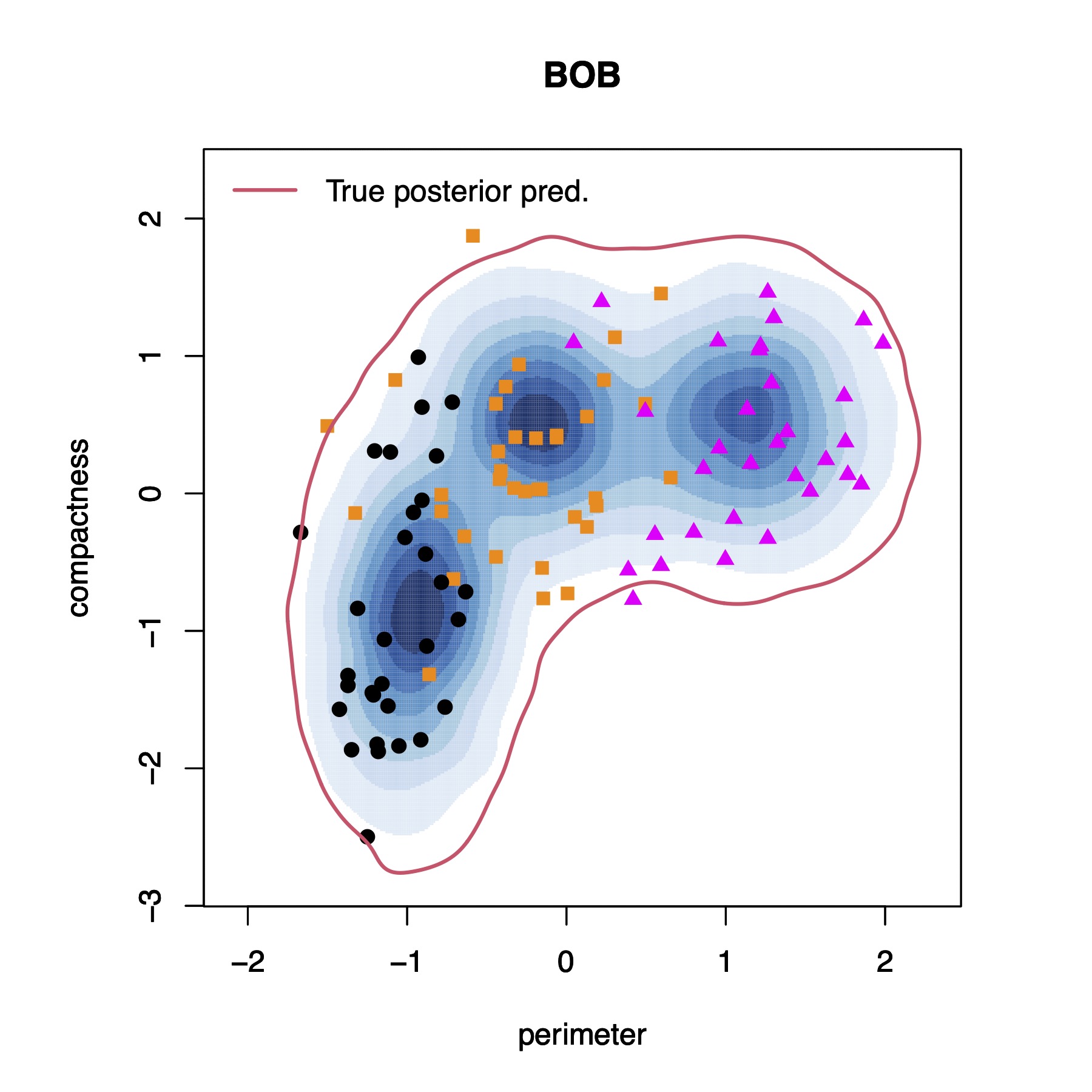}
\end{subfigure}
\hfill
\begin{subfigure}{0.45\textwidth}
    \includegraphics[width=\textwidth]{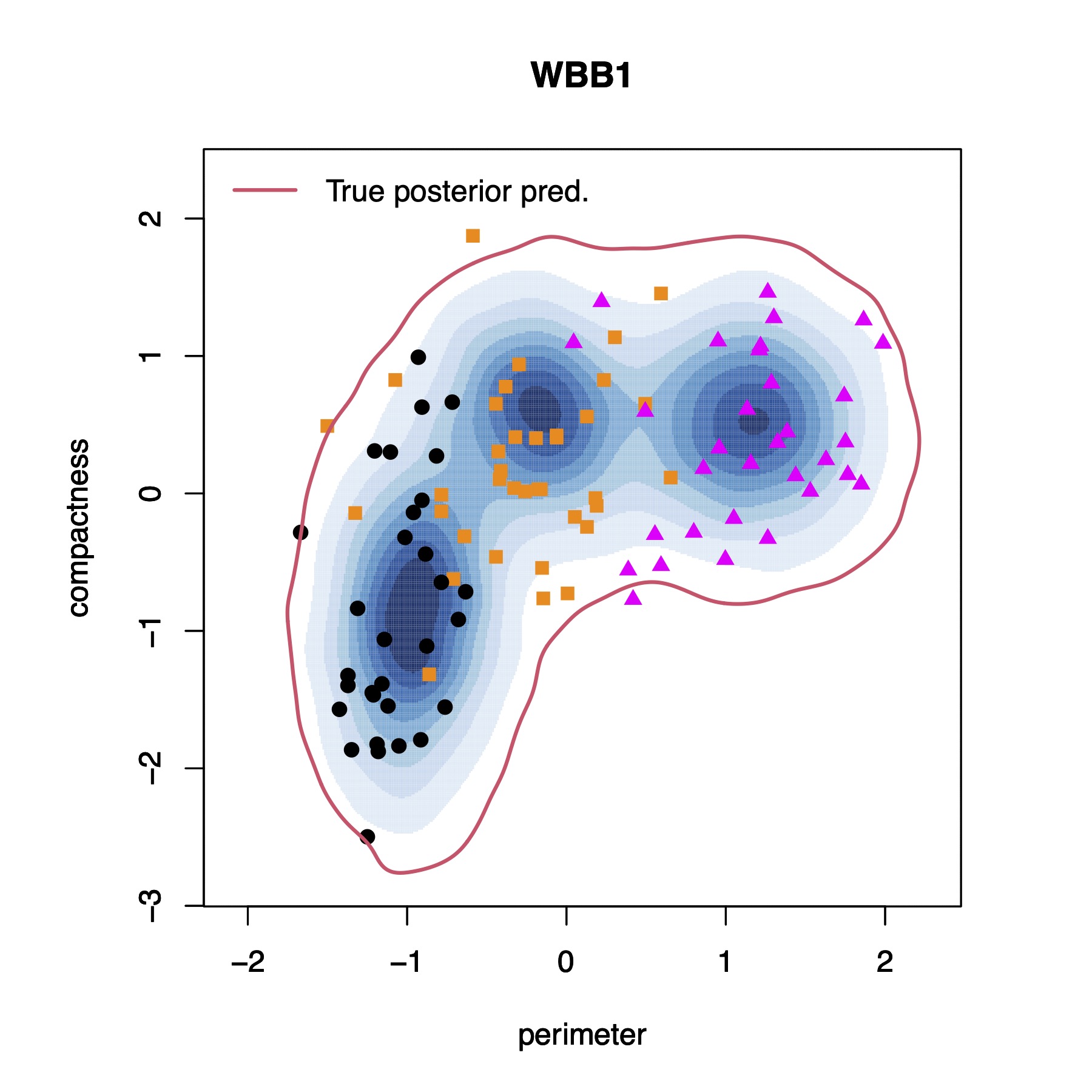}
\end{subfigure}
\hfill
\begin{subfigure}{0.45\textwidth}
    \includegraphics[width=\textwidth]{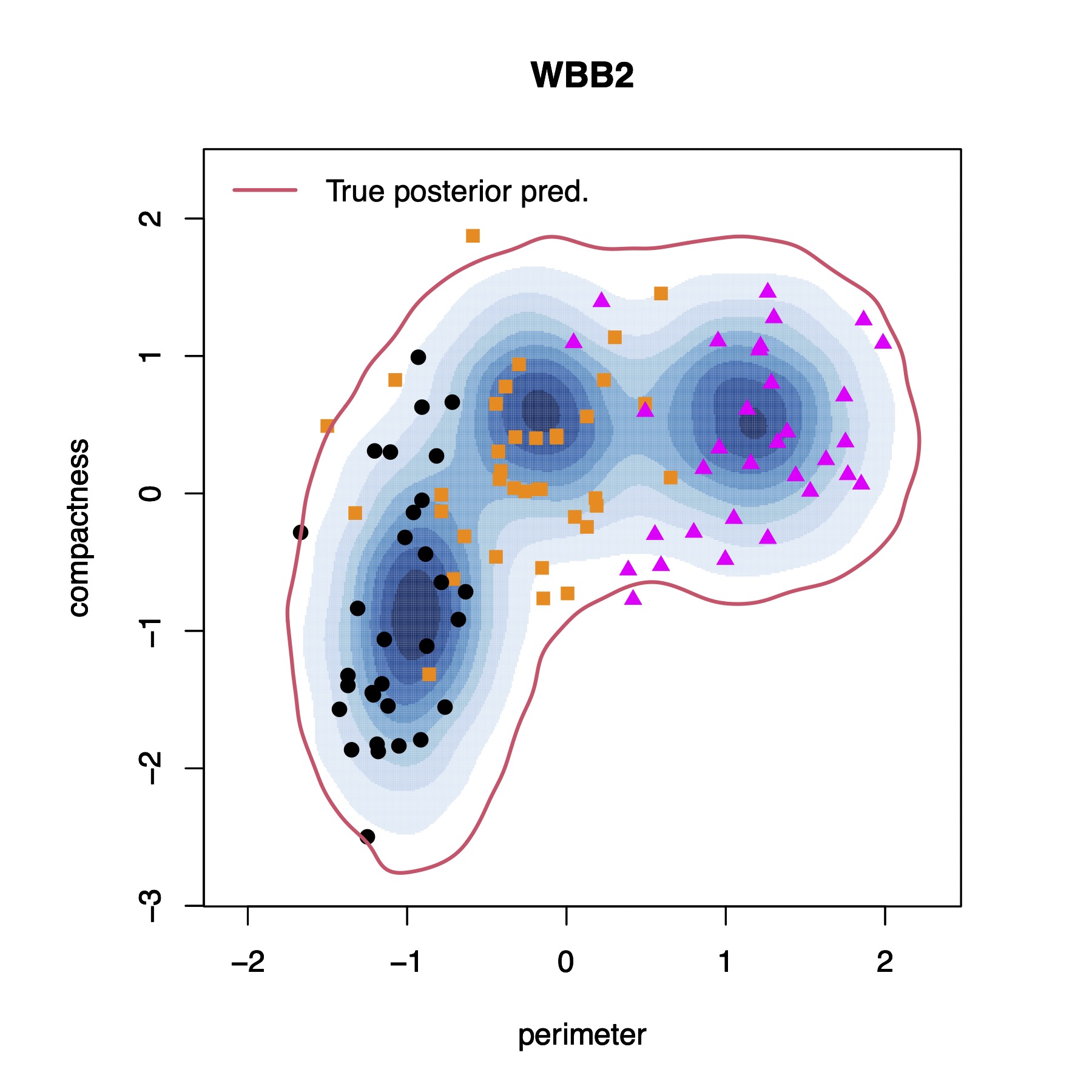}
\end{subfigure}
\hfill
\begin{subfigure}{0.45\textwidth}
    \includegraphics[width=\textwidth]{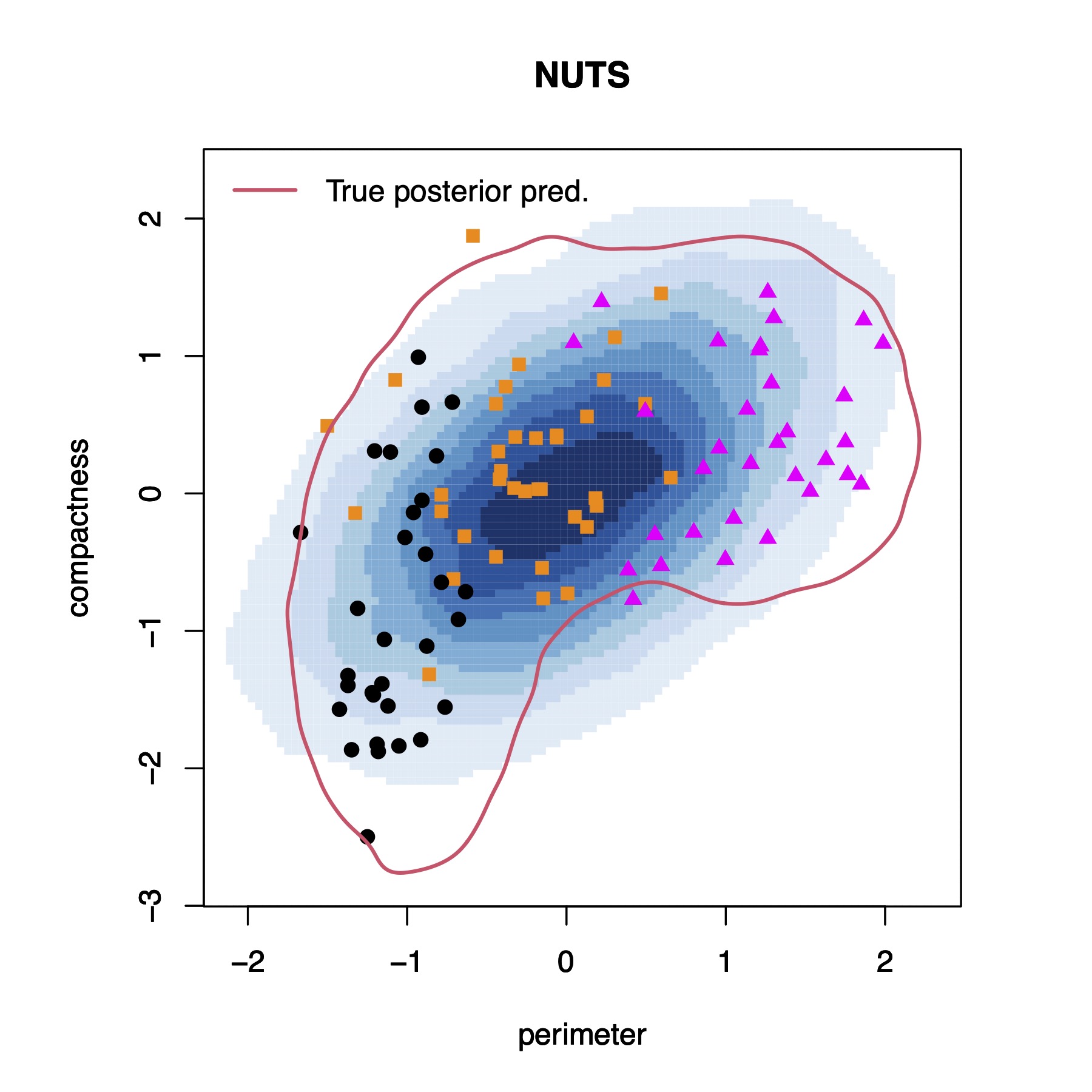}
\end{subfigure}
\hfill
\begin{subfigure}{0.45\textwidth}
    \includegraphics[width=\textwidth]{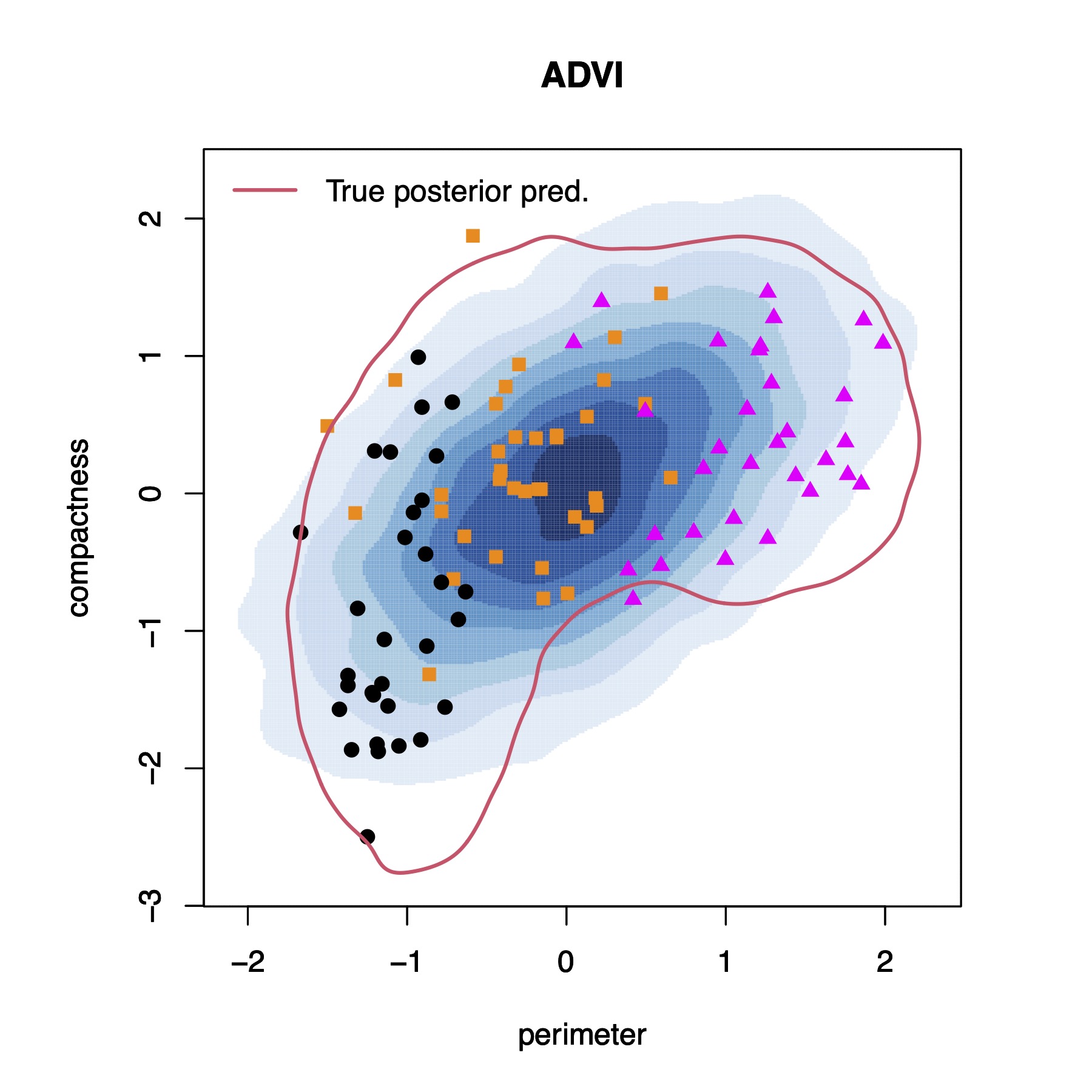}
\end{subfigure}
\caption{True Bayesian posterior predictive density (red contours) and its approximations (blue KDEs) obtained via BOB, WBB1, WBB2, NUTS, and ADVI, for additional variables from the seeds data. Contours and KDEs were computed using only the training data. Scatter plots depict the held-out data. Orange squares, pink triangles and dark circles represent  Kama, Rosa, and Canadian wheat kernels, respectively.}
\label{fig:supp_densities_seeds}
\end{figure*}

Figures \ref{fig:supp_densities_wine} and \ref{fig:supp_densities_seeds} present posterior density plots for additional variables from the wine and seeds datasets, respectively. Similar as in the main document, we can observe that both versions of WBB correctly capture the location of the three clusters. However, both versions of WBB produce overconfident posterior predictive distributions and do not capture the dispersion of the Bayesian posterior. NUTS and ADVI are unable to identify the three clusters in the data. Clearly, BOB produces the closest approximation to the Bayesian posterior predictive distribution and the best uncertainty quantification. 

\bibliographystyleSupp{sn-chicago.bst}
\bibliographySupp{references-supp.bib}

\end{document}